\documentclass[sigconf]{acmart}
\usepackage[linesnumbered,ruled,vlined]{algorithm2e}
\usepackage{tabularx}
\usepackage{booktabs}
\usepackage{array}
\usepackage{makecell}
\usepackage{multirow}
\usepackage{colortbl}
\usepackage{xcolor}
\usepackage{graphicx}
\usepackage{subcaption}
\usepackage{caption}
\usepackage{makecell}
\AtBeginDocument{%
  }

\setcopyright{acmlicensed}
\copyrightyear{2018}
\acmYear{2018}
\acmDOI{XXXXXXX.XXXXXXX}
\acmConference[Conference acronym 'XX]{Make sure to enter the correct
  conference title from your rights confirmation email}{June 03--05,
  2018}{}

\acmISBN{978-1-4503-XXXX-X/2018/06}

\begin{document}

\title{Less Is More — Until It Breaks: Security Pitfalls of Vision Token Compression in Large Vision-Language Models}
\author{Xiaomei Zhang}
\affiliation{%
  \institution{Griffith University}
  \city{}
  \country{}}
\email{xiaomei.zhang@griffithuni.edu.au}

\author{Zhaoxi Zhang}
\affiliation{%
  \institution{University of Technology Sydney}
  \city{}
  \country{}}
\email{zhaoxi.zhang-1@student.uts.edu.au}

\author{Leo Yu Zhang}
\affiliation{%
  \institution{Griffith University}
  \city{}
  \country{}}
\email{leo.zhang@griffith.edu.au}

\author{Yanjun Zhang}
\affiliation{%
  \institution{Griffith University}
  \city{}
  \country{}}
\email{yanjun.zhang@griffith.edu.au}

\author{Guanhong Tao}
\affiliation{%
  \institution{University of Utah}
  \city{}
  \country{}}
\email{guanhong.tao@utah.edu}

\author{Shirui Pan}
\affiliation{%
  \institution{Griffith University}
  \city{}
  \country{}}
\email{s.pan@griffith.edu.au}

\renewcommand{\shortauthors}{xx et al.}

\begin{abstract}
Visual token compression is widely adopted to improve the inference efficiency of Large Vision–Language Models (LVLMs), enabling their deployment in latency-sensitive and resource-constrained scenarios. However, existing work has mainly focused on efficiency and performance, while the security implications of visual token compression remain largely unexplored. In this work, we first reveal that visual token compression substantially degrades the robustness of LVLMs: models that are robust under uncompressed inference become highly vulnerable once compression is enabled. These vulnerabilities are state-specific; failure modes emerge only in the compressed setting and completely disappear when compression is disabled, making them particularly hidden and difficult to diagnose.
By analyzing the key stages of the compression process, we identify instability in token importance ranking as the primary cause of this robustness degradation. Small and imperceptible perturbations can significantly alter token rankings, leading the compression mechanism to mistakenly discard task-critical information and ultimately causing model failure. Motivated by this observation, we propose a Compression-Aware Attack (CAA) to systematically study and exploit this vulnerability. CAA directly targets the token selection mechanism and induces failures exclusively under compressed inference. We further extend this approach to more realistic black-box settings and introduce Transfer CAA (T-CAA), where neither the target model nor the compression configuration is accessible. Experimental results show that compression-induced security risks persist even under these practical settings. We further evaluate potential defenses and find that they provide only limited protection. Extensive experiments across multiple models, datasets, and compression methods consistently demonstrate that visual token compression significantly undermines model robustness, exposing a previously overlooked trade-off between efficiency and security.
\end{abstract}






\maketitle

\section{Introduction}
\begin{figure}[t]
    \centering
    \includegraphics[width=0.90\linewidth]{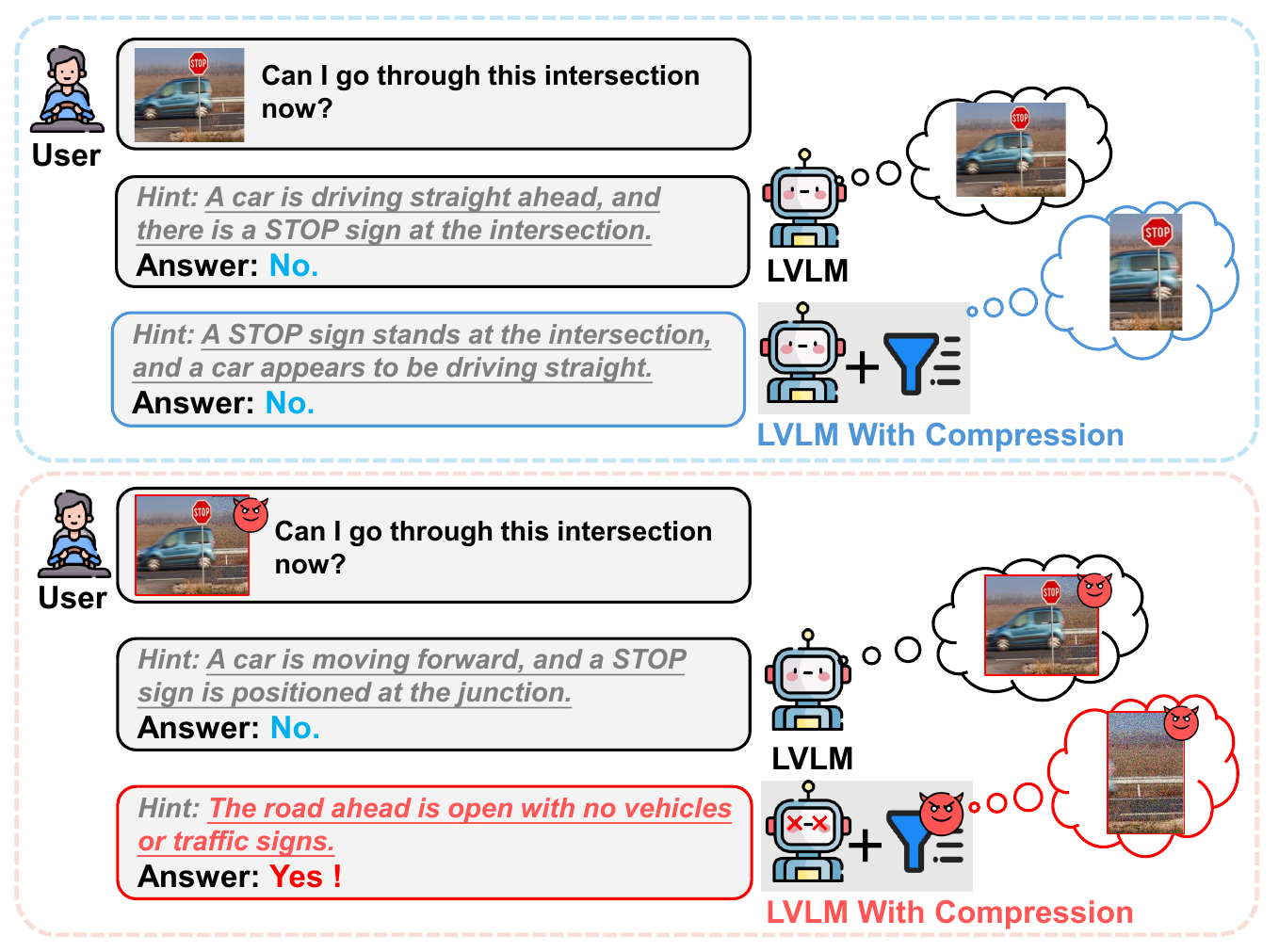}
    \caption{
    Compression-induced safety risk in autonomous driving.
    For clean inputs, both the compressed and uncompressed models attend to critical cues and make safe decisions. Under adversarial input (bottom), the uncompressed model remains correct, whereas compression discards critical visual cues and produces an unsafe “Yes” response, leading to a severe accident.
    }
    \label{fig:caa_senario}
\end{figure}

Recent years have witnessed rapid progress in Large Vision-Language Models (LVLMs), such as LLaVA~\cite{liu2023visual} and the Qwen-VL~\cite{bai2025qwen2} series, which integrate pretrained vision encoders with Large Language Models (LLMs) to powerful multimodal perception and reasoning. These models achieve state-of-the-art performance on a wide range of tasks, including image captioning and visual question answering, and are increasingly deployed in real-world applications such as autonomous driving~\cite{cui2024survey, cao2024maplm}, embodied agents~\cite{ma2024survey}, and multimodal assistants~\cite{hurst2024gpt, team2024gemini}.
To align with the next-token prediction paradigm of LLMs, visual inputs are encoded into sequences of visual tokens and concatenated with text tokens for joint modeling \cite{bai2025qwen2, zhu2024minigpt, dai2023instructblip, liu2023visual, chen2024far}.
While this design effectively leverages the reasoning capabilities of LLMs, it also inherits the \textit{quadratic computational and memory cost} of self-attention with respect to the input token length~\cite{Yang2024VisionZipLI, He_2025_ICCV}. This challenge is particularly acute for LVLMs, where high-resolution images and long videos can generate tens of thousands of visual tokens. For instance, Qwen-2.5VL~\cite{bai2025qwen2} supports up to 16k tokens for multi-image or video inputs, while LongVA~\cite{Zhang2024LongCT} encodes a 128-frame video into over 18k visual tokens. Such token explosion severely hinders the deployment of LVLMs in latency-sensitive and resource-constrained environments~\cite{yao2025efficient, cui2024survey}.

To address this challenge, visual token compression has emerged as a promising inference-time solution. This paradigm is motivated by two key observations:
(i) visual data exhibits substantial spatial redundancy, where neighboring regions often convey highly similar information \cite{Yang2024VisionZipLI}; 
and (ii) vision-language tasks often require only a small subset of task-relevant visual evidence for a given text prompt \cite{ Chen2024AnII, Zhang2024SparseVLMVT, Xing2024PyramidDropAY, zhao2024stitch, lin2025boosting, ye2025fit, Zhuang2025ST3AM}. 
Accordingly, token compression methods rank visual tokens by estimated importance and retain only the most informative ones, discarding redundant tokens without retraining the model \cite{ Chen2024AnII, Zhang2024SparseVLMVT, Xing2024PyramidDropAY, zhao2024stitch, lin2025boosting, ye2025fit, Zhuang2025ST3AM, Yang2024VisionZipLI}. 
By adjusting the token retention rate (i.e., the percentage of tokens preserved during inference), these methods offer a flexible trade-off between efficiency and accuracy, making them well suited for real-time inference, edge deployment, and dynamic workload adaptation \cite{jie2024token, zhu2025fastcache}.

Despite its effectiveness in improving efficiency, the impact of visual token compression on the \emph{robustness} of LVLMs remains largely unexplored. 
In this work, we uncover a concerning phenomenon: \textit{visual token compression fundamentally alters the robustness behavior of LVLMs}. While models are generally robust to small random perturbations under full-token inference, they become highly vulnerable once compression is activated, with vulnerability increasing as the token retention rate decreases.
This effect poses severe risks in practical systems, such as dynamic systems, where  the visual token retention rate is often automatically adjusted according to system load, or computational budgets \cite{jie2024token, zhu2025fastcache, fan2025timebill}. 
Under high-load and low-retention settings, compressed inference can exhibit severe performance failures that do not occur under full-token inference.
This discrepancy complicates debugging and robustness evaluation, as adversarial inputs may appear benign during standard (uncompressed) offline testing yet reliably induce failures when the system operates under resource constraints. Such \emph{compression-specific failures} introduce a new and stealthy attack surface in the LVLM inference pipeline.


To study compression-specific vulnerabilities, adversarial attacks must decouple failures caused by compressed inference from model robustness.
However, existing adversarial attacks~\cite{madry2018towards, bailey2024image, qi2024visual} are designed for fixed, standard inference settings and do not account for the presence of token compression. Consequently, they fail to capture failures that arise exclusively under compressed inference. Moreover, the non-differentiable token selection decisions introduced by compression further limit the applicability of classical end-to-end gradient-based attacks~\cite{madry2018towards, bailey2024image}.

To bridge this critical gap, we propose Compression-Aware Attack (CAA), a new adversarial framework designed to expose vulnerabilities introduced by visual token compression.
Our key insight is that compression relies on importance-based token selection, and that the \textit{relative ordering of token importance is inherently fragile}. Small, imperceptible perturbations can alter this ordering, leading compression to preserve less relevant tokens while discarding task-critical evidence, thereby inducing failures under compressed inference.
Building on this insight, CAA targets the compression mechanism through three tightly coupled components.
First, CAA employs a \textit{selective perturbation} strategy that constrains adversarial perturbation to regions originally assigned low importance, reshaping compression behavior while minimally affecting uncompressed inference.
Second, a \textit{hierarchical ranking objective} explicitly enforces adversarial reordering of token importance, enabling fine-grained manipulation of compression decisions. Finally, a \textit{semantic erasure module} corrupts the semantic content of tokens that survive adversarial compression, further amplifying performance degradation under compressed inference.

We further extend CAA to realistic black-box settings, termed T-CAA, where neither the target model nor the compression configuration is known.
Adversarial examples are generated on a surrogate model and then transferred to the target model.
In this setting, T-CAA optimizes adversarial perturbations to induce consistent manipulation of token importance across a range of plausible compression layers.
To address model uncertainty, we introduce an uninformative image border as the perturbation region, mitigating reliance on model-specific least-important regions.
Using two jointly optimized universal templates that elevate border-token importance while mildly suppressing the original image content, T-CAA reliably forces the retention of uninformative tokens, demonstrating that compression-induced risks persist  under practical black-box constraints.

We conduct extensive experiments across multiple state-of-the-art LVLMs, diverse benchmarks, and representative token compression paradigms.
Our results show that existing vision token compression methods suffer severe robustness degradation under CAA, revealing an inherent \emph{efficiency-security trade-off}. 
We further evaluate potential defense strategies and find that they provide limited protection.
Our main contributions are summarized as follows:
\begin{itemize}
    \item 
   We show that visual token compression fundamentally degrades the robustness of LVLMs: models that are robust under full-token inference become highly vulnerable once compression is enabled. We identify token importance ranking instability as a key mechanism underlying this vulnerability.

    \item 
   We propose Compression-Aware Attack (CAA), a new adversarial framework that manipulates token importance ranking to induce failures only under compressed inference.
    We further extend CAA to realistic black-box settings, demonstrating that compression-induced vulnerabilities persist even when the target model and compression configuration are unknown.

    \item 
    Through extensive evaluations across models, benchmarks, and compression methods, we demonstrate severe and consistent compression-specific robustness degradation. CAA induces an average Compression Sensitivity Gap (CSG) of 47.61\%, compared to only 2.36\% induced by baseline attacks. These results expose a previously overlooked efficiency-security trade-off in visual token compression and highlight the need for more robust compression designs\footnote{Our code: \url{https://anonymous.4open.science/r/Compression_Aware_Attack-368D}}.
\end{itemize}

\section{Background}
\label{sec:background}
\subsection{Large Vision Language Models}
\label{sec:lvlm}

Large Vision-Language Models (LVLMs), such as LLaVA~\cite{liu2023visual} and Qwen-VL\footnote{Throughout this work, we use Qwen and Qwen-VL interchangeably to refer to the Qwen2.5-VL model.}~\cite{bai2025qwen2}, integrate visual perception and language reasoning within a unified framework~\cite{chen2024far}, enabling multimodal understanding and generation across a broad spectrum of tasks.

\noindent\textbf{Input Processing.}
An LVLM $f$ takes an image $I$ and a text prompt $T$ as input, and generates a textual output sequence $Y=f(I,T)$. 
The vision encoder $f_{\text{vis}}$ partitions the image $I$ into visual patches $P=\{P_j\}_{j=1}^{n_V}$ and encodes them into visual features, which are further mapped by a projection module $f_{\text{proj}}$ into vision tokens $V = \{v_j\}_{j=1}^{n_V} = f_{\text{proj}}(f_{\text{vis}}(I))$, 
where each token $v_j \in \mathbb{R}^d$ corresponds to a patch $P_j$, and $d$ is the hidden dimension of the language model $f_\text{llm}$.
The text prompt\footnote{For simplicity, we use $T$ to denote both the textual input and the corresponding text tokens.} is tokenized into $T = \{t_i\}_{i=1}^{n_T}$.
The vision tokens and text tokens are concatenated to form the initial input sequence
$X^{(0)} = [V, T]$, with total length $ n = n_V + n_T$. 

\noindent\textbf{Attention Computation.}
The language model processes the input through $L$ transformer layers. At layer $l$, given hidden states $X^{(l)}=\{x_i^{(l)}\}_{i=1}^{n}$ with $x_i^{(l)}\in\mathbb{R}^d$, the query and key vectors are computed as
\begin{equation}
q_i^{(l)} = x_i^{(l)} W_Q^{(l)}, \quad
k_i^{(l)} = x_i^{(l)} W_K^{(l)},
\label{eq:qk_vector}
\end{equation}
where $W_Q^{(l)}, W_K^{(l)} \in \mathbb{R}^{d \times d_k}$ are learnable projection matrices and $d_k$ is the per-head attention dimension.
The attention weight from token $i$ to token $i'$ is
\begin{equation}
a_{i,i'}^{(l)} =
\operatorname{softmax}_{i'}\!\left(
\frac{q_i^{(l)} (k_{i'}^{(l)})^\top}{\sqrt{d_k}}
\right).
\label{eq:attn_score}
\end{equation}
The attention output is computed as
$o_i^{(l)} = \sum_{i'=1}^{n} a_{i,i'}^{(l)} \, x_{i'}^{(l)} W_V^{(l)}$,
where $W_V^{(l)} \in \mathbb{R}^{d \times d_k}$.
The multi-head attention applies this operation in parallel across multiple heads, followed by linear projection and a feed-forward network to produce the next-layer hidden states $X^{(l+1)}$.
After $L$ layers, the final hidden states $X^{(L)}$ are used for autoregressive decoding to generate the output sequence $Y$ until an end-of-sequence (EOS) token or a maximum length is reached.


\noindent \textbf{Inference Complexity.}
Self-attention dominates the computational cost of transformer-based LVLMs, with a quadratic complexity of $\mathcal{O}(n^2)$ in the input length $n$. In multimodal inference, this cost is largely driven by visual tokens since $n_V > n_T$. Empirically, the number of vision tokens $n_V$ often accounts for a substantial portion of the input sequence (e.g., around 80\% in common multimodal reasoning tasks~\cite{shao2025tokens}), making them the primary bottleneck for inference efficiency.

\begin{figure}[t]
    \centering
    \includegraphics[width=0.98\linewidth]{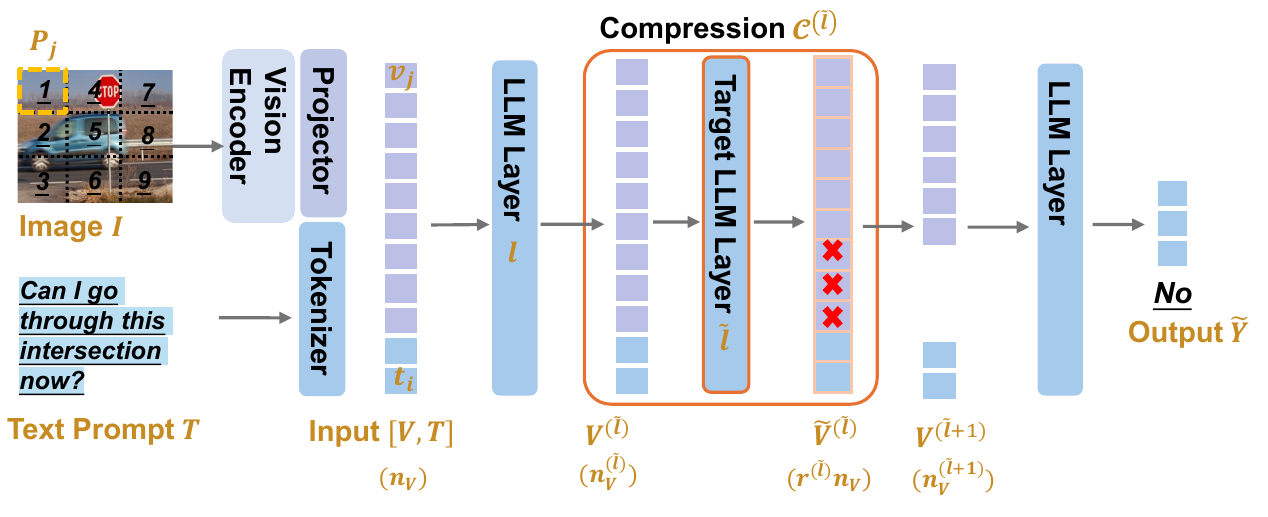}
    \caption{
    LVLM inference with vision token compression. Multimodal input $(I,T)$ enters the LLM. At compression layer $\mathcal{C}^{(\tilde{l})}$, the module evaluates token importance ($F_s$) and retains tokens based on retention rate $r^{(\tilde{l})}$ to produce the compressed sequence $\tilde{V}^{(\tilde{l})}$, which forms the input for the subsequent layer $V^{(\tilde{l}+1)}$. The bottom row shows the visual token count.
    }
    \label{fig:inference_comp}
\end{figure}

\subsection{Vision Token Compression}
\label{sec:vision_comp}
Visual token compression has emerged as a promising approach for alleviating the substantial computational burden caused by the large number of visual tokens in LVLMs without requiring retraining. 
Motivated by the substantial spatial redundancy in visual data, \textit{text-agnostic} compression methods prune or merge visual tokens based on attention scores from the vision encoder or embedding similarity~\cite{radford2021learning, Yang2024VisionZipLI, zoudon, shang2025llava}. However, these methods do not account for the task-specific visual requirements induced by language prompts.
A more effective approach is \textit{text-guided} compression, which leverages textual context to perform more aggressive yet accuracy-preserving token reduction \cite{Chen2024AnII, Zhang2024SparseVLMVT, Xing2024PyramidDropAY, zhao2024stitch, lin2025boosting, ye2025fit, Zhuang2025ST3AM}. Accordingly, we focus on this compression in this work. More related work is provided in Appendix~\ref{sec:related_work}.

A compression mechanism $\mathcal{C}$ is characterized by three components:
(i) \emph{where to compress}, specified by a set of target LLM layers
$\tilde{L} = \{\tilde{l}_1, \tilde{l}_2, \cdots\}$;
(ii) \emph{how to score visual tokens}, given by an importance function $F_s$;
and (iii) \emph{how many tokens to retain}, specified by a retention schedule
$R = { r^{(\tilde{l})} \in (0,1] }_{\tilde{l} \in \tilde{L}}$, where $r^{(\tilde{l})}$ denotes the fraction of visual tokens retained at layer $\tilde{l}$ relative to the original vision token count $n_V$.
Given input $(I, T)$ and a compression mechanism $\mathcal{C}=\{\tilde{L}, R, F_s\}$, as illustrated in Fig.~\ref{fig:inference_comp}, the LVLM inference with compression is
$\tilde{Y} = f(I, T; \mathcal{C}).$

At a compression layer $\tilde{l} \in \tilde{L}$, the input sequence is
$ [V^{(\tilde{l})}, T]$,
where $V^{(\tilde{l})} = \{ v_j^{(\tilde{l})} \}_{j=1}^{n_V^{(\tilde{l})}}$ denotes the visual tokens at that layer.
Given an importance function $F_s$ and a retention rate $r^{(\tilde{l})}$, the compression operator $\mathcal{C}^{(\tilde{l})}$ retains the top $\lceil r^{(\tilde{l})} n_V \rceil$ visual tokens according to their importance scores:
\begin{equation}
\mathcal{C}^{(\tilde{l})}:(V^{(\tilde{l})}, F_s, r^{(\tilde{l})}) \rightarrow \tilde{V}^{(\tilde{l})}.
\label{eq:compression_operator_clean}
\end{equation}
Text tokens are left unchanged. 
Attention is the most widely adopted approach for estimating visual token importance in text-guided compression, as it directly reflects each visual token’s contribution to the model’s response under a given text query \cite{Chen2024AnII, lin2025boosting, zhao2024stitch, Zhang2024SparseVLMVT, tzachristas2025mathematical}. 
Token compression of this form is theoretically guaranteed to preserve attention distributions and downstream model outputs \cite{tzachristas2025mathematical}, and have been adopted in practical systems~\cite{deepseek32, deepseek32f}. 
We focus on attention-based importance estimation in this work.
Let $\mathcal{I}_{\text{ref}} \subseteq \{1, \cdots, n_T\}$ denote the indices of reference text tokens, the importance score of $v_j^{(\tilde{l})}$ is:
\begin{equation}
s_j^{(\tilde{l})} = F_s(v_j^{(\tilde{l})}) = \frac{1}{|\mathcal{I}_{\text{ref}}|} \sum_{i \in \mathcal{I}_{\text{ref}}} a_{n_V^{(\tilde{l})}+i, j}^{(\tilde{l})},
\label{eq:importance_score}
\end{equation}
where $a_{n_V^{(\tilde{l})}+i, j}^{(\tilde{l})}$ is the attention weight from the $i$-th text token (at position $(n_V^{(\tilde{l})}+i)$ in the current layer's sequence) to the $j$-th visual token.
The compressed visual sequence is obtained by selecting the top-ranked tokens and restoring their original order:
\begin{equation}
\tilde V^{(\tilde{l})}
= \mathrm{Sort}_{\text{index}}
  \left(
    \mathrm{TopK}_{\lceil r^{(\tilde{l})} n_V \rceil}
    (V^{(\tilde{l})},\, s^{(\tilde{l})})
  \right).
\label{eq:compressed_sequence_clean}
\end{equation}
$\tilde V^{(\tilde{l})}$ is used to form the input to the next layer $V^{(\tilde{l}+1)}$. If multiple layers apply compression, each operation uses the already-compressed tokens from the previous layer.

\section{Demystifying Robustness Degradation Induced by Compression}
\label{sec:comp_and_roubst}


While vision token compression is widely used to accelerate LVLM inference, its impact on robustness remains underexplored. We first present empirical evidence that compression degrades robustness, and then analyze the compression pipeline to identify token ranking as the primary source of this degradation.
\subsection{Compression Compromises Model Robustness}
\label{sec:comp_affect_robust}
   
\begin{figure}[t]
    \centering
    \begin{subfigure}{0.46\linewidth}
        \centering
        \includegraphics[width=\linewidth]{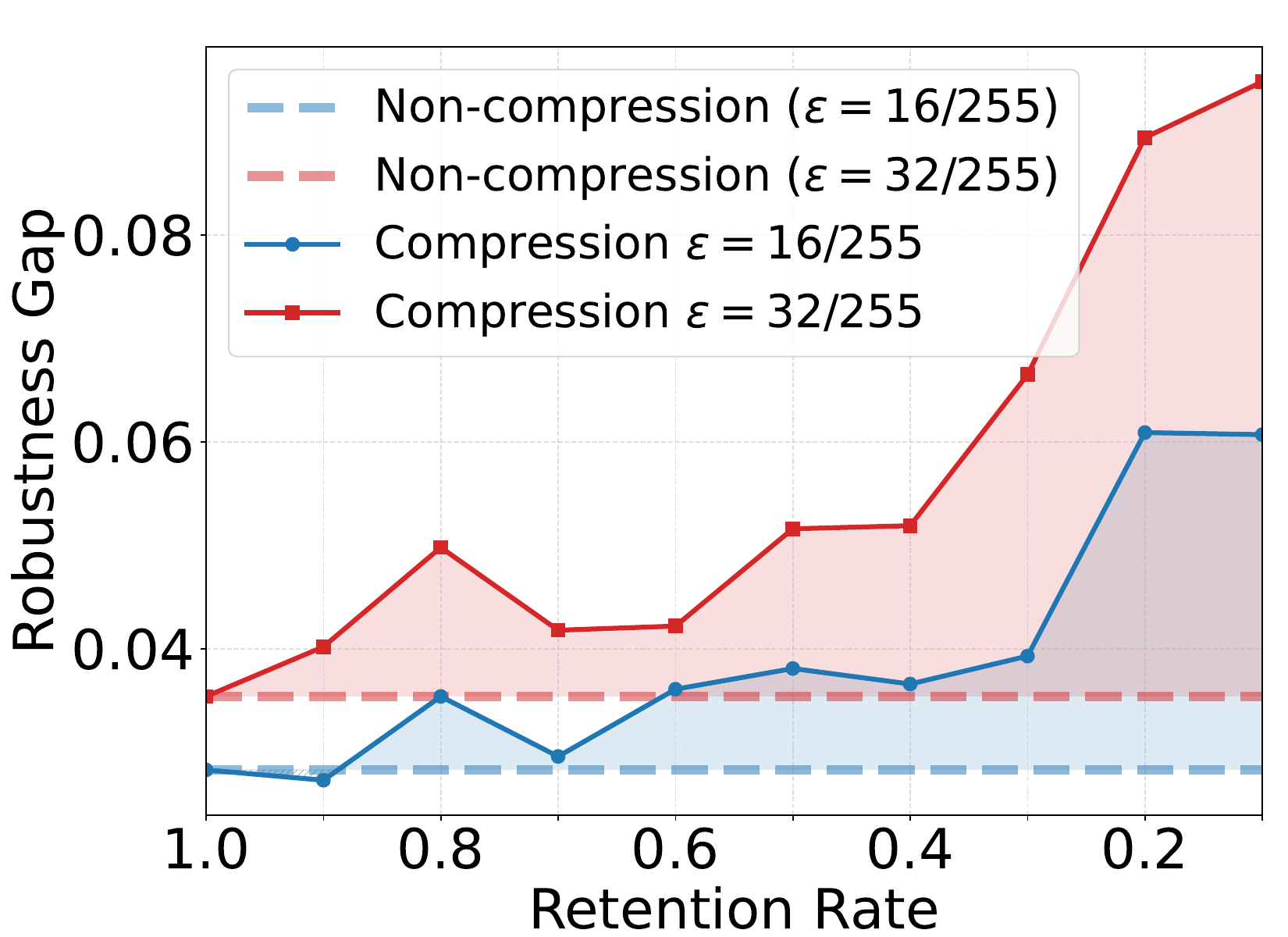} 
        \caption{LLaVA}
        \label{fig:o1_llava_gap}
    \end{subfigure}
    \begin{subfigure}{0.46\linewidth}
        \centering
        \includegraphics[width=\linewidth]{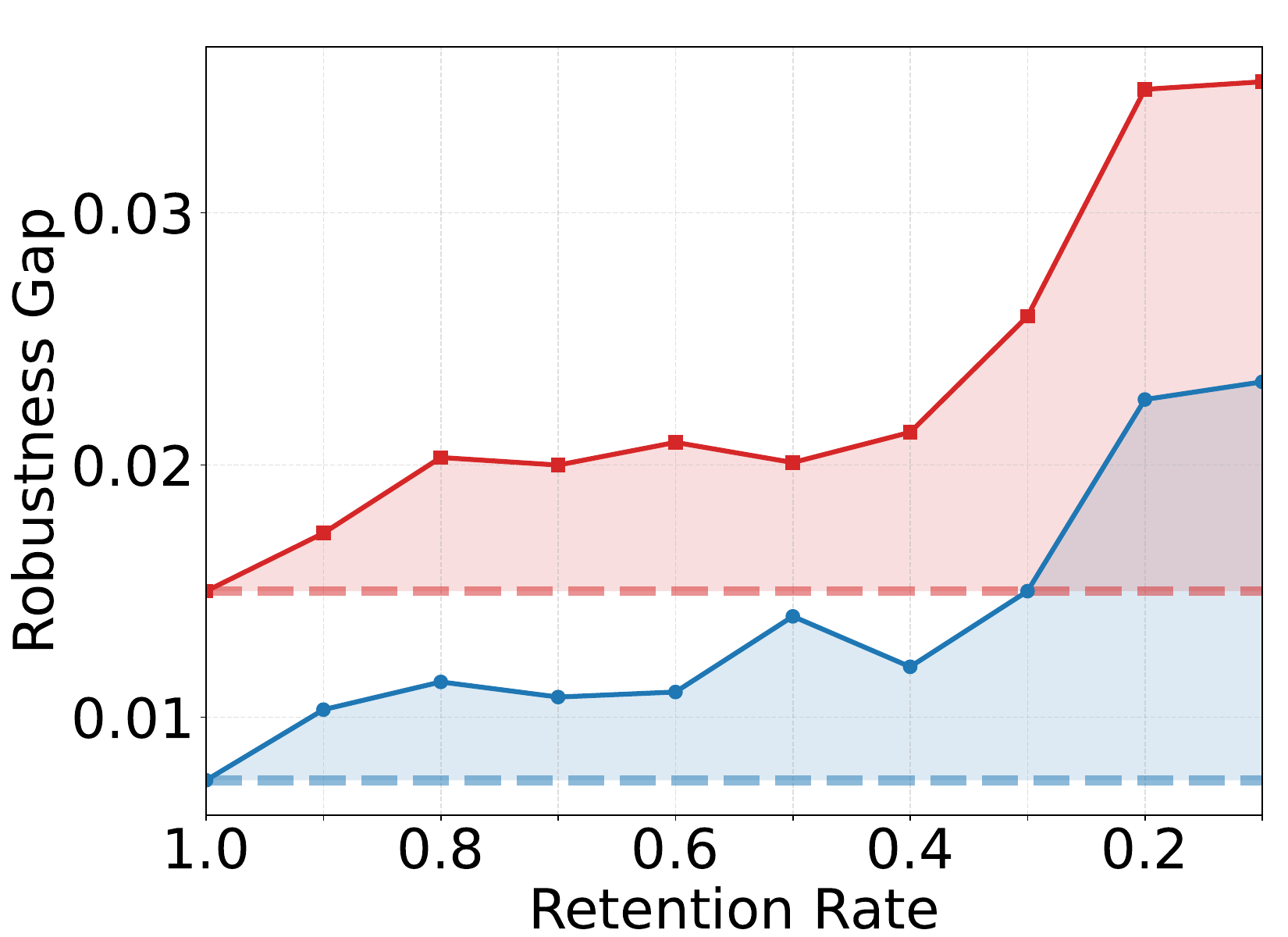}
        \caption{Qwen-VL}
        \label{fig:o1_qwen_gap}
    \end{subfigure}
    \caption{Robustness gap under different token retention rates on LLaVA and Qwen-VL. 
    The solid curves (compressed, retention rate $< 1$) consistently lie above the dashed baseline (uncompressed), demonstrating that the introduction of compression incurs additional robustness degradation.
    }
    \label{fig:o1_performance_gap}
\end{figure}

To empirically assess how compression affects model robustness, we conduct controlled experiments on the POPE dataset~\cite{li2023evaluating}. Following common practice, we adopt a single-layer compression setup with compression applied at the second LLM layer~\cite{Chen2024AnII}. 
We generate perturbed inputs by injecting Gaussian noise into the image, bounded by an $\ell_\infty$ norm $\epsilon \in {16/255, 32/255}$, yielding samples $(I+\delta^{\text{rand}}, T)$.
We measure the \emph{robustness gap}, defined as the performance difference between clean and perturbed inputs, under a wide range of token retention rates from $1.0$ (no compression) down to $0.1$.
The results on LLaVA and Qwen-VL are shown in Fig.~\ref{fig:o1_performance_gap}.
The dashed lines represent the robustness gap of the no-compression baseline (retention rate = 1), while the solid curves show the gaps under different token retention rates.
Across models and perturbation magnitudes, token compression consistently enlarges the robustness gap compared to the uncompressed baseline, with the gap increasing as the retention rate decreases. These results indicate that visual token compression amplifies sensitivity to input perturbations and introduces additional robustness degradation.

\subsection{Mechanism of Robustness Degradation Under Compression}
\label{sec:mechanism_of_robust}

To understand why compression increases model vulnerability, we analyze its core mechanism: discrete token ranking and selection.

\noindent\textbf{The Critical Role of Token Importance Ranking in Robustness Degradation.}
\begin{figure}[t]
    \centering
    \begin{subfigure}{0.43\linewidth}
        \centering
        \includegraphics[width=\linewidth]{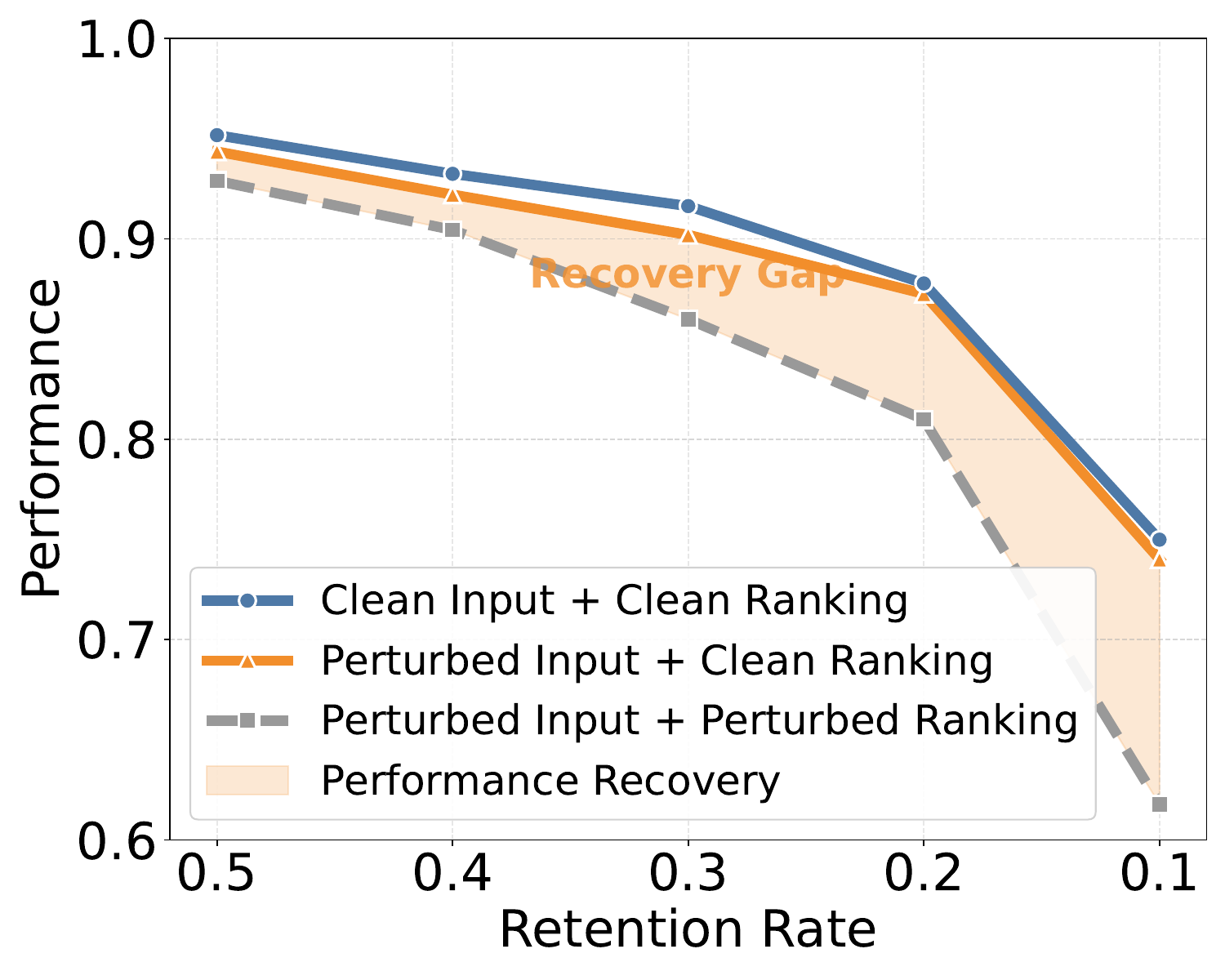} 
        \caption{LLaVA $ \epsilon = 32/255$}
        \label{fig:semantic_rank_role_a}
    \end{subfigure}
        \begin{subfigure}{0.43\linewidth}
        \centering
        \includegraphics[width=\linewidth]{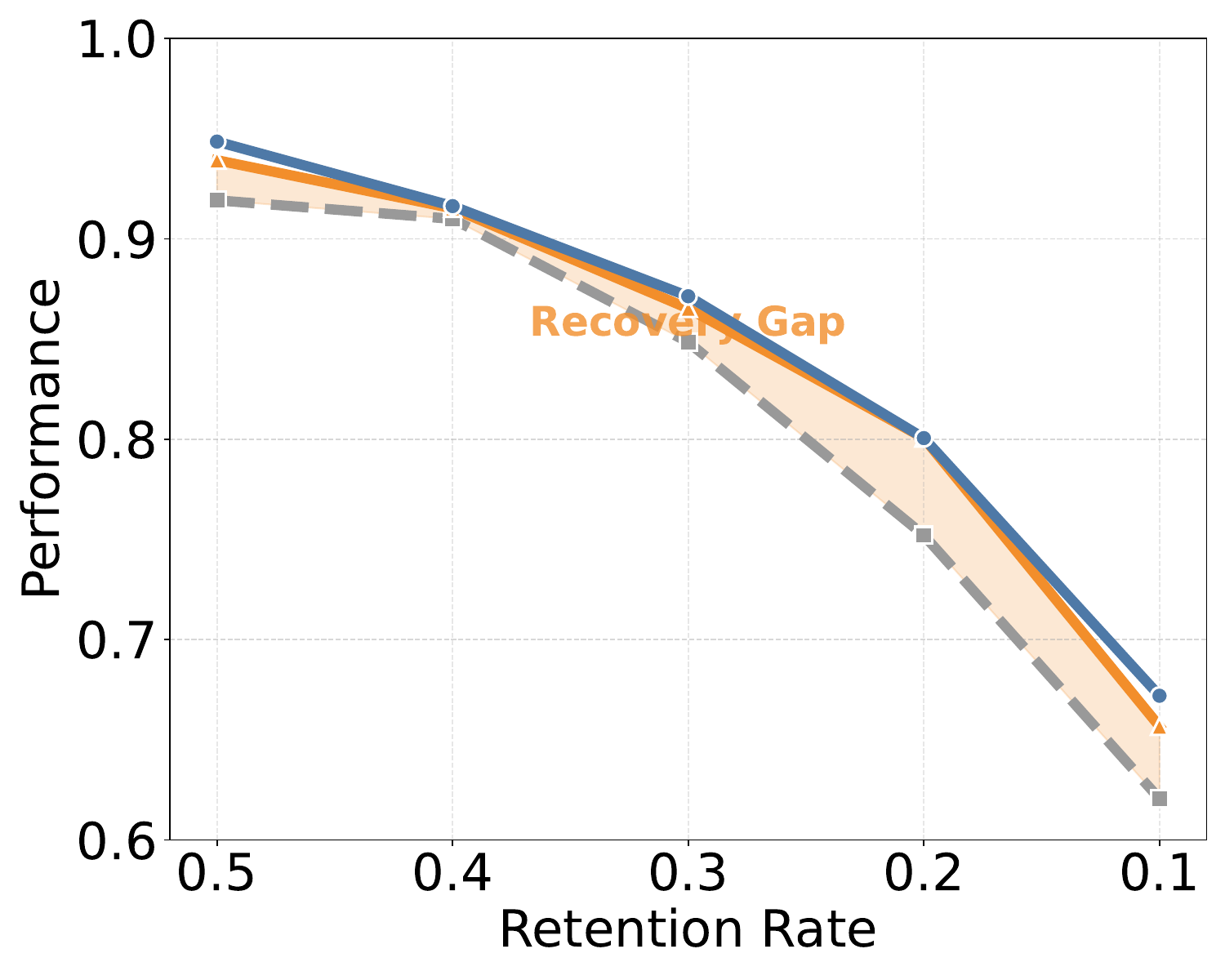}
        \caption{Qwen-VL $\epsilon = 32/255$}
        \label{fig:semantic_rank_role_c}
    \end{subfigure}
    \caption{
    Effect of token importance ranking on robustness under compressed inference.
    The shaded region highlights the significant performance recovery achieved by restoring correct rankings. 
    }
    \label{fig:semantic_rank_role}
\end{figure}
Under compressed inference with perturbed inputs $(I+\delta^{\text{rand}}, T)$, robustness degradation arises from two factors: (i) perturbations may destabilize importance ranking, leading to suboptimal token selection; and (ii) perturbations may corrupt the semantic content of the retained tokens.
To disentangle these effects, Fig.~\ref{fig:semantic_rank_role} and Fig.\ref{fig:semantic_rank_role_64} in Appendix \ref{subsec:critical_ranking_role_supp} (for $\epsilon$ =64/255) compare the model performance under three inference configurations:
(a) \emph{Clean input + clean ranking} (blue solid line), serving as the performance upper bound;
(b) \emph{Perturbed input + perturbed ranking} (gray dashed line), where the input is perturbed and compression is performed using rankings derived from the perturbed input;
and
(c) \emph{Perturbed input + clean ranking} (orange solid line), a controlled setting in which the input is perturbed but the compression module is forced to use a fixed ranking derived from the clean input.
Fig.~\ref{fig:semantic_rank_role} shows that restoring the clean ranking (from (b) to (c)) substantially recovers performance, bringing it close to the clean baseline in (a), as highlighted by the shaded region.
These results indicate that robustness degradation under compression is primarily driven by instability in token importance ranking: identical perturbations largely preserve token semantics while substantially altering their relative ordering. 


\noindent\textbf{Importance Ranking Is Unstable Under Input Perturbations.}
We next quantitatively compare the token rankings induced by the clean input $(I, T)$ and the perturbed input $(I+\delta^{\text{rand}}, T)$ through 3 metrics:
(i) \emph{Rank correlation}, measured by Kendall’s $\tau$ and Spearman’s $\rho$\footnote{They are classical rank correlation metrics that quantify the agreement between two orderings by comparing pairwise consistency and monotonic relationships~\cite{kendall1938new}, respectively.
}, to characterize the overall agreement between the two rankings;
(ii) \emph{Top-100 preservation rate}, defined as the fraction of tokens that appear in the Top-100 under the clean input and remain in the Top-100 after perturbation; 
and (iii) \emph{Bottom-100 infiltration rate}, defined as the fraction of tokens that belong to the Bottom-100 under the clean input but enter the Top-100 after perturbation. 

The results are shown in Fig.~\ref{fig:o2_ranking_llava} for LLaVA and Fig.~\ref{fig:o2_ranking_llava_qwen} for Qwen-VL in Appendix~\ref{subsec:ranking_instab_supp}. 
Increasing perturbation magnitude destabilizes token-importance rankings, as reflected by declining Kendall’s $\tau$ and Spearman’s $\rho$, reduced Top-100 preservation, and increased Bottom-100 infiltration, ultimately degrading model predictions.

\begin{figure}[t]
    \centering
    \begin{subfigure}{0.43\linewidth}
        \centering
        \includegraphics[width=\linewidth]{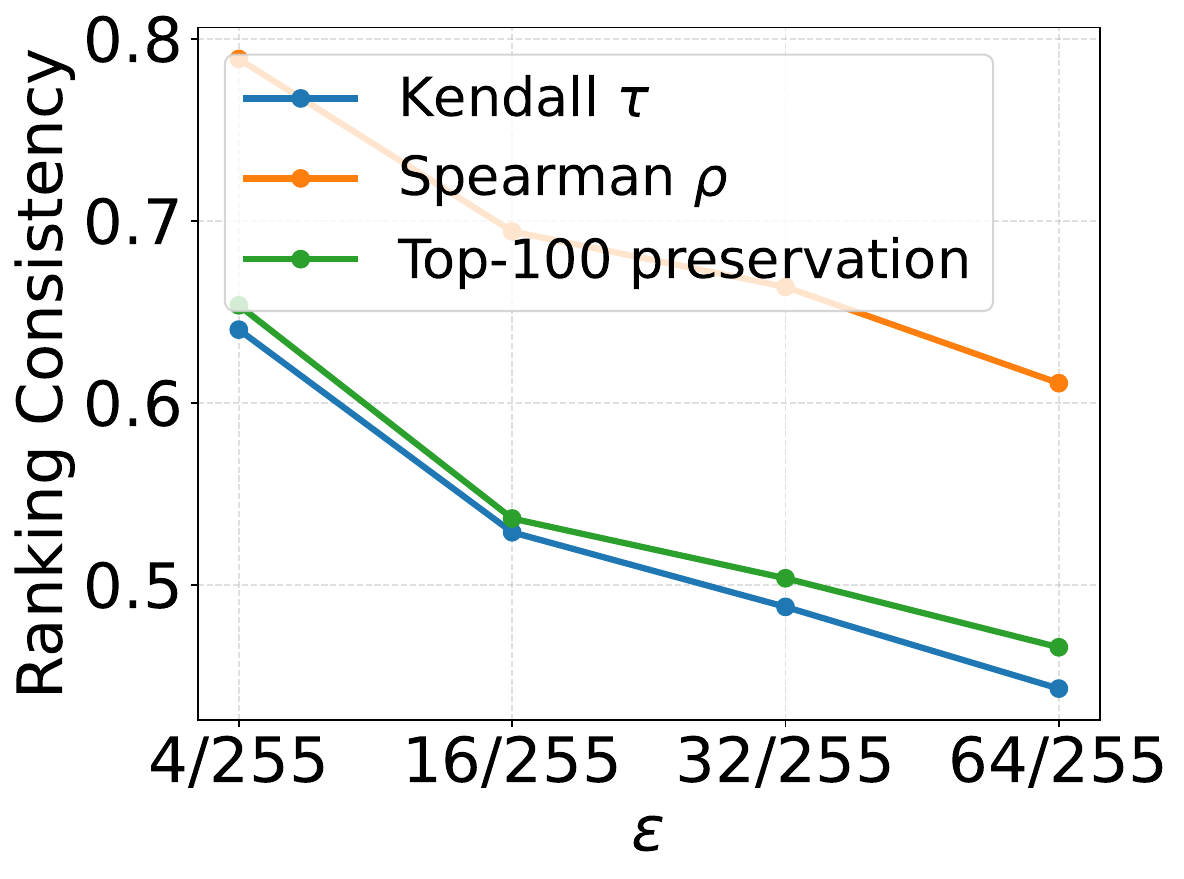} 
        \caption{LLaVA}
        \label{fig:o2_1_ranking_llava}
    \end{subfigure}
    \begin{subfigure}{0.43\linewidth}
        \centering
        \includegraphics[width=\linewidth]{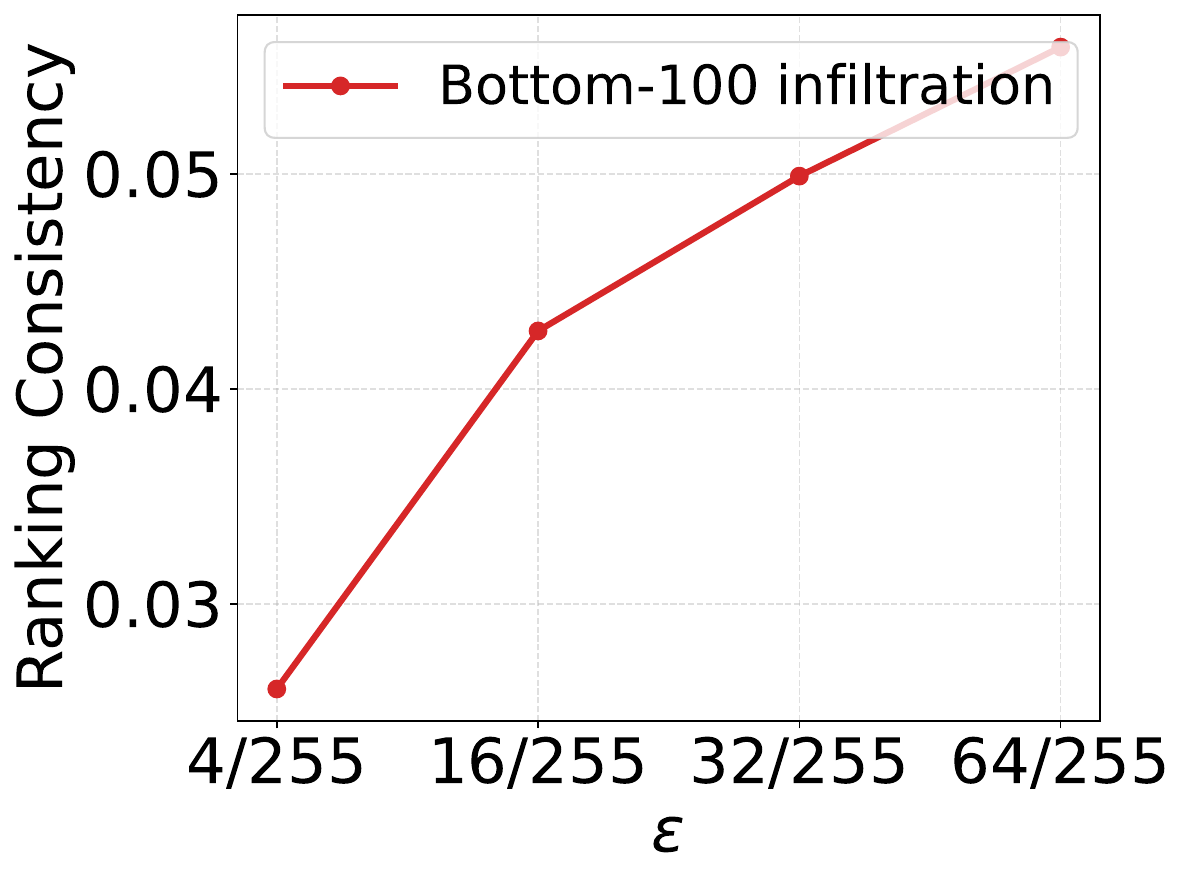}
        \caption{LLaVA}
        \label{fig:02_ranking_bot_llava}
    \end{subfigure}

    \caption{
    Ranking stability under random $\ell_\infty$-bounded noise with different budgets.
    Global stability is measured using Kendall's~$\tau$ and Spearman's~$\rho$.
    Local stability is evaluated via Top-100 preservation and Bottom-100 infiltration rate. 
    }
    \label{fig:o2_ranking_llava}
\end{figure}

\begin{figure}[t]
    \centering
    \begin{subfigure}{0.43\linewidth}
        \centering
        \includegraphics[width=\linewidth]{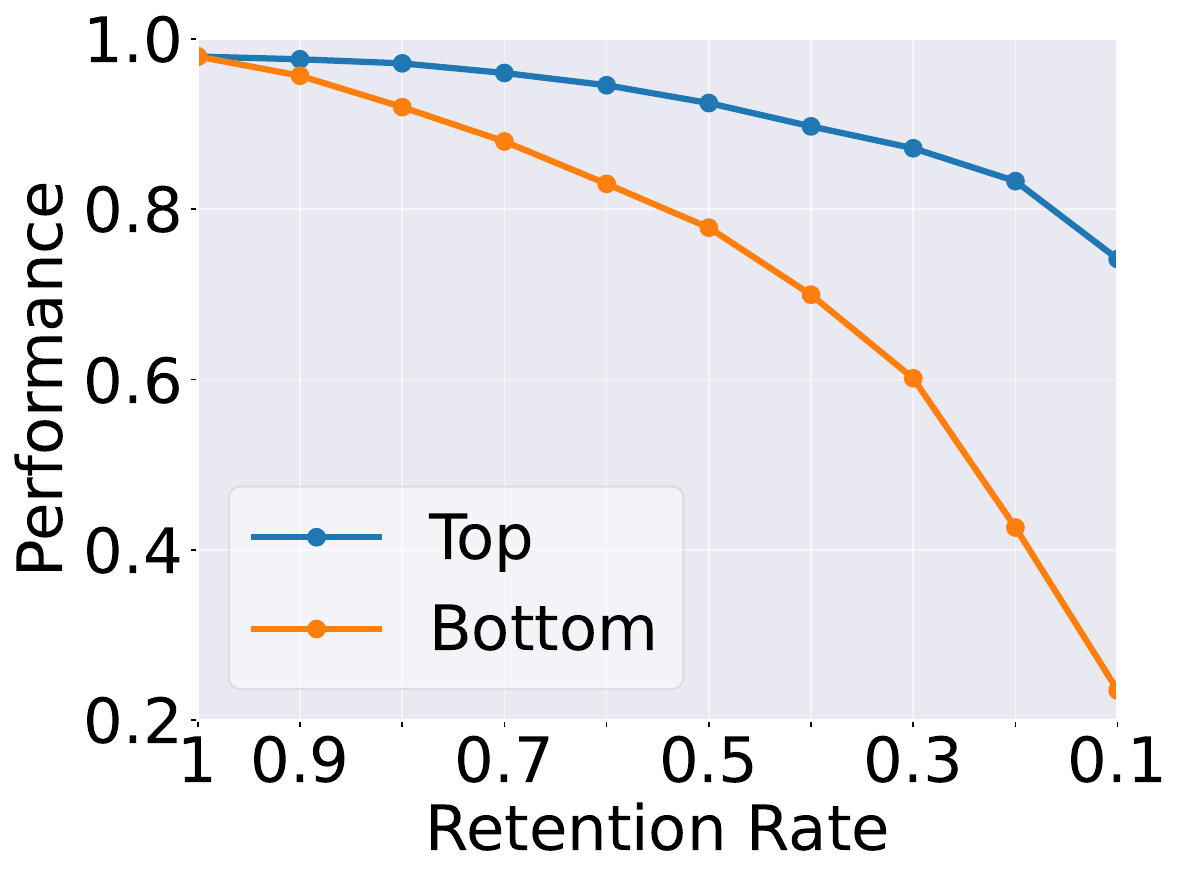} 
        \caption{LLaVA}
        \label{fig:o2_2_llava}
    \end{subfigure}
    \begin{subfigure}{0.43\linewidth}
        \centering
        \includegraphics[width=\linewidth]{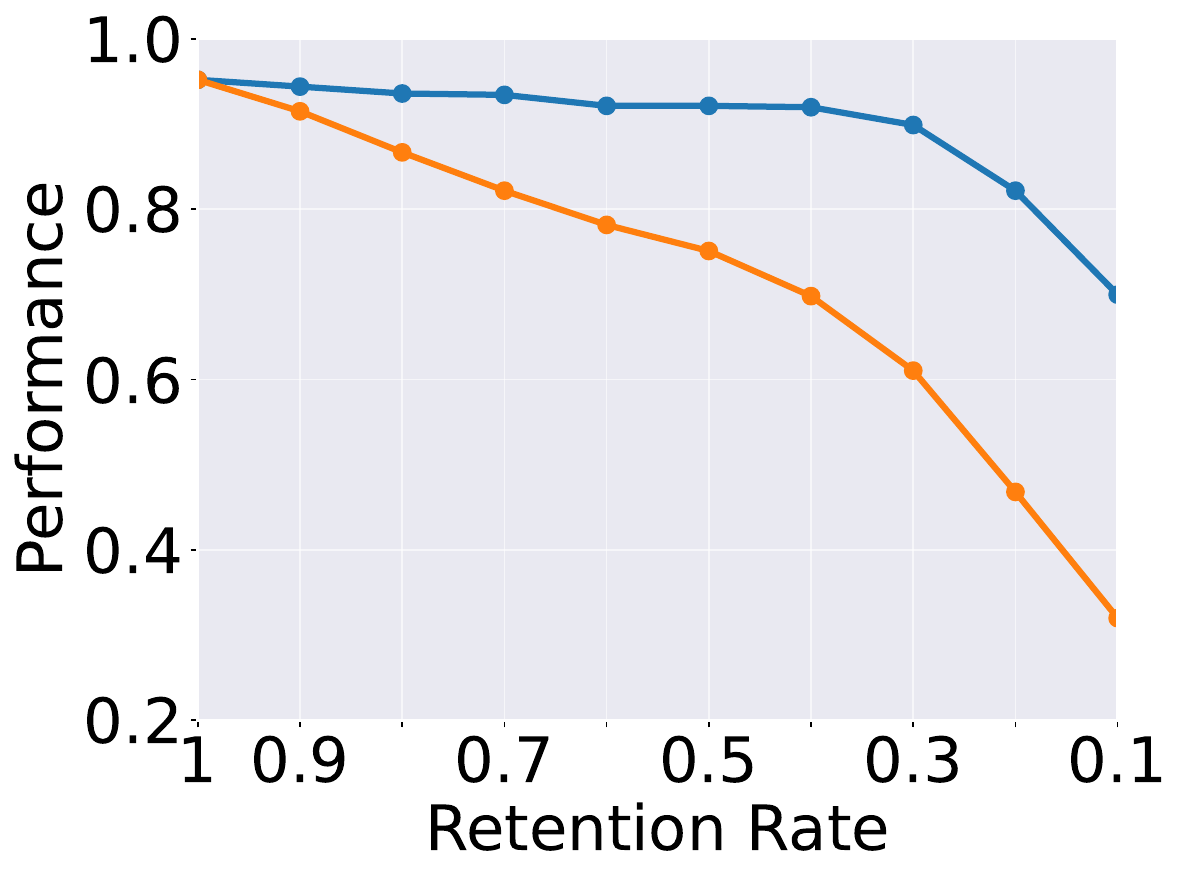}
        \caption{Qwen-VL}
        \label{fig:o2_2_qwen}
    \end{subfigure}

    \caption{
    Impact of token selection strategy under compression. Retaining the Top-$k$ most important tokens preserves performance, while retaining the Bottom-$k$ least important tokens leads to a drastic performance collapse. 
    }
    \label{fig:o2_2_llava_qwen}
\end{figure}

\noindent\textbf{Extreme Misranking Leads to Complete Model Collapse.}
To further illustrate the critical role of correct token selection, we consider an extreme scenario: instead of retaining the Top-$k$ tokens based on importance scores, we intentionally keep the Bottom-$k$ tokens. As shown in Fig.~\ref{fig:o2_2_llava_qwen}, this reversed selection leads to a complete performance collapse, demonstrating the model’s strong dependence on selecting the correct subset of visual tokens. 
This impact is amplified in multi-layer compression, such early misselection is irreversible and propagates across layers.



\section{The Proposed Compression-Aware Attack}
\subsection{Motivation}
\label{sec:motivation}


Section~\ref{sec:comp_and_roubst} shows that visual token compression introduces a compression specific robustness degradation driven by instability in importance ranking.
As compression is increasingly adopted to reduce inference cost in practical deployments, understanding the security implications of this mechanism becomes essential. 
For example, in dynamic systems, compression is enabled and the retention rate is adjusted to satisfy latency or resource constraints \cite{jie2024token, zhu2025fastcache, fan2025timebill}. As a consequence, the same input may be processed under different compression states at different times, and failures that occur only in compressed settings are unlikely to surface during standard security inspection, 
which is typically conducted under resource-sufficient, non-compressed inference.
These state-dependent failures remain hidden, rendering reliable reproduction and attribution difficult.

To reveal and measure such compression-induced vulnerabilities, it is essential to disentangle them from the model's inherent adversarial weaknesses. 
This necessitates attacks that preserve full-sequence behavior while inducing failure only when compression is enabled. Traditional adversarial attacks fail this requirement, as they are designed to induce failures in the full-sequence (non-compressed) model and are unaware of the compression mechanism, thereby conflating compression-induced failures with inherent model vulnerabilities ~\cite{madry2018towards, bailey2024image, qi2024visual}. Moreover, compression introduces discrete, non-differentiable operations, which render gradient-based optimization ineffective. These challenges call for a new class of attacks that are explicitly aware of the compression.

\subsection{Threat Model}
\label{sec:threat_model}

\paragraph{Attacker Goal.}
The attacker aims to craft visual adversarial input that induce failures through the vision token compression mechanism, causing the model to break down under compressed inference while remaining unaffected in the non-compressed setting.
Formally, the compression-aware attack objective is:
\begin{equation}
\begin{alignedat}{2}
\max_{\delta} \quad &
\mathcal{L}(f(I+\delta, T; \mathcal{C}),\, f(I, T))
-\lambda\, \mathcal{L}(f(I+\delta, T),\, f(I, T))
\\
\text{s.t.} \quad &
\|\delta\|_{\infty} \le \epsilon .
\end{alignedat}
\label{eq:attack_goal}
\end{equation}
where $f(I+\delta, T; \mathcal{C})$ and $f(I+\delta, T)$ are the outputs on the adversarial input with and without compression, respectively, and $\mathcal{L}(\cdot,\cdot)$ measures the divergence between two outputs.



\paragraph{Attacker Capabilities.}
We consider two attack settings: 
\begin{itemize}
    \item \textbf{White-box Setting.} The attacker has full access to the target model parameters, gradients, and the details of the compression mechanism. 
    This setting enables direct gradient-based optimization and represents a worst-case scenario for assessing the security limits of compressed inference.

    \item \textbf{Black-box Setting.} 
    The attacker lacks access to target model parameters, gradients, and compression configurations (e.g., layer placement, retention rates).
    Furthermore, modern LVLM systems typically deploy access control mechanisms to mitigate malicious behavior, rendering query-intensive probing attacks infeasible~\cite{openai2024Terms, deepseek2025Terms, gemini2024Terms}.
\end{itemize}

\subsection{White-box Attack}
\label{sec:white_box}

\begin{figure*}[t]
    \centering
    \includegraphics[width=0.90\linewidth]{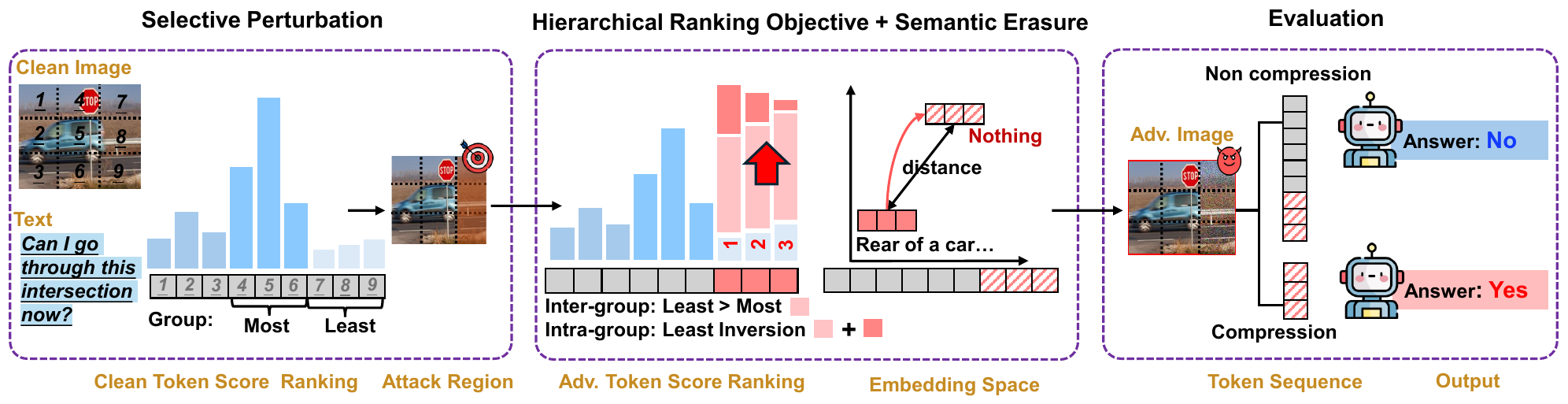}
    \caption{
    CAA selectively perturbs low-importance regions to preserve correct behavior under non-compressed inference.
A hierarchical ranking objective manipulates token importance ordering, promoting low-importance tokens over the most important ones and reversing their relative order, while a semantic erasure objective removes informative cues from them.
As a result, the adversarial image induces failures only under compressed inference.
    }
    \label{fig:caa_white}
\end{figure*}

Section~\ref{sec:comp_and_roubst} shows that ranking instability drives compression-induced robustness degradation. Thus, directly manipulating importance scores offers a principled strategy for compression-aware attacks.

\noindent\textbf{Attack Analysis.}
In the white-box setting, the attacker has full knowledge of the target model and the compression algorithm, including the first compression layer $\tilde{l}_1$ and its retention rate $r^{(\tilde{l}_1)}$. 
The attack therefore targets this stage to disrupt importance ranking, causing task-irrelevant tokens to be retained.
This strategy applies to both single- and multi-layer compression, as token selection at $\tilde{l}_1$ is irreversible and errors propagate to all subsequent stages.
\noindent\textbf{Overview.}
Guided by this analysis, our attack perturbs selected image regions to manipulate token-importance rankings, preserving uncompressed behavior while degrading compressed inference. The attack comprises three components (Fig.~\ref{fig:caa_white}): 
(i) \emph{Selective perturbation}, which restricts modifications to low-importance regions to preserve uncompressed inference;  
(ii) \emph{Hierarchical ranking objective}, which effectively promotes low-importance tokens into the retained set; and  
(iii) \emph{Semantic erasure}, which corrupts retained tokens to further impair compressed inference.

\noindent\textbf{Selective Perturbation.}
To satisfy the compression-aware objective in Eq.~\ref{eq:attack_goal}, the attack must maximize the discrepancy between compressed and uncompressed inference. 
This requires avoiding perturbations to high-importance visual tokens, which encode the core semantic evidence for correct predictions.
Consequently, we restrict perturbations exclusively to the least important regions.

Given the visual tokens $V=\{v_j\}_{j=1}^{n_V}$ from image $I$, we compute their importance scores $s_j^{(\tilde{l})}$ (Eq.~\ref{eq:importance_score}) at layer $\tilde{l}$ and sort them in descending order.  
Let $\Omega_{\text{most}}$ denote the indices of the highest-scoring tokens and $\Omega_{\text{least}}$ denote those of the lowest-scoring tokens.  
Perturbations are applied only to patches corresponding to $\Omega_{\text{least}}$:
\begin{equation}
\hat{P}_j =
\begin{cases}
P_j+\delta_j, & j \in \Omega_{\text{least}},\;\|\delta_j\|_\infty\le\epsilon, \\
P_j, & j\notin\Omega_{\text{least}},
\end{cases}
 \text{~~~and~~~}
\hat{I}=[\hat{P}_j]_{j=1}^{n_V}.
\label{eq:adv_image_patch}
\end{equation}
where $[\hat{P}_j]_{j=1}^{n_V}$ reconstructs the full image, and $\delta = [\delta_j]_{j=1}^{n_V}$ satisfies $\|\delta\|_{\infty} \leq \epsilon$.
Under uncompressed inference, ranking perturbations may temporarily divert attention to less informative tokens, but core semantic content remains intact, allowing later layers to re-attend to informative tokens and maintain stable predictions. In contrast, under compressed inference, the same ranking shift causes irreversible token removal, forcing reliance on perturbed low-importance tokens and leading to severe performance degradation.

\begin{figure*}[t]
    \centering
    \includegraphics[width=0.90\linewidth]{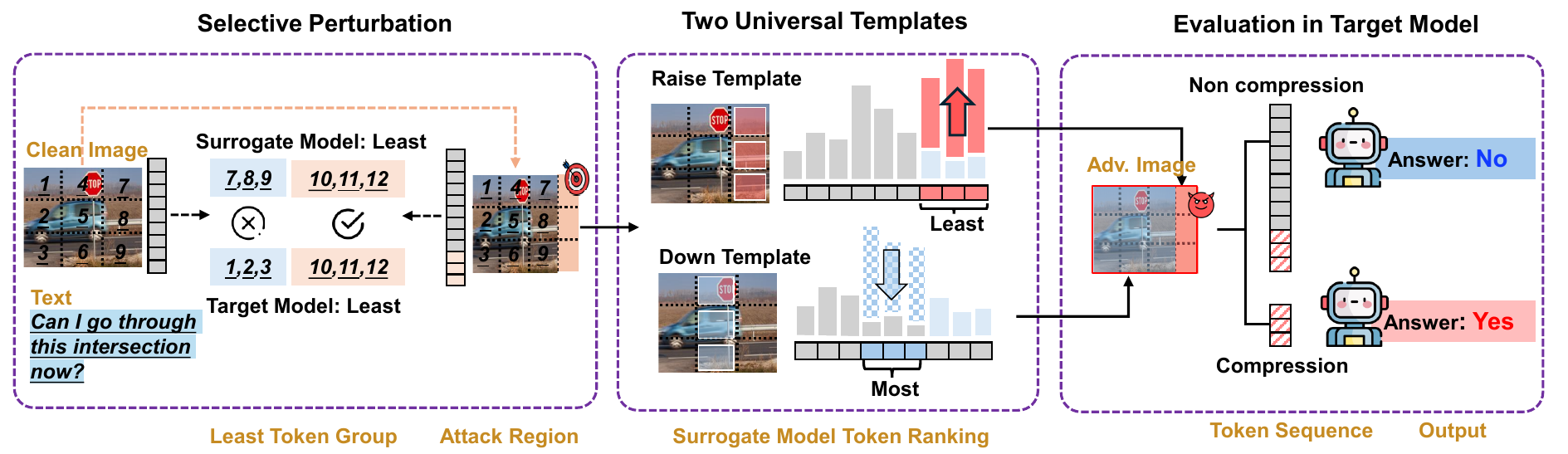}
    \caption{
T-CAA employs a border perturbation to avoid reliance on model-specific least-informative regions.
Two self-universal templates are optimized: a raise template that amplifies border importance and a down template that suppresses the original image.
As a result, the attack transfers across models, preserving uncompressed utility while failing under compression in the target model.
    }
    \label{fig:caa_transfer}
\end{figure*}

\noindent\textbf{Hierarchical Ranking Objective.}
Under a perturbation $\delta$, the perturbed image $\hat{I}$ yields a set of visual tokens $\hat{V}$ at the attack layer, each associated with a perturbed importance score $\hat{s}_j^{(\tilde{l})}$ (Eq.~\ref{eq:importance_score}).
To manipulate relative importance ordering, we adopt Bayesian Personalized Ranking (BPR)~\cite{rendle2009bpr}, extending it to the token set setting to enforce least-important tokens to outrank most-important ones.
However, this \textit{inter-group} constraint leaves the internal ordering of $\Omega_{\text{least}}$ unconstrained, allowing the model to retain the most informative tokens within $\Omega_{\text{least}}$.
To prevent this, we introduce \textit{intra-group reversal}. We partition $\Omega_{\text{least}}$ into $n_g$ subgroups $\Omega_{\text{least}}^{(1)}, \cdots, \Omega_{\text{least}}^{(n_g)}$ according to their clean importance scores, where smaller indices correspond to higher clean informativeness. The resulting \textit{hierarchical ranking objective} comprises two components:
(i) \textbf{Inter-group manipulation:} all tokens in $\Omega_{\text{least}}$ are ranked above those in $\Omega_{\text{most}}$;
(ii) \textbf{Intra-group reversal:} the ordering within $\Omega_{\text{least}}$ is inverted, such that tokens with lower clean importance receive higher perturbed scores.
These conditions are encoded via two sets of pairwise preferences:
\begin{equation}
\begin{aligned}
\mathcal{P}_{\text{LM}}^{(g)} &= \{(\ell, m) \mid \ell \in \Omega_{\text{least}}^{(g)},\; m \in \Omega_{\text{most}}\}, \\
\mathcal{P}_{\text{LL}}^{(b,a)} &= \{(\ell_b, \ell_a) \mid \ell_a \in \Omega_{\text{least}}^{(a)},\; \ell_b \in \Omega_{\text{least}}^{(b)},\; a < b\},
\end{aligned}
\label{eq:pair_sets}
\end{equation}
The hierarchical ranking loss is:
\begin{equation}
\begin{aligned}
\mathcal{L}_{\text{rank}}(\delta) &= -\alpha \sum_{g=1}^{n_g} \sum_{(\ell,m) \in \mathcal{P}_{\text{LM}}^{(g)}} \log\sigma\big(\hat{s}_\ell^{(\tilde{l})} - \hat{s}_m^{(\tilde{l})}\big) \\
&\quad - \beta \sum_{a<b} \sum_{(\ell_b, \ell_a) \in \mathcal{P}_{\text{LL}}^{(b,a)}} \log\sigma\big(\hat{s}_{\ell_b}^{(\tilde{l})} - \hat{s}_{\ell_a}^{(\tilde{l})}\big),
\end{aligned}
\label{eq:hierarchical_rank}
\end{equation}
where $\alpha$ and $\beta$ balance the inter- and intra-group terms. 
These constraints enforce the ordering
$\Omega_{\text{least}}^{(n_g)} > \cdots > \Omega_{\text{least}}^{(1)} > \Omega_{\text{most}}$,
causing compression to retain minimally informative tokens.

To stabilize and accelerate optimization, we exploit the fact that importance scores are proportional to the dot product of vision keys and text queries (Section \ref{sec:vision_comp}). Consequently, we explicitly encourage the alignment of perturbed vision keys with reference text queries and define the query-guided alignment loss as:
\begin{equation}
\mathcal{L}_{\mathrm{key}}(\delta)
=
-\frac{1}{|\Omega_{\mathrm{least}}|\,|\mathcal{I}_{\mathrm{ref}}|}
\sum_{\ell \in \Omega_{\mathrm{least}}}
\sum_{i \in \mathcal{I}_{\mathrm{ref}}}
\hat{q}_i^{(\tilde{l})\top}\hat{k}_\ell^{(\tilde{l})},
\label{eq:key_align}
\end{equation}
where $\hat{q}_i^{(\tilde{l})} = \hat{t}_i^{(\tilde{l})} W_Q^{(\tilde{l})}$ and
$\hat{k}_\ell^{(\tilde{l})} = \hat{v}_\ell^{(\tilde{l})} W_K^{(\tilde{l})}$ are computed from the perturbed input $(\hat{I}, T)$, and $\mathcal{I}_{\mathrm{ref}}$ denotes the reference text indices.

\noindent\textbf{Semantic Erasure.} While the hierarchical objective controls token selection, retained tokens may still preserve residual semantics that support inference. To eliminate such information, we introduce a \emph{semantic erasure} objective that corrupts the representations of the retained tokens.
Let $v_{\ell}^{(\tilde{l})}$ and $\hat{v}_{\ell}^{(\tilde{l})}$ denote the visual token representations at layer $\tilde{l}$ under the clean image $I$ and the perturbed image $\hat{I}$, respectively. We encourage maximal deviation between them:
\begin{equation}
\mathcal{L}_{\text{erase}}(\delta)
=
-\frac{1}{|\Omega_{\text{least}}|}
\sum_{\ell \in \Omega_{\text{least}}}
\left\|
\hat{v}_{\ell}^{(\tilde{l})} - v_{\ell}^{(\tilde{l})}
\right\|_2^2 .
\label{eq:semantic_erasure}
\end{equation}
This objective complements the ranking manipulation by ensuring that the tokens selected by compression are both incorrectly ranked and semantically uninformative.

\noindent\textbf{Total Objective.}
The total objective is formulated as:
\begin{equation}
\begin{aligned}
\min_{\delta}\quad 
& \mathcal{L}_{\mathrm{rank}}(\delta)
+ \lambda_{e}\,\mathcal{L}_{\mathrm{erase}}(\delta)
+ \lambda_{k} \mathcal{L}_{\mathrm{key}}(\delta)
\quad 
\text{s.t.}\quad 
\|\delta\|_{\infty}\le\epsilon.
\end{aligned}
\label{eq:final_loss}
\end{equation}
Here, $\delta$ denotes the effective perturbation applied to the image, obtained by aggregating per-patch perturbations $\{\delta_j\}$. 
We solve this optimization problem using projected gradient descent (PGD)~\cite{madry2018towards}, and summarize the attack procedure in Algorithm~\ref{alg:caa} in Appendix \ref{sec:white_box_attack_supp}.


\subsection{Transfer Attack Under Black-box Setting}
\label{sec:transfer_attack}

\noindent\textbf{Attack Analysis.}
Compared to the white-box setting, the black-box attacker follows the same high-level attack strategy: targeting the layer where token compression is applied and optimizing perturbations to induce incorrect token selection under compression.
However, due to limited access to the deployed system, the attacker faces two fundamental gaps: 
\begin{itemize}
    \item \textbf{Unknown Compression Configuration}:
    The attacker has no knowledge of the compression mechanism’s internal configuration, including the placement of compression layers and their retention rates, which prevents precise identification of the first compression stage or tailor attacks to specific retention settings.
    \item \textbf{Limited Target Model Access}: The attacker has no access to the target model’s parameters, internal states, or gradients, and is restricted to interacting with the model through a query-limited API, making gradient-based or query-intensive attacks infeasible.
\end{itemize}

\subsubsection{Handling Unknown Compression Configurations.}
Although the attacker has no knowledge of the exact compression configuration, it can estimate a range of compression layers ${\tilde{L}_c}$ over which the attack is expected to be effective, and jointly manipulate token-importance rankings across all layers within this range. This enforces consistent adversarial token orderings throughout the candidate layers and ensures attack effectiveness despite configuration uncertainty.
Consequently, if the target model applies token compression at least once within the estimated range, the attack induces erroneous token selection at that layer.

To estimate the desired range, the attacker first infers the target model's average retention rate $\bar{r}$ defined as the average fraction of visual tokens retained per layer across the model, from public documentation~\cite{gemini3flash, gptmini, deepseek32} or limited API latency measurements by exploring differences in inference time between compressed and non-compressed settings.
Given this estimate, the attacker enumerates plausible compression configurations consistent with common design practices \cite{Zhang2024SparseVLMVT, Xing2024PyramidDropAY, Chen2024AnII, Yang2024VisionZipLI}
(e.g., multi-stage compression with progressively decreasing retention rate), and derives the corresponding range of candidate compression layers over which the attack is applied (An example is provided in the Appendix \ref{subsec:enum_example}).


\subsubsection{Handling Limited Target Model Access.}
To address the lack of access to target model, we adopt transfer-based attacks.
However, model mismatches introduce two key challenges:
(i) adversarial perturbations tend to decay during transfer \cite{schaeffer2024failures, he2025few}, weakening their ability to elevate least-important tokens over salient ones in the target model, resulting in ineffective attacks;
(ii) different LVLMs often identify different image regions as the least important for the same input \cite{Guo2025Focus, wu2020boosting}, as illustrated in Fig.~\ref{fig:transer_least_example} in Appendix \ref{subsec:least_important_mismatch}, making region selection non-transferable.


To address these challenges, we introduce a common least informative region by augmenting the input image with an uninformative border, inspired by \cite{gu2024agent, lu2024test}.
Based on this design, we optimize two self-universal perturbation templates with complementary roles to construct transferable adversarial examples.
(i) A \emph{raise} template $\delta_R$ is applied to the border region to consistently amplify the importance of border tokens, forming a shared low-salience region across models.
(ii) A \emph{down} template $\delta_D$ is applied to the original image content to mildly suppress its attention and reduce the dominance of inherently salient tokens.
Both templates are jointly optimized across the candidate compression-layer range, ensuring robustness to unknown compression configurations. An overview of Transferable Compression-Aware Attack (T-CAA) is illustrated in Fig.~\ref{fig:caa_transfer}.

\noindent\textbf{Self-Universal Perturbation Templates.} 
Prior work has shown that self-universal perturbations can mitigate the effectiveness decay commonly observed in transfer attacks~\cite{wei2023enhancing}.
Motivated by this, we jointly optimize two self-universal perturbation templates, $\delta_R$ and $\delta_D$, over the candidate compression-layer range to enable consistent attention manipulation across models.

Before optimization, for each candidate compression layer $\tilde{l}$ of the surrogate model $f_{\text{sur}}$, we extract the visual token sequence
$V^{(\tilde{l})}=\{v_j^{(\tilde{l})}\}_{j=1}^{n_V}$ from the clean image $I$ and sort the tokens by descending importance.
We then select a set of shared lowest-ranked tokens $\Omega_{\text{least}}$ and partition it into two subsets
$\Omega_{\text{least}}^{\text{high}}$ and $\Omega_{\text{least}}^{\text{low}}$, such that
$\Omega_{\text{least}}=\Omega_{\text{least}}^{\text{high}}\cup\Omega_{\text{least}}^{\text{low}}$.
Similarly, we define the set of highest-ranked tokens $\Omega_{\text{most}}$ and partition it into
$\Omega_{\text{most}}^{\text{high}}$ and $\Omega_{\text{most}}^{\text{low}}$.

The raise template $\delta_R$ is applied to regions corresponding to tokens in $\Omega_{\text{least}}^{\text{low}}$ to increase their importance above that of unperturbed highest-ranked tokens, thereby altering the token ranking used by compression. 
Formally, the loss at layer $\tilde{l}$ is defined as
\begin{equation} 
\begin{aligned} 
\mathcal{L}_{\text{raise}}(\delta_R, \tilde{l}) =\;
& -\alpha_{\text{raise}} 
\sum_{\ell \in \Omega_{\text{least}}^{\text{low}}} 
\sum_{m \in \Omega_{\text{most}}^{\text{high}}} 
\log \sigma\!\left(\hat{s}^{(\tilde{l})}_\ell - \hat{s}^{(\tilde{l})}_m\right) \\[3pt]
& -\gamma_{\text{raise}} 
\frac{1}{|\Omega_{\text{least}}^{\text{low}}|\,|\mathcal{I}_{\text{ref}}|}
\sum_{\ell \in \Omega_{\text{least}}^{\text{low}}}
\sum_{i \in \mathcal{I}_{\text{ref}}}
\hat{q}^{(\tilde{l})\top}_{i}\,\hat{k}^{(\tilde{l})}_{\ell},
\end{aligned} 
\label{eq:transfer_loss_raise} 
\end{equation}
where $\hat{s}^{(\tilde{l})}$ and $\hat{k}^{(\tilde{l})}_\ell$ denote the importance score and key vector of token $\hat{v}_\ell$ at layer $\tilde{l}$, and $\{\hat{q}^{(\tilde{l})}_{i}\}_{i \in \mathcal{I}_{\text{ref}}}$ denote the query vectors of the reference text tokens used for compression, all computed from the perturbed input $(\hat{I}, T)$.
For the down template $\delta_D$, we apply $\delta_D$ uniformly to tokens in $\Omega_{\text{most}}^{\text{low}}$ to suppress their importance, ensuring their scores fall below those of $\Omega_{\text{least}}^{\text{high}}$. 
Formally, the loss at layer $\tilde{l}$ is defined as
\begin{equation} 
\begin{aligned} 
\mathcal{L}_{\text{down}}(\delta_D, \tilde{l}) =\;
& -\alpha_{\text{down}} 
\sum_{m \in \Omega_{\text{most}}^{\text{low}}} 
\sum_{\ell \in \Omega_{\text{least}}^{\text{high}}} 
\log \sigma\!\left(\hat{s}^{(\tilde{l})}_m - \hat{s}^{(\tilde{l})}_\ell\right) \\[3pt]
& +\gamma_{\text{down}} 
\frac{1}{|\Omega_{\text{most}}^{\text{low}}|\,|\mathcal{I}_{\text{ref}}|}
\sum_{m \in \Omega_{\text{most}}^{\text{low}}}
\sum_{i \in \mathcal{I}_{\text{ref}}}
\hat{q}^{(\tilde{l})\top}_{i}\,\hat{k}^{(\tilde{l})}_{m}.
\end{aligned} 
\label{eq:transfer_loss_down} 
\end{equation}

\noindent\textbf{Final Perturbation and Transfer Objective.} 
We jointly optimize $\delta_R$ and $\delta_D$ across all candidate layers ${\tilde{L}_c}$ on the surrogate model via gradient-based updates. The transfer objective is 
\begin{equation}
\begin{aligned}
\min_{\delta_R,\,\delta_D}\quad & \mathcal{L}_{\text{transfer}} 
= \frac{1}{|\tilde{L}_c|}\sum_{\tilde{l} \in \tilde{L}_c}\mathcal{L}_{\text{raise}}(\delta_R, \tilde{l}) 
+ \,\mathcal{L}_{\text{down}}(\delta_D, \tilde{l}), \\
\text{s.t.}\quad 
& \|\delta_R\|_\infty \le \epsilon_R,\quad 
\|\delta_D\|_\infty \le \epsilon_D, 
\end{aligned}
\label{eq:transfer_total}
\end{equation}

\noindent\textbf{Border-Based Perturbation.} 
After optimization, we construct the final adversarial image by applying $\delta_R$ to all border patches and $\delta_D$ to all patches belonging to the original image:
\begin{equation}
\hat{P}_j=
\begin{cases}
P_j+\delta_R, & j\in\mathcal{B},\\
P_j+\delta_D, & \text{otherwise},
\end{cases}
\qquad
\hat{I}_{\text{border}} = \{\hat{P}_j\}_{j=1}^{n_V + |\mathcal{B}|}.
\label{eq:transfer_perturbation}
\end{equation}
$\mathcal{B}$ is the set of border-aligned patches; $|\mathcal{B}|$ is the number of border tokens; and $\hat{I}_{\text{border}}$ the border-augmented image.
We adopt asymmetric perturbation budgets with $\epsilon_D < \epsilon_R$, weakly modifying the original image to preserve uncompressed performance while strongly perturbing the border to ensure its retention under compression.
As the border consists of the down template that lack task-relevant semantics. It establishes a consistent low-importance prior across models, enabling transferable compression-induced degradation

\section{Experiments}
\subsection{Experimental Settings}
This section presents the experimental settings common to both the white-box and black-box evaluations, while experiment-specific configurations are described in their respective sections.

\noindent \textbf{Victim Models and Datasets.}
Following prior work~\cite{zhao2025q, Zhang2024SparseVLMVT, Xing2024PyramidDropAY}, we evaluate our attacks on three state-of-the-art LVLMs:
\textbf{LLaVA}~\cite{liu2023visual}, \textbf{LLaVA-Next}~\cite{liu2024llavanext}, and \textbf{Qwen2.5-VL}~\cite{bai2025qwen2}.
We conduct experiments on three widely used benchmarks:
\textbf{POPE}~\cite{li2023evaluating},
\textbf{MME}~\cite{liu2024mmbench}, and
\textbf{TextVQA}~\cite{singh2019towards}.
Following standard practice, we report \textbf{accuracy (\%)} on POPE and MME, and the standard \textbf{VQA accuracy} metric~\cite{antol2015vqa} on TextVQA.
Detailed information is provided in Appendix~\ref{sec:victim_model_dataset}.


\subsection{Evaluation Metrics}
\label{subsec:eval_metric}
To evaluate compression-aware behavior, we define three metrics based on model performance $\mathcal{M}(\text{input}, \text{state})$, where $\text{input}\in\{\text{cl},\text{adv}\}$ (clean or adversarial) and $\text{state}\in\{\text{nc},\text{c}\}$ (non-compressed or compressed). These metrics measure an attack’s ability to preserve accuracy without compression while selectively degrading performance under compression:

\begin{itemize}

    \item \textbf{Uncompressed Performance Retention (UPR)}\;($\uparrow$):  
     We define $
        \text{UPR} = 
        \frac{\mathcal{M}{(\text{adv},~\text{nc})}}{\mathcal{M}{(\text{cl},~\text{nc})}}$. UPR quantifies the extent to which adversarial examples preserve model performance under uncompressed inference.

    \item \textbf{Compressed Attack Effectiveness (CAE)}\;($\uparrow$):  
    We define $\text{CAE} = 
    1 - \frac{\mathcal{M}{(\text{adv},~\text{c})}}{\mathcal{M}{(\text{cl},~\text{c})}}$, which quantifies the relative performance degradation induced by the attack under compression. 

    \item \textbf{Compression Sensitivity Gap (CSG)}\;($\uparrow$):  
       We define $ \text{CSG} = 
        \frac{\mathcal{M}{(\text{adv},~\text{nc})}}{\mathcal{M}{(\text{cl}, ~\text{nc})}} \;-\;
        \frac{\mathcal{M}{(\text{adv},~\text{c})}}{\mathcal{M}{(\text{cl},~\text{c})}}$ to captures the performance gap induced by compression. Larger values indicate benign behavior without compression and harmful effects under compression, reflecting state-dependent behavior.

\end{itemize}

\noindent \textbf{Vision Token Compression Methods.}
We evaluate our attacks on three representative compression methods \cite{Yu2025ASO, yao2025efficient_mllm_token_compression}.
\textbf{FastV}~\cite{Chen2024AnII} applies single-layer vision token compression at early LLM layers, 
with token importance guided by the final text token.
\textbf{PDrop}~\cite{Xing2024PyramidDropAY} employs progressive multi-layer compression, gradually pruning visual tokens across depth based on importance signals from the final text token.
\textbf{SparseVLM}~\cite{Zhang2024SparseVLMVT} performs multi-layer vision token compression guided by text tokens relevant to visual features, merging low-information visual tokens.

\begin{table*}[htbp]
  \centering
  \caption{Comparison between our Compression-Aware Attack (CAA) and two baselines.
CAA causes a sharp performance drop only in the compressed model while leaving the uncompressed model largely unaffected, resulting in a stable CSG that quantifies the additional risk introduced by the compression mechanism.}
  \resizebox{0.98\linewidth}{!}{
    \begin{tabular}{ll|cc|ccccc|ccccc|ccccc}
    \toprule
    \multicolumn{1}{c}{\multirow{2}[2]{*}{\textbf{Dataset}}} & \multicolumn{1}{c|}{\multirow{2}[2]{*}{\textbf{Model}}} & \multirow{2}[2]{*}{\textbf{M(cl,nc)}} & \multirow{2}[2]{*}{\textbf{M(cl,c)}} & \multicolumn{5}{c|}{\textbf{Vanilla Attack}} & \multicolumn{5}{c|}{\textbf{Random Attack}} & \multicolumn{5}{c}{\textbf{Compression-Aware Attack}} \\
          &       &       &       & \textbf{M(adv,nc)} & \textbf{M(adv,c)} & \textbf{UPR} & \textbf{CAE} & \textbf{CSG} & \textbf{M(adv,nc)} & \textbf{M(adv,c)} & \textbf{UPR} & \textbf{CAE} & \textbf{CSG} & \textbf{M(adv,nc)} & \textbf{M(adv,c)} & \textbf{UPR} & \textbf{CAE} & \textbf{CSG} \\
    \midrule
    \rowcolor[rgb]{ .929,  .929,  .929} \multicolumn{19}{c}{\textbf{FastV}} \\
    \midrule
    \multirow{3}[2]{*}{POPE} & LLaVA & 0.9775 & 0.8842 & 0.1210 & 0.1274 & 0.1238 & 0.8559 & -0.0203 & 0.9550 & 0.7910 & 0.9770 & 0.1054 & 0.0824 & 0.9196 & 0.3633 & 0.9408 & 0.5891 & \textbf{0.5299} \\
          & LLaVA-NEXT & 0.9003 & 0.8199 & 0.2866 & 0.2293 & 0.3183 & 0.7203 & -0.1559 & 0.8650 & 0.7492 & 0.9608 & 0.0862 & 0.0470 & 0.9045 & 0.3631 & 1.0047 & 0.5571 & \textbf{0.5618} \\
          & Qwen-VL & 0.9550 & 0.8199 & 0.1465 & 0.1656 & 0.1534 & 0.7980 & -0.0782 & 0.9260 & 0.7524 & 0.9696 & 0.0823 & 0.0520 & 0.8746 & 0.0611 & 0.9158 & 0.9255 & \textbf{0.8413} \\
    \midrule
    \multirow{3}[2]{*}{TextVQA} & LLaVA & 0.8817 & 0.7428 & 0.1306 & 0.1962 & 0.1481 & 0.7359 & -0.1404 & 0.7473 & 0.6331 & 0.8476 & 0.1477 & -0.0047 & 0.7354 & 0.2158 & 0.8341 & 0.7095 & \textbf{0.5435} \\
          & LLaVA-NEXT & 0.6318 & 0.5820 & 0.1541 & 0.1503 & 0.2439 & 0.7418 & -0.1345 & 0.5695 & 0.4977 & 0.9014 & 0.1448 & 0.0462 & 0.6000 & 0.3382 & 0.9497 & 0.4189 & \textbf{0.3686} \\
          & Qwen-VL & 0.9135 & 0.6662 & 0.3261 & 0.3076 & 0.3570 & 0.5383 & -0.3379 & 0.8556 & 0.6196 & 0.9366 & 0.0699 & 0.0066 & 0.8585 & 0.1916 & 0.9398 & 0.7124 & \textbf{0.6522} \\
    \midrule
    \multirow{3}[2]{*}{MME} & LLaVA & 0.8714 & 0.8360 & 0.4713 & 0.4650 & 0.5409 & 0.4438 & -0.4324 & 0.8489 & 0.7685 & 0.9742 & 0.0807 & 0.0549 & 0.8071 & 0.4823 & 0.9262 & 0.4231 & \textbf{0.3493} \\
          & LLaVA-NEXT & 0.8617 & 0.8264 & 0.4459 & 0.4331 & 0.5175 & 0.4759 & -0.4003 & 0.8424 & 0.8006 & 0.9776 & 0.0312 & 0.0088 & 0.7500 & 0.4000 & 0.9262 & 0.5160 & \textbf{0.3863} \\
          & Qwen-VL & 0.9085 & 0.7451 & 0.2051 & 0.3269 & 0.2258 & 0.5613 & -0.3149 & 0.8954 & 0.7340 & 0.9856 & 0.0149 & 0.0005 & 0.8489 & 0.2572 & 0.9344 & 0.6548 & \textbf{0.5892} \\
    \midrule
    \rowcolor[rgb]{ .929,  .929,  .929} \multicolumn{19}{c}{\textbf{SparseVLM}} \\
    \midrule
    \multirow{3}[2]{*}{POPE} & LLaVA & 0.9775 & 0.6785 & 0.1210 & 0.1274 & 0.1238 & 0.8122 & -0.0640 & 0.9550 & 0.6624 & 0.9770 & 0.0237 & 0.0007 & 0.9003 & 0.3215 & 0.9210 & 0.5262 & \textbf{0.4472} \\
          & LLaVA-NEXT & 0.9003 & 0.6913 & 0.2866 & 0.1783 & 0.3183 & 0.7421 & 0.0604 & 0.8650 & 0.6629 & 0.9608 & 0.0411 & 0.0019 & 0.8981 & 0.3248 & 0.9210 & 0.5302 & \textbf{0.5277} \\
          & Qwen-VL & 0.9550 & 0.7781 & 0.1720 & 0.3057 & 0.1801 & 0.6071 & -0.2395 & 0.9260 & 0.6116 & 0.8936 & 0.2140 & 0.1076 & 0.8854 & 0.3611 & 0.9210 & 0.5359 & \textbf{0.3336} \\
    \midrule
    \multirow{3}[2]{*}{TextVQA} & LLaVA & 0.8817 & 0.6154 & 0.1306 & 0.2306 & 0.1481 & 0.6253 & -0.2266 & 0.7473 & 0.5206 & 0.8476 & 0.1540 & 0.0016 & 0.7217 & 0.1548 & 0.8185 & 0.7485 & \textbf{0.5670} \\
          & LLaVA-NEXT & 0.6318 & 0.5042 & 0.1541 & 0.1834 & 0.2439 & 0.6363 & -0.1198 & 0.5695 & 0.4539 & 0.9014 & 0.0998 & 0.0012 & 0.6338 & 0.2726 & 1.0032 & 0.4593 & \textbf{0.4625} \\
          & Qwen-VL & 0.9135 & 0.6714 & 0.3299 & 0.3000 & 0.3249 & 0.5532 & -0.1219 & 0.8556 & 0.6299 & 0.9342 & 0.0618 & -0.0016 & 0.8331 & 0.3376 & 0.9120 & 0.4972 & \textbf{0.4092} \\
    \midrule
    \multirow{3}[2]{*}{MME} & LLaVA & 0.8714 & 0.7556 & 0.4713 & 0.4395 & 0.5409 & 0.4183 & -0.0408 & 0.8489 & 0.7358 & 0.9742 & 0.0262 & 0.0004 & 0.7500 & 0.4157 & 0.8607 & 0.4498 & \textbf{0.3105} \\
          & LLaVA-NEXT & 0.8617 & 0.8135 & 0.4459 & 0.4459 & 0.5175 & 0.4519 & -0.0307 & 0.8424 & 0.7878 & 0.9776 & 0.0316 & 0.0092 & 0.7833 & 0.4667 & 0.9090 & 0.4263 & \textbf{0.3353} \\
          & Qwen-VL & 0.9085 & 0.7614 & 0.2051 & 0.3205 & 0.2258 & 0.5791 & -0.1952 & 0.8424 & 0.7009 & 0.9272 & 0.0795 & 0.0067 & 0.8706 & 0.3495 & 0.9583 & 0.5410 & \textbf{0.4993} \\
    \bottomrule
    \end{tabular}%
    }
  \label{tab:white_main_result}%
\end{table*}%

\begin{table}[htbp]
  \centering
  \caption{CAA performance under different token retention rates on POPE.
  }
\resizebox{0.95\linewidth}{!}{
    \begin{tabular}{lc|ccccccc}
    \toprule
    \multicolumn{1}{c}{\textbf{Model}} & \textbf{Ratio} & \textbf{M(cl, nc)} & \textbf{M(cl,c)} & \textbf{M(adv,nc)} & \textbf{M(adv,c)} & \textbf{UPR} & \textbf{CAE} & \textbf{CSG} \\
    \midrule
    \multirow{5}[2]{*}{LLaVA} & 0.5   & \multirow{5}[2]{*}{0.9775 } & 0.9518  & 0.9003  & 0.6049  & 0.9210  & 0.3645  & 0.2855  \\
          & 0.4   &       & 0.9404  & 0.9102  & 0.5618  & 0.9312  & 0.4026  & 0.3337  \\
          & 0.3   &       & 0.9164  & 0.9228  & 0.4759  & 0.9440  & 0.4807  & 0.4247  \\
          & 0.2   &       & 0.8842  & 0.9196  & 0.3633  & 0.9408  & 0.5891  & 0.5299  \\
          & 0.1   &       & 0.7395  & 0.9646  & 0.1593  & 0.9868  & 0.7846  & 0.7714  \\
    \midrule
    \multirow{5}[2]{*}{Qwen-VL} & 0.5   & \multirow{5}[2]{*}{0.9550 } & 0.9325  & 0.8167  & 0.3794  & 0.8552  & 0.5931  & 0.4483  \\
          & 0.4   &       & 0.9165  & 0.8578  & 0.2941  & 0.8982  & 0.6791  & 0.5773  \\
          & 0.3   &       & 0.9003  & 0.8746  & 0.2219  & 0.9158  & 0.7535  & 0.6693  \\
          & 0.2   &       & 0.8199  & 0.8746  & 0.0611  & 0.9158  & 0.9255  & 0.8413  \\
          & 0.1   &       & 0.7846  & 0.9108  & 0.1083  & 0.9537  & 0.8620  & 0.8157  \\
    \bottomrule
    \end{tabular}%
    }
  \label{tab:caa_other_rate}%
\end{table}%

\begin{table}[t]
  \centering
  \caption{Ablation on Selective Perturbation. "CAA w/ Full" perturbs the entire image. "Most-only" targets high-importance regions to simultaneously corrupt semantics and ensure retention during compression.}
    \resizebox{0.92\linewidth}{!}{
    \begin{tabular}{l|ccc|ccc|ccc}
    \toprule
    \multirow{2}[2]{*}{\textbf{Model}} & \multicolumn{3}{c|}{\textbf{Most-Only}} & \multicolumn{3}{c|}{\textbf{CAA-Full}} & \multicolumn{3}{c}{\textbf{CAA}} \\
          & \textbf{UPR} & \textbf{CAE} & \textbf{CSG} & \textbf{UPR} & \textbf{CAE} & \textbf{CSG} & \textbf{UPR} & \textbf{CAE} & \textbf{CSG} \\
    \midrule
    \rowcolor[rgb]{ .906,  .902,  .902} \multicolumn{10}{c}{\textbf{POPE}} \\
    \midrule
    LLaVA & 0.8818  & 0.1893  & 0.0712  & 0.8235  & 0.6146  & 0.4381  & 0.9408  & 0.5891  & \textbf{0.5299 } \\
    Qwen-VL & 0.6936  & 0.7427  & 0.4363  & 0.7272  & 0.7137  & 0.4410  & 0.9158  & 0.9255  & \textbf{0.8413 } \\
    \midrule
    \rowcolor[rgb]{ .906,  .902,  .902} \multicolumn{10}{c}{\textbf{TextVQA}} \\
    \midrule
    LLaVA & 0.7116  & 0.2971  & 0.0087  & 0.6769  & 0.6060  & 0.2828  & 0.8341  & 0.7095  & \textbf{0.5435 } \\
    Qwen-VL & 0.6895  & 0.5650  & 0.2545  & 0.7695  & 0.6288  & 0.3982  & 0.9398  & 0.7124  & \textbf{0.6522 } \\
    \bottomrule
    \end{tabular}%
    }
  \label{tab:ab_selective_pert}%
\end{table}%

\subsection{Compression Aware-Attack Evaluation}
\label{sec:white_eval}

\subsubsection{Experimental Setup}
\label{sec:exp_setup_caa}
We evaluate CAA against two baseline attacks. 
(i) \textbf{Vanilla Attack}:
Existing adversarial attacks rely on gradient-based optimization and designed for non-compressed models~\cite{bailey2024image, qi2024visual}, therefore they cannot be directly applied to compressed inference which involves discrete operations (e.g., ranking and pruning).
We thus evaluate vanilla perturbations crafted on the non-compressed model under compression to test whether they can trigger compression-specific failures.
Implementation details of vanilla adversarial attacks are provided in Appendix~\ref{subsec:vanilla_attack_supp}. 
(ii) \textbf{Random Attack}:
This baseline applies random noise to the entire input, without targeting model predictions or the compression mechanism. 
The uncontrolled perturbation cannot serve as a reliable attack for quantifying compression-induced robustness degradation.

Following prior work~\cite{Zhang2024SparseVLMVT, Xing2024PyramidDropAY, zoudon, zhao2024stitch}, we focus on compression settings with initial retention rates in the range of 0.1 to 0.5 to balance efficiency and performance. For multi-layer compression, we target the first compression layer and its retention rate. 
Unless otherwise specified, all white-box experiments are conducted on FastV~\cite{Chen2024AnII} with compression applied at the second layer and a retention rate of 0.2, using a perturbation budget of $32/255$. We also evaluate different perturbation budgets in Appendix~\ref{subsec:pert_budget}.

\subsubsection{Attack Performance}
\label{sec:main_white_box}

Results for the commonly used retention rate of 0.2 are reported in Table~\ref{tab:white_main_result} and Table~\ref{tab:caa_mian_reulst_supp} (Appendix~\ref{subsec:main_caa_supp}), while additional results for other retention rates are provided in Table~\ref{tab:caa_other_rate} and Table~\ref{tab:caa_other_rate_textvqa} in Appendix \ref{subsec:main_caa_supp}.
Under CAA, compressed models suffer substantial performance degradation (average drop = 56.60\%), while uncompressed models largely preserve accuracy (UPR = 91.99\%), yielding a high average CSG of 47.61\%. 
In contrast, vanilla adversarial attacks exhibit highly unstable and often negative CSG values, indicating that they primarily disrupt uncompressed inference, and their adversarial effects are frequently attenuated by compression.
Random attacks yield consistently positive but marginal CSG (2.36\%), reflecting their inability to reliably exploiting this vulnerability.

\begin{figure}[t]
    \centering
    \begin{subfigure}{0.42\linewidth}
        \centering\includegraphics[width=\linewidth]{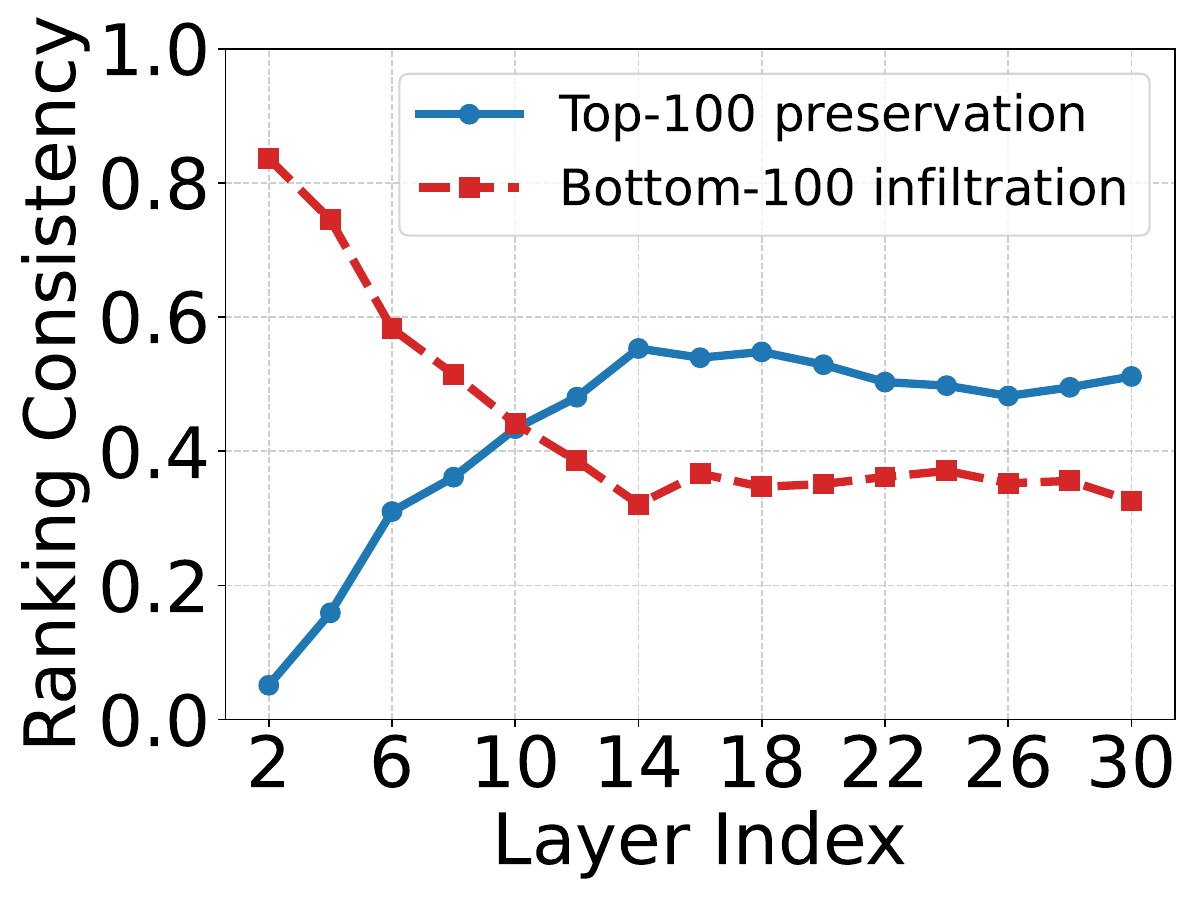} 
        \caption{LLaVA}
    \end{subfigure}
    \
    \begin{subfigure}{0.42\linewidth}
        \centering\includegraphics[width=\linewidth]{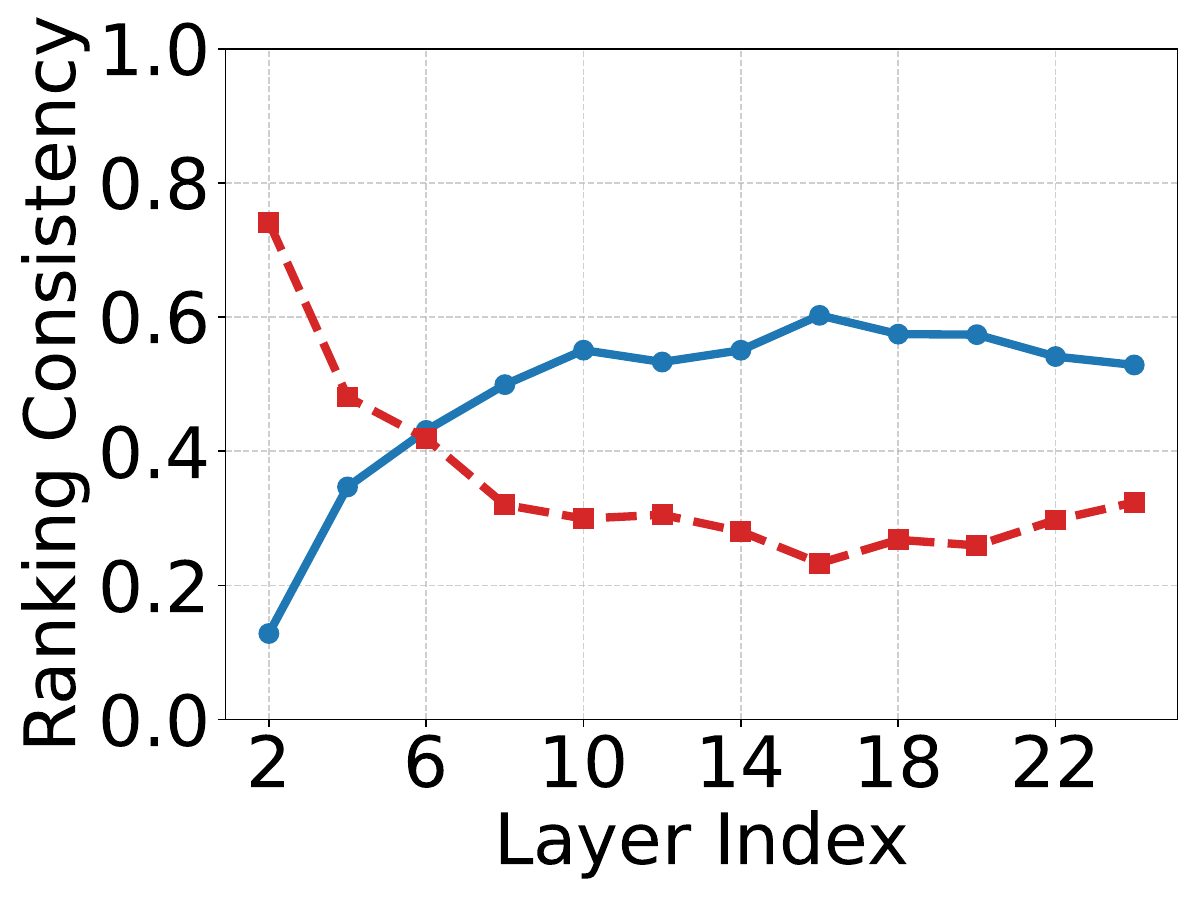}
        \caption{Qwen-VL}
    \end{subfigure}

    \caption{
    Layer-wise Top-100 preservation and Bottom-100 infiltration rates under adversarial attack (attack at Layer 2).
    }
    \label{fig:ranking_trace}
\end{figure}

To understand why CAA induces compression-specific failures, we analyze how perturbations alter token-importance rankings across layers. Figure~\ref{fig:ranking_trace} tracks the evolution of Top-100 preservation and Bottom-100 infiltration rates, defined in Section~\ref{sec:mechanism_of_robust}. At the attack layer (layer 2, retention rate 0.2), CAA disrupts the ranking by promoting originally irrelevant Bottom-100 tokens into the Top-100 set. Under uncompressed inference, this distortion is transient, as subsequent layers reallocate attention to correct the misaligned focus. Under compression, however, token selection becomes irreversible: noisy tokens are locked into the Top-100 set, permanently discarding essential visual evidence and causing inevitable performance collapse.
\begin{table}[htbp]
  \centering
  \caption{Ablation of CAA Components. Three variants are formed by progressively removing hierarchical ranking, semantic erasure, and query-guided term.
}
   \resizebox{0.95\linewidth}{!}{
    \begin{tabular}{l|rrr|rrr|rrr}
    \toprule
    \multirow{2}[2]{*}{\textbf{Attack}} & \multicolumn{3}{c|}{\textbf{POPE}} & \multicolumn{3}{c|}{\textbf{TextVQA}} & \multicolumn{3}{c}{\textbf{MME}} \\
          & \multicolumn{1}{c}{\textbf{UPR}} & \multicolumn{1}{c}{\textbf{CAE}} & \multicolumn{1}{c|}{\textbf{CSG}} & \multicolumn{1}{c}{\textbf{UPR}} & \multicolumn{1}{c}{\textbf{CAE}} & \multicolumn{1}{c|}{\textbf{CSG}} & \multicolumn{1}{c}{\textbf{UPR}} & \multicolumn{1}{c}{\textbf{CAE}} & \multicolumn{1}{c}{\textbf{CSG}} \\
    \midrule
    CAA w/o Hierarchy & 0.9433  & 0.5532  & 0.4965  & 0.7738  & 0.5485  & 0.3223  & 0.9705  & 0.3282  & 0.2987  \\
    CAA w/o Erasure & 0.9357  & 0.5883  & 0.5240  & 0.7728  & 0.6236  & 0.3964  & 0.9445  & 0.3713  & 0.3157  \\
    CAA w/o Query & 0.9597  & 0.4896  & 0.4493  & 0.8240  & 0.5866  & 0.4105  & 0.8937  & 0.4390  & 0.3327  \\
    CAA   & 0.9408  & 0.5891  & \textbf{0.5299}  & 0.8341  & 0.7095  & \textbf{0.5435}  & 0.9262  & 0.4231  & \textbf{0.3493}  \\
    \bottomrule
    \end{tabular}%
    }
  \label{tab:ab-key-components}%
\end{table}%

\subsubsection {The Impact of Selective Perturbation}
\label{sec:selective_pert}

To assess the necessity of selective perturbation, we compare CAA with two alternative attack variants.  
(i) \textit{CAA-Full}, which perturbs the entire image, and (ii) \textit{Most-Only}, which perturbs only high-importance tokens to corrupt semantics while preserving their ranks for retention under compression.
As shown in Table~\ref{tab:ab_selective_pert}, \textit{CAA-Full} substantially degrades non-compressed inference (Avg. UPR $0.9152 \rightarrow 0.7691$),  resulting in a low CSG ($0.3475$). Meanwhile, the \textit{Most-Only attack} attains high CAE (up to $0.7427$) but exhibits a low average CSG ($0.1919$), as conflicting objectives between semantic corruption and rank preservation limit its effectiveness.
\subsubsection{Impact of the BPR Ranking Loss}
\label{sec:ab_bpr_loss}

\begin{figure}[t]
    \centering
    \begin{subfigure}{0.4\linewidth}
        \centering\includegraphics[width=\linewidth]{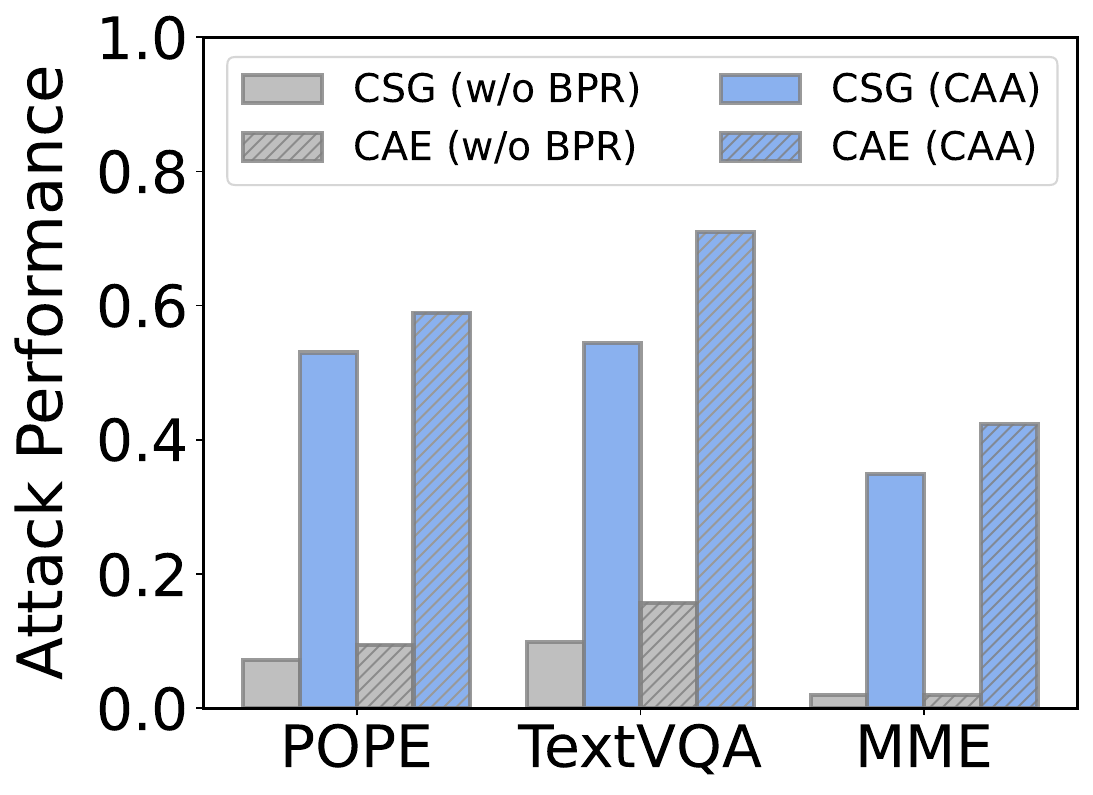} 
        \caption{LLaVA}
        \label{fig:ab_bpr_llava}
    \end{subfigure}
    \
    \begin{subfigure}{0.4\linewidth}
        \centering\includegraphics[width=\linewidth]{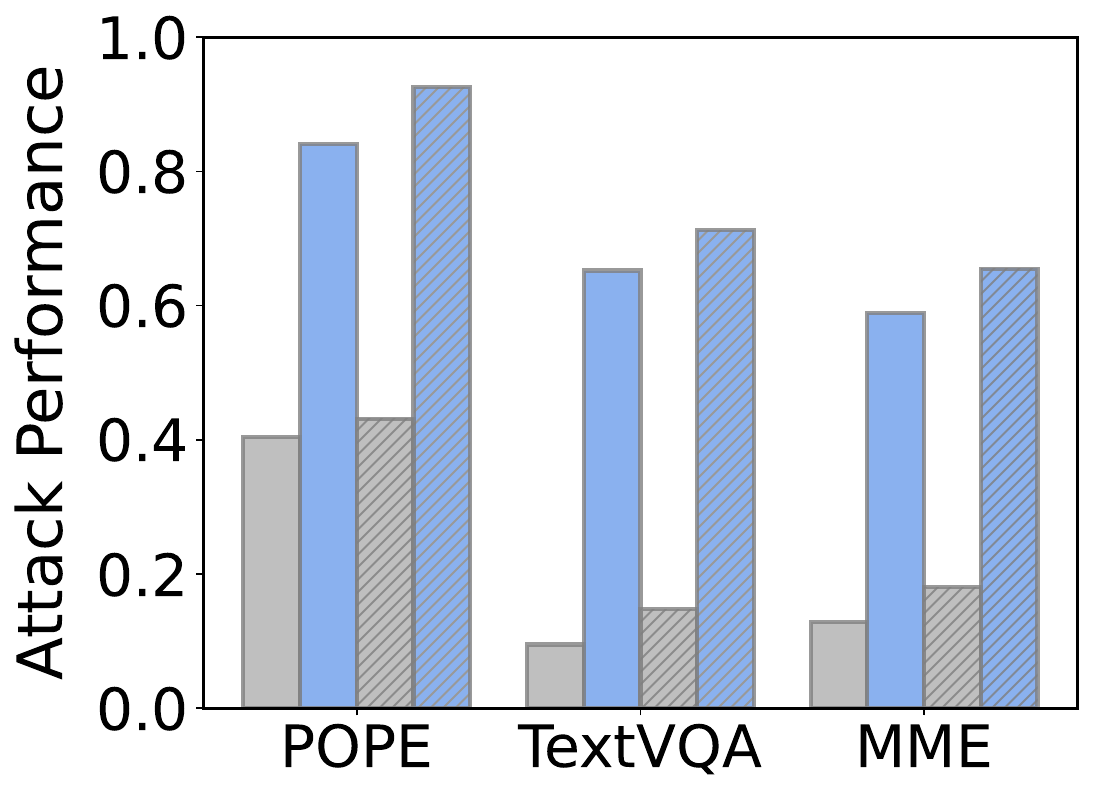}
        \caption{Qwen-VL}
        \label{fig：ab_bpr_qwen}
    \end{subfigure}

    \caption{Comparison between CAA, which explicitly optimizes relative token ordering via BPR, and a variant that maximizes only absolute importance scores.
    }
    \label{fig:ab_bpr}
\end{figure}

Comparing the BPR-based objective with a variant that directly optimizes absolute importance scores (\textit{CAA w/o BPR}) highlights the critical role of BPR. Fig.~\ref{fig:ab_bpr} shows that omitting BPR collapses the average CAE from $0.6691$ to $0.1715$ and reduces CSG by $0.4482$. This confirms that increasing absolute scores is insufficient for Top-$k$ inclusion, as compression relies on relative ranking. BPR addresses this by enforcing pairwise constraints for stable rank manipulation.
\subsubsection{Impact of Key Components.}
\label{sec:ac_white_key_components}

We ablate three key components of CAA, with results summarized in Table~\ref{tab:ab-key-components} for LLaVA and Table \ref{tab:ab-key-components_qwen} for Qwen-VL (Appendix \ref{subsec:ab_caa_key_qwen}).

\noindent \textbf{Hierarchical Ranking Constraint} (\textit{CAA w/o Hierarchy}).  
Removing the hierarchical ranking constraint substantially weakens the attack, reducing CAE from $0.6691$ to $0.5714$ and CSG to $0.4954$. Without explicit intra-group ranking optimization, informative tokens can still remain highly ranked, enabling correct reasoning.

\noindent \textbf{Semantic Erasure} (\textit{CAA w/o Erasure}).  
Omitting semantic erasure also degrades attack effectiveness, lowering average CAE ($0.6691 \to 0.5807$) and CSG ($0.5842 \to 0.4887$). We adopt semantic erasure instead of fixed “meaningless” targets, as the latter conflicts with the query-guided alignment required for effective ranking manipulation.

\noindent \textbf{Query-Guided Optimization} (\textit{CAA w/o Query}).  
Removing query guidance leads to a clear drop in CSG ($0.5842 \to 0.4864$), as token importance is determined by key–query dot products, making ranking manipulation less stable and effective without explicit guidance.

\subsubsection{Effectiveness to Configuration Mismatch}
\label{sec:attack_mismatch}

We study CAA under mismatched attack-time and test-time compression seetings.

\begin{figure}[t]
    \centering
    \begin{subfigure}{0.4\linewidth}
        \centering
   \includegraphics[width=\linewidth]{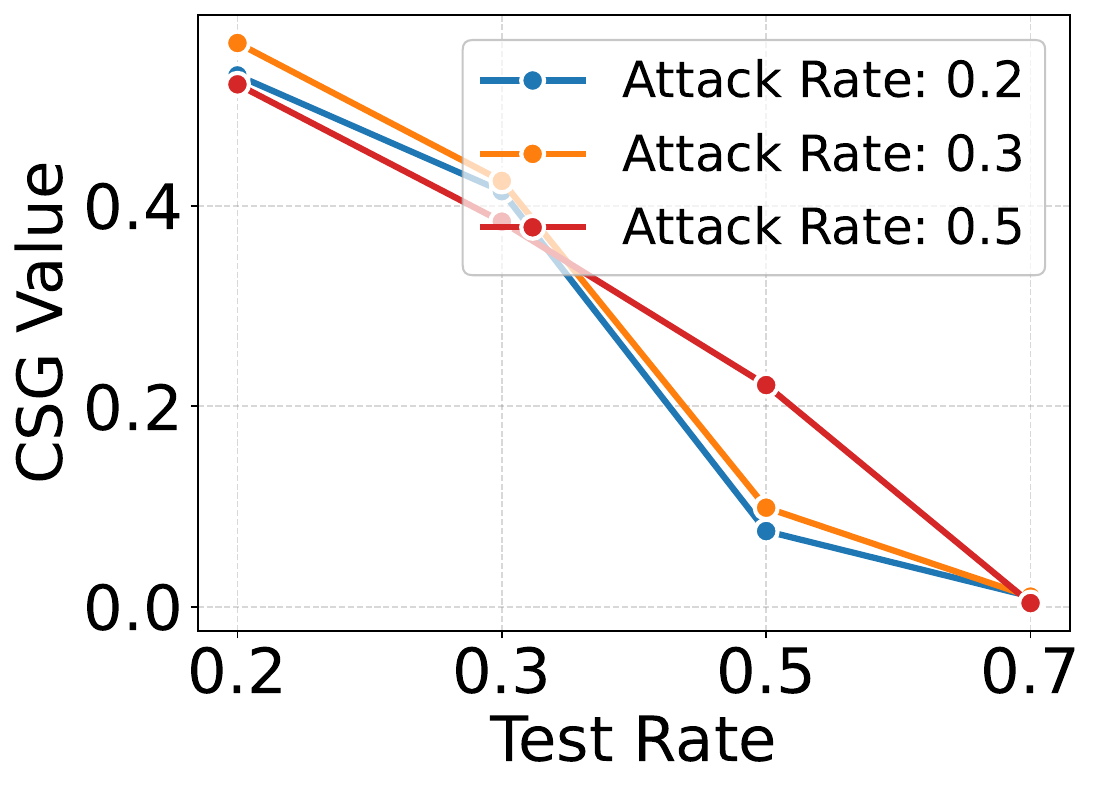} 
        \caption{Rate Mismatch}
        \label{fig:ab_attack_num_llava}
    \end{subfigure}
    \begin{subfigure}{0.4\linewidth}
        \centering
        \includegraphics[width=\linewidth]{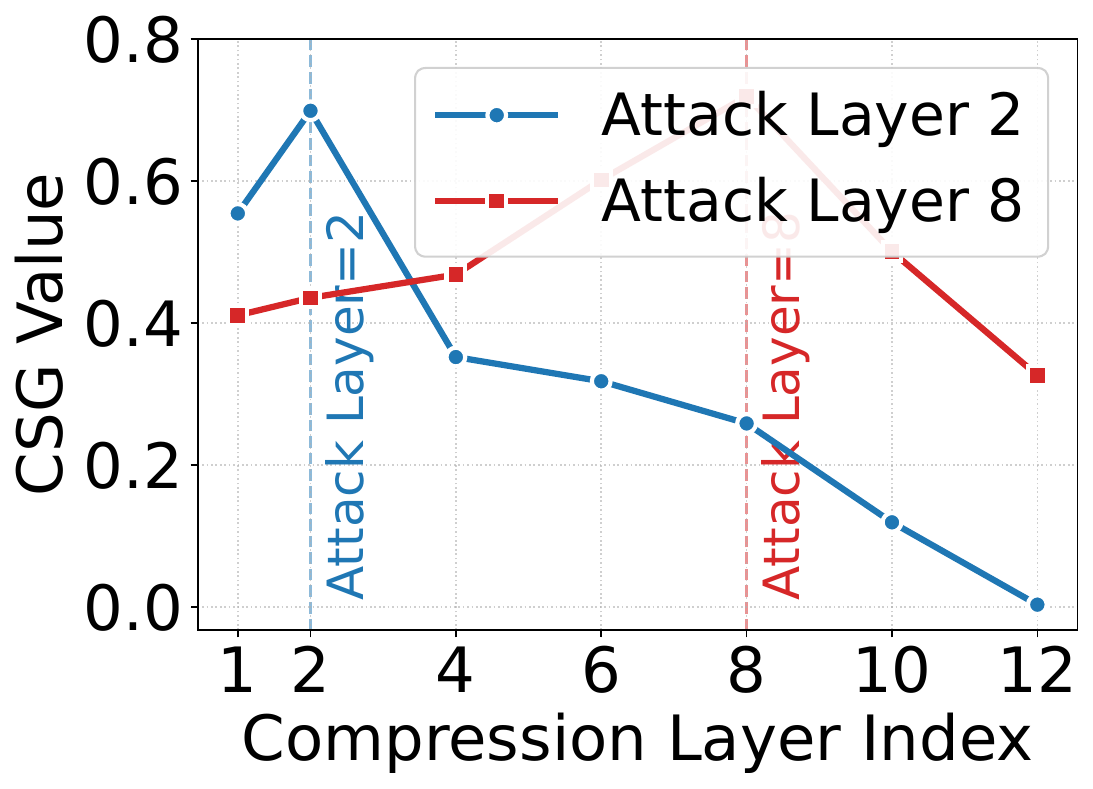} 
        \caption{Layer Mismatch}
        \label{fig:ab_attack_layer_llava}
    \end{subfigure}

    \caption{
   (a) Mismatch between attack and test retention rates. (b) Mismatch between the attack layer and the compression layer. Evaluated on LLaVA with the POPE dataset.  
    }
    \label{fig:ab_attack_num_and_layer}
\end{figure}

\noindent \textbf{Retention Rate Mismatch.}
We generate adversarial examples with attack retention rates in $\{0.5, 0.3, 0.2\}$ and evaluate them under test retention rates in $\{0.7, 0.5, 0.3, 0.2\}$. As shown in Fig.~\ref{fig:ab_attack_num_llava} and Fig.~\ref{fig:ab_attack_num_and_layer_Qwen} (Appendix~\ref{subsec:rate_layer_mismatch}), CAA sustains strong attack impact under moderate  mismatch: when the test retention rate is higher, retained clean tokens dilute the attack effect, whereas when it is lower or equal, retained tokens are dominated by the perturbed set.

\noindent \textbf{Layer Mismatch.}
We vary the attack layer around representative compression layers (Layers 2 and Layer 8) and report results in Fig.~\ref{fig:ab_attack_layer_llava} and Fig.~\ref{fig:ab_attack_num_and_layer_qwen} (Appendix~\ref{subsec:rate_layer_mismatch}). Attacks are remain effective under moderate layer mismatch. For example, on LLaVA-POPE with compression at Layer 2, attacking the same layer achieves a CSG of 0.6986, while attacking Layer 8 still yields 0.4106.

Our analysis suggests an intuitive strategy when compression configurations are unknown: perturb a larger set of low-importance tokens and target later layers, thereby maximizing perturbed token overlap and preserving promoted tokens across layers.

\subsection{Transfer Attack Evaluation}
\label{sec:transfer_eval}

\subsubsection{Experimental Setup}
\label{sec:exp_setup_tcaa}

In the black-box setting, the attacker does not know the exact compression layers or retention rates, and therefore enforces the desired token ranking over a plausible range of layers. The attack remains effective as long as this range includes the earliest compression layer within the attack-effective regime, since ranking errors introduced there propagate through subsequent compression stages. As shown in Section~\ref{sec:white_eval}, CAA remains effective across a broad range of retention rates ($r\in[0.1,0.4]$), offering a practical margin against configuration uncertainty.
To derive such a candidate range, the attacker first obtains a coarse estimate of the target model’s \emph{average retention rate} $\bar{r}$ (i.e., the average fraction of visual tokens retained per layer), for example from public documentation or a small number of API-side latency probes. 
Given $\bar{r}$, the attacker enumerates compression configurations consistent with common deployment practices reported in prior work~\cite{Zhang2024SparseVLMVT, Xing2024PyramidDropAY, Chen2024AnII, shang2025llava, zhao2024stitch}.
Specifically, following the literature, we summarize two practical constraints:
(i) \textbf{Rate constraints:} the retention rate at the first compression layer is drawn from $\{0.1, 0.2, \ldots, 0.9\}$, and at subsequent stages decreases progressively as 
$r^{(\tilde{l}_i)} = \max\!\left(\tfrac{1}{2} r^{(\tilde{l}_{i-1})},\, 0.1\right)$;
(ii) \textbf{Layer constraints:} successive compression layers, indexed by $\tilde{l}_i$ and $\tilde{l}_{i+1}$, satisfy $\tilde{l}_{i+1} - \tilde{l}_i \ge 4$ and compression typically starts after the first layer.
Under these constraints, the attacker enumerates all feasible compression configurations $(\tilde{L}, R)$ via depth-first search, where $\tilde{L}$ denotes compression layers and $R$ their retention rates.
An example for  $\bar{r} \le 0.2$  is provided in Appendix~\ref{subsec:enum_example}.

Following prior implementations~\cite{Zhang2024SparseVLMVT, Xing2024PyramidDropAY, Zhuang2025ST3AM, Yang2024VisionZipLI}, 
we focus on a commonly used regime with average retention rate $\bar{r}\in[0.1,0.4]$, where compression is applied at most three times.
Enumerating all configurations within this regime, we select layers $1$-$11$ as the candidate attack range $\tilde{L}_c$.
This choice covers over $80\%$ of feasible configurations, measured by the fraction for which the earliest compression layer with retention rate in the attack-effective regime ($r\in[0.1,0.4]$) falls within $\tilde{L}_c$.
Perturbation templates that increase token importance are assigned a larger budget ($255/255$), as they are applied to the border region, whereas down templates use a smaller budget ($16/255$). Covering approximately $30\%$ of the visual patches after border augmentation, the border is sufficient to dominate the compressed representation, given that the maximum attack-effective retention rate is $0.4$. 
Visually, this $30\%$ border mimics natural margins or background regions, ensuring the adversarial examples remain plausible and unobtrusive in realistic settings.

\subsubsection{Attack Performance for T-CAA}
\label{sec:main_black_box}

\begin{table}[htbp]
  \centering
  \caption{T-CAA performance across different models and compression configurations. Rate means retention rate.
  }
  \resizebox{0.99\linewidth}{!}{
    \begin{tabular}{ccc|cc|cc|cc}
    \toprule
    \multicolumn{2}{c}{\textbf{Configuration}} & \multirow{2}[2]{*}{\textbf{Avg. Rate}} & \multicolumn{2}{c|}{\textbf{LLaVA->LLaVA-NEXT}} & \multicolumn{2}{c|}{\textbf{LLaVA->Qwen-VL}} & \multicolumn{2}{c}{\textbf{Qwen-VL->LLaVA}} \\
    \textbf{Layer} & \textbf{Rate} &       & \textbf{CAE} & \textbf{CSG} & \textbf{CAE} & \textbf{CSG} & \textbf{CAE} & \textbf{CSG} \\
    \midrule
    \multirow{3}[2]{*}{2} & 0.3   & 0.344  & 0.2795  & 0.1694  & 0.3491  & 0.2095  & 0.4061  & 0.2207  \\
          & 0.2   & 0.250  & 0.5461  & 0.4361  & 0.5338  & 0.3943  & 0.5791  & 0.3937  \\
          & 0.1   & 0.156  & 0.6212  & 0.5111  & 0.6591  & 0.5195  & 0.6686  & 0.4833  \\
    \midrule
    \multirow{2}[2]{*}{8} & 0.2   & 0.400  & 0.4237  & 0.3136  & 0.3498  & 0.2102  & 0.4853  & 0.2999  \\
          & 0.1   & 0.325  & 0.5295  & 0.4194  & 0.3303  & 0.1908  & 0.5515  & 0.3662  \\
    \midrule
    10    & 0.1   & 0.381  & 0.4014  & 0.2914  & 0.4014  & 0.2618  & 0.3868  & 0.2014  \\
    \midrule
    \multirow{4}[2]{*}{2,8} & 0.5,0.25 & 0.344  & 0.2641  & 0.1540  & 0.3064  & 0.1668  & 0.4183  & 0.2329  \\
          & 0.4,0.2 & 0.288  & 0.3845  & 0.2745  & 0.4239  & 0.2843  & 0.4183  & 0.2330  \\
          & 0.3,0.15 & 0.231  & 0.4688  & 0.3587  & 0.4573  & 0.3177  & 0.4504  & 0.2650  \\
          & 0.2,0.1 & 0.175  & 0.4999  & 0.3899  & 0.5861  & 0.4465  & 0.6230  & 0.4376  \\
    \midrule
    \multirow{4}[2]{*}{2,10} & 0.5,0.25 & 0.359  & 0.2501  & 0.1400  & 0.3286  & 0.1891  & 0.4091  & 0.2238  \\
          & 0.4,0.2 & 0.300  & 0.4233  & 0.3133  & 0.4468  & 0.3073  & 0.4437  & 0.2583  \\
          & 0.3,0.15 & 0.241  & 0.4662  & 0.3561  & 0.5050  & 0.3654  & 0.4391  & 0.2537  \\
          & 0.2,0.1 & 0.181  & 0.5077  & 0.3977  & 0.5966  & 0.4570  & 0.5109  & 0.3255  \\
    \midrule
    \multirow{2}[2]{*}{8,16} & 0.3,0.15 & 0.400  & 0.3455  & 0.2355  & 0.3313  & 0.1917  & 0.3818  & 0.1965  \\
          & 0.2,0.1 & 0.350  & 0.3954  & 0.2853  & 0.3715  & 0.2319  & 0.3817  & 0.1963  \\
    \midrule
    \multirow{5}[2]{*}{2,8,16} & 0.7,0.35,0.175 & 0.369  & 0.2477  & 0.1377  & 0.2896  & 0.1500  & 0.3934  & 0.2080  \\
          & 0.5,0.25,0.125 & 0.281  & 0.2664  & 0.1564  & 0.3136  & 0.1741  & 0.4115  & 0.2261  \\
          & 0.4,0.2,0.1 & 0.238  & 0.4308  & 0.3207  & 0.3551  & 0.2155  & 0.4265  & 0.2411  \\
          & 0.3,0.15,0.1 & 0.206  & 0.4688  & 0.3587  & 0.4573  & 0.3177  & 0.4843  & 0.2989  \\
          & 0.2,0.1,0.1 & 0.175  & 0.5249  & 0.4149  & 0.5787  & 0.4391  & 0.5787  & 0.3933  \\
    \midrule
    \multirow{2}[2]{*}{8,16,24} & 0.3,0.15,0.1 & 0.388  & 0.3235  & 0.2134  & 0.2973  & 0.1577  & 0.3950  & 0.2096  \\
          & 0.2,0.1,0.1 & 0.350  & 0.3333  & 0.2233  & 0.3028  & 0.1632  & 0.4351  & 0.2497  \\
    \bottomrule
    \end{tabular}%
    }
  \label{tab:caa_transfer}%
\end{table}%


\noindent \textbf{ Transferability Across Models and Compression Configurations.}
We focus on the commonly used average retention-rate regime $\bar{r}\in[0.1,0.4]$, and evaluate representative single-layer and multi-layer compression configurations that realize similar average retention rates\footnote{We omit multi-layer settings with a first compression layer retention rate of 0.1, as under the setup in Section~\ref{sec:exp_setup_tcaa} this degenerates into a single-layer compression with retention rate 0.1.}. To hedge against uncertainty in the target compression configuration, adversarial examples are generated once on a surrogate model using a fixed retention rate $r=0.4$ across layers $1$-$11$, 
and reused across all compression settings.
Consequently, UPR is fixed per transfer pair: 0.8899 (LLaVA $\rightarrow$ LLaVA-Next), 0.8604 (LLaVA $\rightarrow$ Qwen), and 0.8146 (Qwen $\rightarrow$ LLaVA).
As shown in Table~\ref{tab:caa_transfer} and Table~\ref{tab:caa_transfer_other_conf} for POPE (Appendix~\ref{subsec:transfer_attack_supp}), T-CAA demonstrates robust transferability, achieving an average CSG of 29.97\%. These results confirm that T-CAA reliably exposes compression vulnerabilities in black-box settings, without requiring precise knowledge of either the target model or its compression configuration.

\begin{table}[t]
  \centering
  \caption{Comparison of single- ("Final") vs. multi-token ("Multi") reference strategies at similar avg. retention rates.}
  \resizebox{0.95\linewidth}{!}{
    \begin{tabular}{cccl|ccc|ccc}
    \toprule
    \multicolumn{2}{c}{\textbf{Configuration}} & \multirow{2}[2]{*}{\textbf{Avg. Rate}} & \multicolumn{1}{c|}{\multirow{2}[2]{*}{\textbf{Model}}} & \multicolumn{3}{c|}{\textbf{Final->Multi}} & \multicolumn{3}{c}{\textbf{Final->Final}} \\
    \textbf{Layer} & \textbf{Rate} &       &       & \textbf{UPR} & \textbf{CAE} & \textbf{CSG} & \textbf{UPR} & \textbf{0.5338 } & \textbf{CSG} \\
    \midrule
    \multirow{2}[2]{*}{2} & \multirow{2}[2]{*}{0.2} & \multirow{2}[2]{*}{0.250 } & LLaVA->Qwen-VL & 0.8388  & 0.5273  & 0.3661  & 0.8604  & 0.5338  & 0.3943  \\
          &       &       & Qwen-VL->LLaVA & 0.8145  & 0.4604  & 0.2750  & 0.8146  & 0.5412  & 0.3558  \\
    \midrule
    \multirow{2}[2]{*}{2,8} & \multirow{2}[2]{*}{0.3,0.15} & \multirow{2}[2]{*}{0.231 } & LLaVA->Qwen-VL & 0.8388  & 0.5388  & 0.3777  & 0.8604  & 0.4573  & 0.3177  \\
          &       &       & Qwen-VL->LLaVA & 0.8145  & 0.4778  & 0.2923  & 0.8146  & 0.4504  & 0.2650  \\
    \midrule
    \multirow{2}[2]{*}{2,8,16} & \multirow{2}[2]{*}{0.4,0.2,0.1} & \multirow{2}[2]{*}{0.238 } & LLaVA->Qwen-VL & 0.8388  & 0.3456  & 0.1845  & 0.8604  & 0.3551  & 0.2155  \\
          &       &       & Qwen-VL->LLaVA & 0.8145  & 0.4256  & 0.2401  & 0.8146  & 0.4265  & 0.2411  \\
    \bottomrule
    \end{tabular}%
    }
  \label{tab:transer_ref_text}%
\end{table}%

\noindent \textbf{Effect of Text-Reference Strategies on Transfer Attacks.}
We examine whether different strategies for computing visual-token importance influence transfer attack performance. Specifically, we compare (i) using the final text token as the reference \cite{Chen2024AnII, Xing2024PyramidDropAY} and (ii) jointly using multiple text tokens \cite{Zhang2024SparseVLMVT}. Results in Table~\ref{tab:transer_ref_text} for POPE show negligible differences between the two strategies. This is attributed to the border-based design of T-CAA, where visually uninformative border tokens are inherently ranked as low-importance regardless of the text prompt. As a result, both strategies yield comparable performance.

\subsubsection{Comparison with Traditional Black-box Adversarial Attacks}
\label{sec:black_basline}


\begin{table}[t]
  \centering
  \caption{Comparison of T-CAA with two representative traditional black-box adversarial attacks.}  \resizebox{0.92\linewidth}{!}{
    \begin{tabular}{l|ccc|ccc|ccc}
    \toprule
    \multicolumn{1}{c|}{\multirow{2}[2]{*}{\textbf{Target Model}}} & \multicolumn{3}{c|}{\textbf{HSJA}} & \multicolumn{3}{c|}{\textbf{RayS}} & \multicolumn{3}{c}{\textbf{T-CAA}} \\
          & \textbf{UPR} & \textbf{CAE} & \textbf{CSG} & \textbf{UPR} & \textbf{CAE} & \textbf{CSG} & \textbf{UPR} & \textbf{CAE} & \textbf{CSG} \\
    \midrule
    \rowcolor[rgb]{ .906,  .902,  .902} \multicolumn{10}{c}{\textbf{POPE}} \\
    \midrule
    LLaVA & 0.9778  & 0.1192  & 0.0970  & 0.9871  & 0.1071  & 0.0942  & 0.8146  & 0.5791  & \textbf{0.3937 } \\
    Qwen-VL & 0.9915  & 0.0610  & 0.0525  & 0.9271  & 0.1610  & 0.0881  & 0.8604  & 0.5338  & \textbf{0.3943 } \\
    \midrule
    \rowcolor[rgb]{ .906,  .902,  .902} \multicolumn{10}{c}{\textbf{TextVQA}} \\
    \midrule
    LLaVA & 0.7146  & 0.2745  & -0.0109 & 0.6689  & 0.3286  & -0.0024 & 0.9256  & 0.5024  & \textbf{0.4280 } \\
    Qwen-VL & 0.5764  & 0.2986  & -0.1251 & 0.9794  & 0.0236  & 0.0030  & 0.8798  & 0.4305  & \textbf{0.3103 } \\
    \bottomrule
    \end{tabular}%
    }
  \label{tab:black_basline}%
\end{table}%

Although query limits and variable compression states typically invalidate traditional black-box attacks, we evaluate T-CAA against representative baselines under a relaxed threat model. This setting allows repeated queries for evaluation, a condition that is rarely met in real-world deployments.
We consider two label-only attacks: HSJA~\cite{chen2020hopskipjumpattack}, which estimates gradients via binary feedback, and RayS~\cite{chen2020rays}, which employs discrete ray search for boundary detection. Both are tested in the untargeted setting (Section~\ref{sec:exp_setup_caa}) with a query budget of 600 and $\epsilon=32/255$.
As shown in Table~\ref{tab:black_basline}, T-CAA substantially outperforms both baselines. All results are obtained under single-layer compression applied at Layer 2 with a retention rate of $0.2$. HSJA and RayS struggle to probe a consistent decision boundary, as perturbations alter the set of retained tokens, effectively changing the compressed model state with each query. Even when successful, they indiscriminately degrade both compressed and uncompressed inference. In contrast, T-CAA explicitly targets the compression mechanism, inducing compression-specific failures and offering deeper insights for safer compression design.

\subsubsection{Ablation Study for T-CAA}
\label{subsubsec:ab_black}
We also perform ablation studies on T-CAA to assess the effects of the perturbation region, optimization strategy, and down-template (Appendix~\ref{subsec:t_ab_attack_component}).
\section{Discussion}
\label{sec:discussion}

\begin{table}[t]
  \centering
  \caption{
  Adaptive Defense Performance. We compare the undefended baseline ("None") against random and importance-based region removal defenses guided by reference LVLMs.
  }
  \resizebox{0.8\linewidth}{!}{
    \begin{tabular}{c|cc|ccc}
    \toprule
    \textbf{Mask type + Ref Model} & \textbf{M(cl,nc)} & \textbf{M(cl,c)} & \textbf{M(adv,nc)} & \textbf{M(adv,c)} & \textbf{CSG} \\
    \midrule
    None (Baseline) & 0.9769  & 0.8408  & 0.9538  & 0.3365  & 0.5761  \\
    Most - LLaVA & 0.8496  & 0.6554  & 0.9346  & 0.6685  & 0.1616  \\
    Least - LLaVA & 0.9538  & 0.8362  & 0.9115  & 0.3423  & 0.5259  \\
    Least - Qwen & 0.9362  & 0.8064  & 0.9231  & 0.3984  & 0.4711  \\
    Random & 0.9423  & 0.8192  & 0.9269  & 0.4035  & 0.4689  \\
    \bottomrule
    \end{tabular}%
    }
  \label{tab:adaptive_defense}%
\end{table}%

\begin{figure}[t]
    \centering
    \begin{subfigure}{0.3\linewidth}
        \centering
        \includegraphics[width=\linewidth]{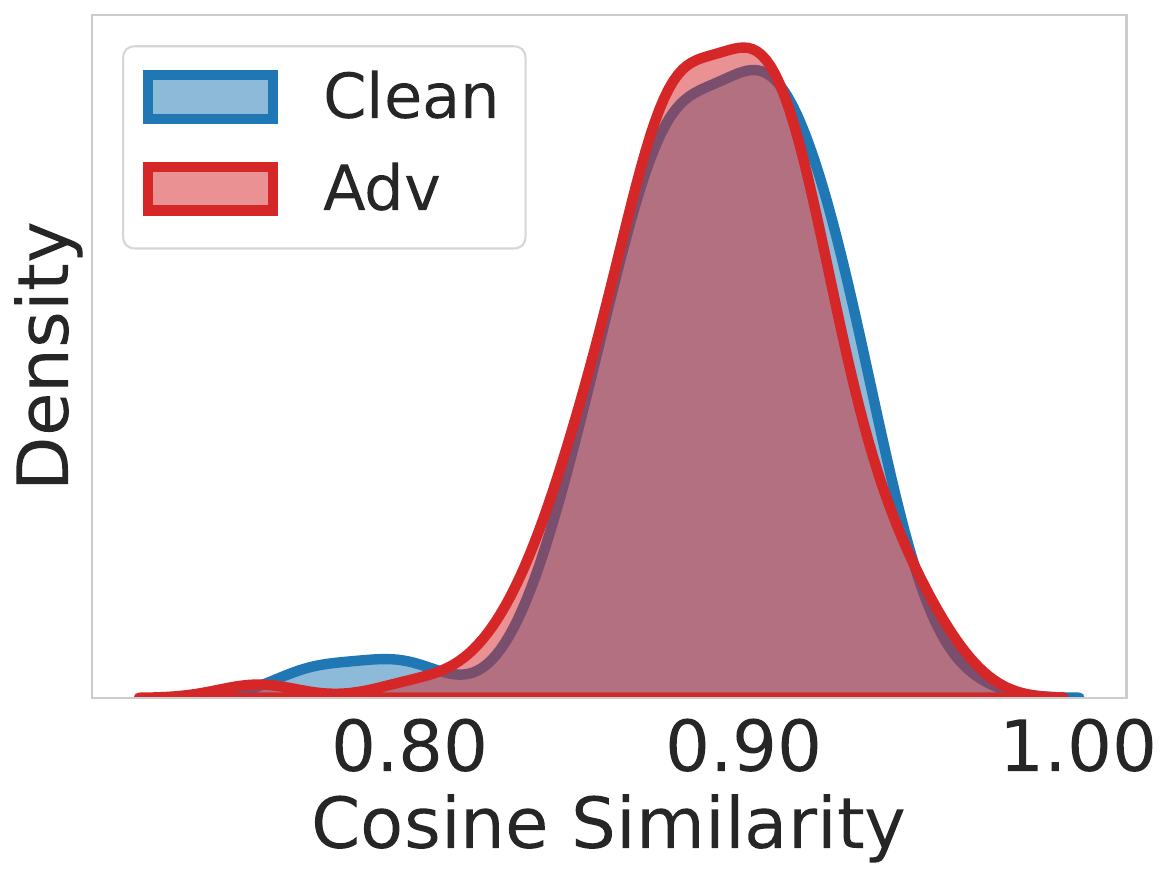}
        \caption{LLaVA-Rotation}
        \label{fig:llava-rotation}
    \end{subfigure}
    \begin{subfigure}{0.3\linewidth}
        \centering
        \includegraphics[width=\linewidth]{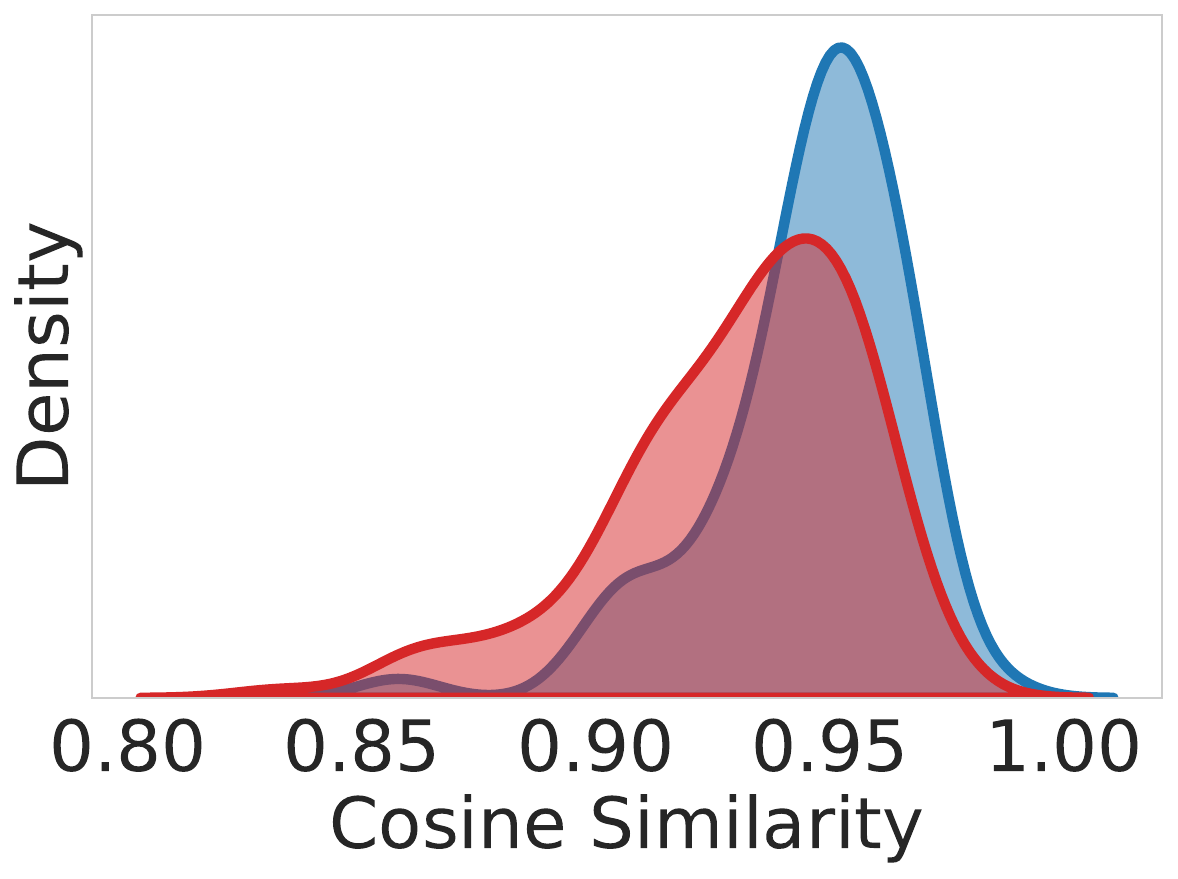}
        \caption{LLaVA-JPEG}
        \label{fig:llava-jpeg}
    \end{subfigure}
     \begin{subfigure}{0.3\linewidth}
        \centering
        \includegraphics[width=\linewidth]{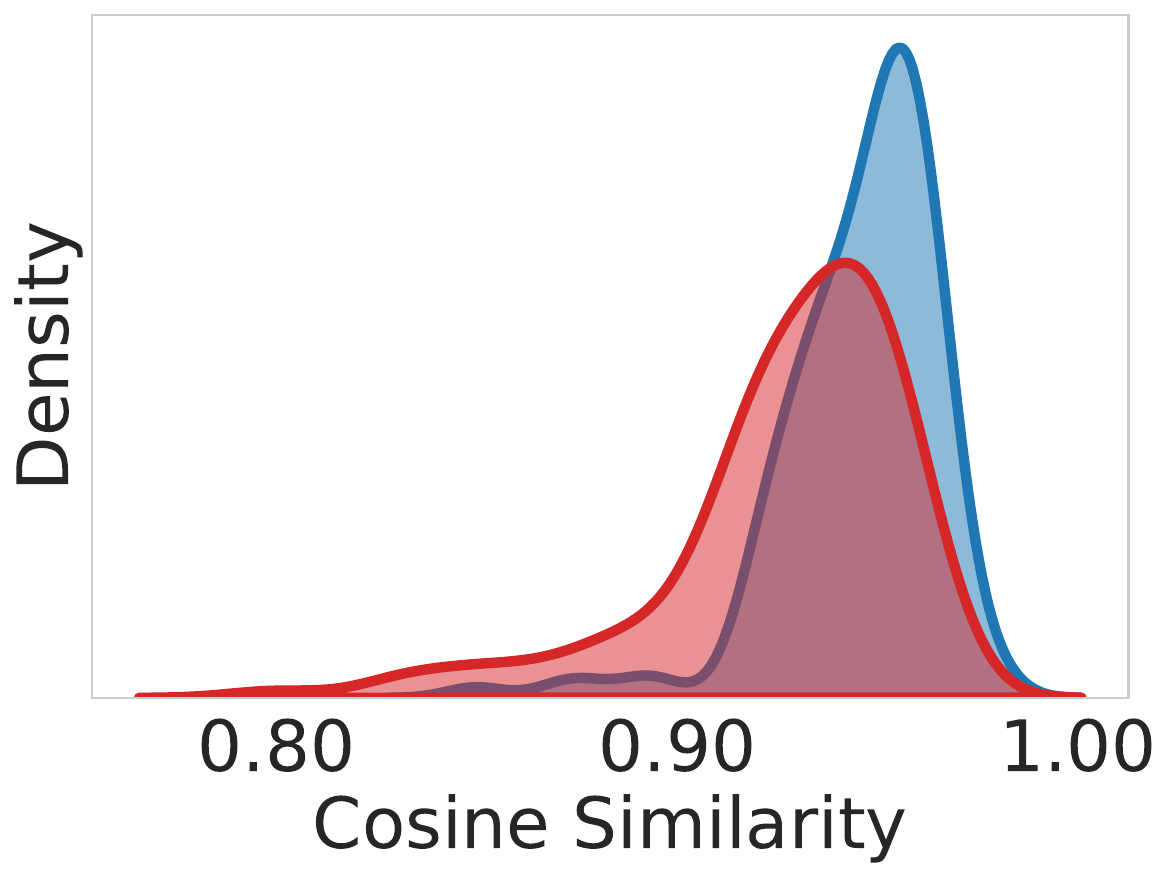}
        \caption{LLaVA-Blur}
        \label{fig:llava-blur}
    \end{subfigure}

     \begin{subfigure}{0.3\linewidth}
        \centering
        \includegraphics[width=\linewidth]{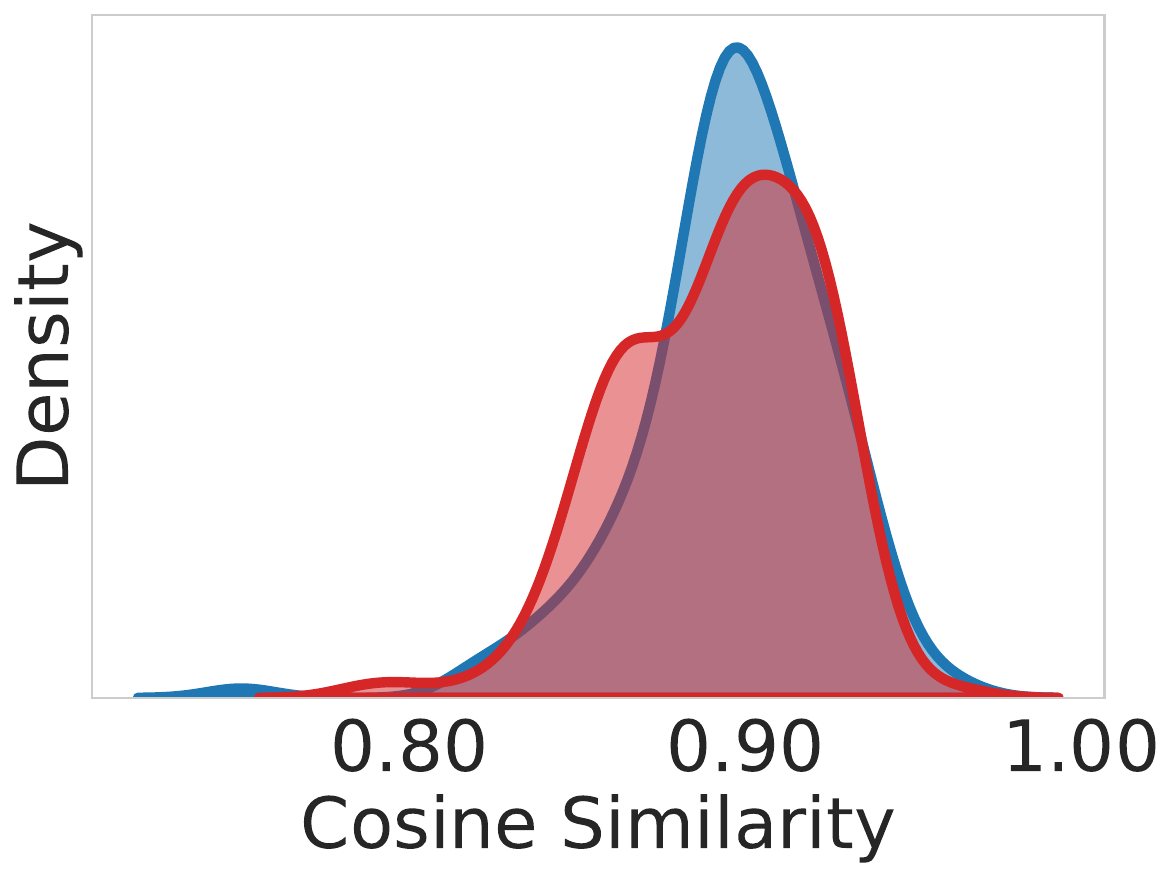}
        \caption{Qwen-Rotation}
        \label{fig:qwen-rotation}
    \end{subfigure}
    \begin{subfigure}{0.3\linewidth}
        \centering
        \includegraphics[width=\linewidth]{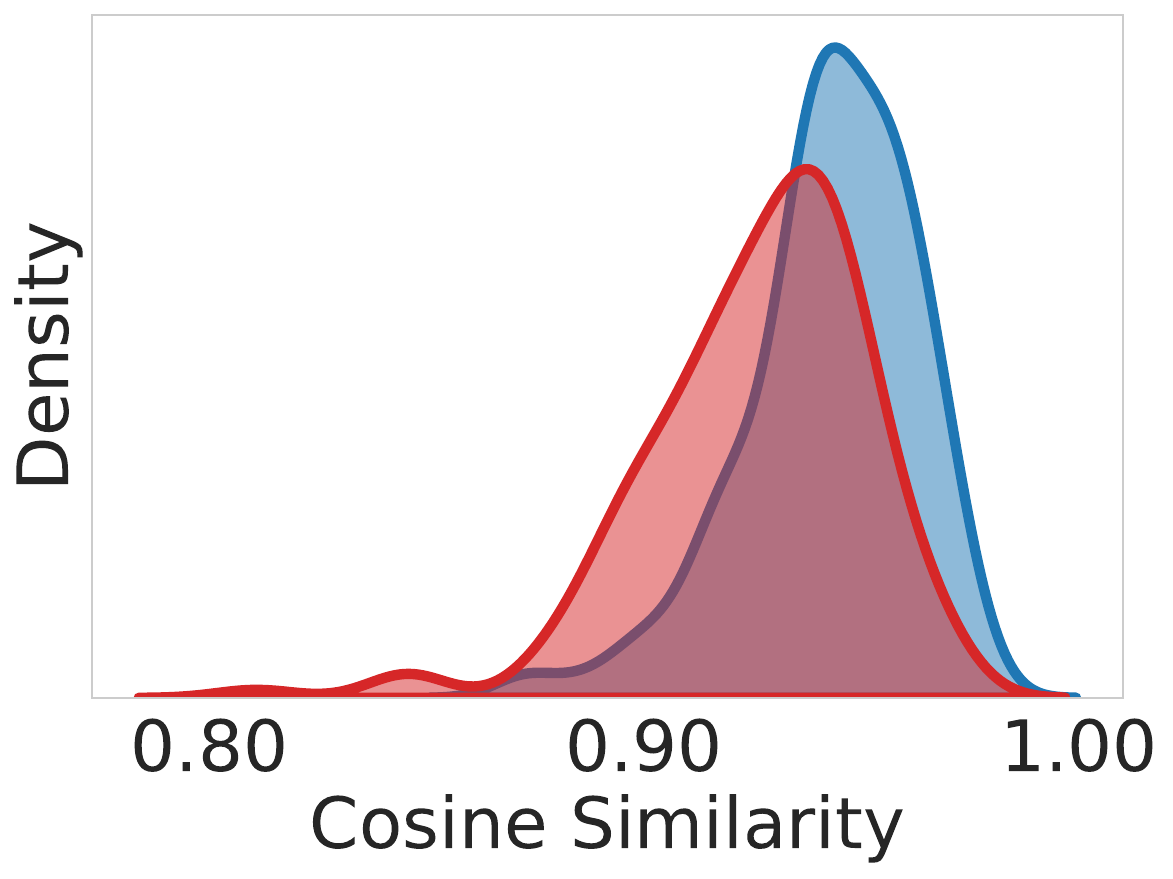}
        \caption{Qwen-JPEG}
        \label{fig:qwen-jpeg}
    \end{subfigure}
     \begin{subfigure}{0.3\linewidth}
        \centering
        \includegraphics[width=\linewidth]{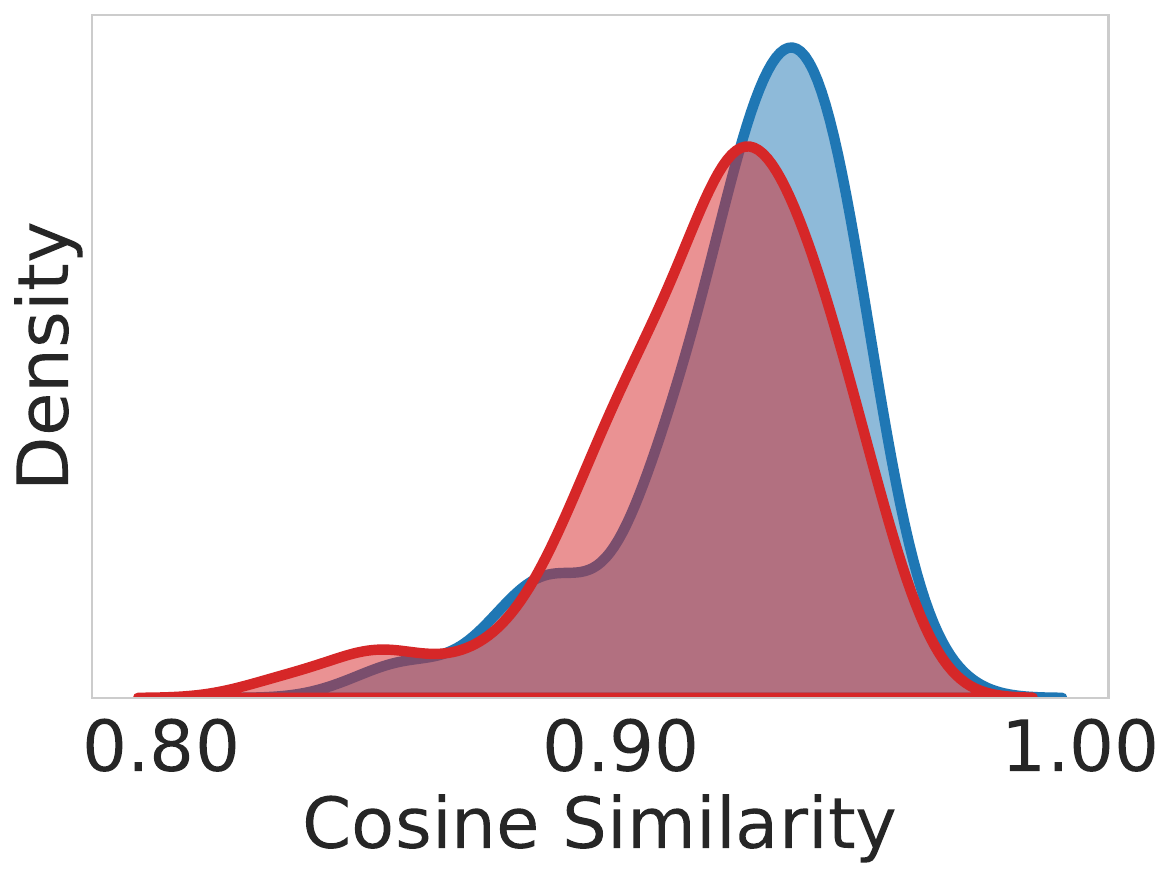}
        \caption{Qwen-Blur}
        \label{fig:qwen-blur}
    \end{subfigure}

    \caption{ CLIP Embedding similarity between the original image and its transformed variants (rotation, JPEG compression, and Gaussian blur) for clean and CAA adversarial samples. 
    }
    \label{fig:defense_clip}
\end{figure}

CAA selectively perturbs least-informative tokens to distort token-importance rankings without altering the overall image semantics.
As a result, CAA induces failures only after token compression, making it inherently stealthy.
This property also renders both adversarial detection and adaptive defenses ineffective against CAA.

\noindent \textbf{Adversarial Detection.}
We analyze adversarial detection from two complementary perspectives: the visual signal level and the end-to-end LVLM behavior. 
At the visual signal level, following prior work~\cite{xu2025one}, we consider test-time detection methods that measure representation shifts between an image and its mildly transformed variants (e.g., JPEG compression or Gaussian blur) in feature space such as CLIP embeddings~\cite{radford2021learning, bagdasaryan2024adversarial}.
While effective against traditional black-box attacks (e.g., HSJA and RayS) that perturb salient semantics, these methods fail to distinguish clean and CAA inputs.
As shown in Fig.~\ref{fig:defense_clip}, clean and adversarial samples exhibit highly overlapping similarity distributions across transformations. 
At the LVLM level, detection based on prediction shifts also fails. 
Because CAA is optimized to preserve uncompressed behavior and trigger failures only after compression, detection performance approaches random guessing (AUC $\approx$ 0.5).

\noindent \textbf{Adaptive Defenses.}
Even when the defender has knowledge of the CAA mechanism, adaptive defenses based on post-hoc token ranking remain ill-posed. 
A natural adaptive strategy is to identify and remove perturbed regions prior to inference. 
We consider two ways to identify suspect visual regions: (i) regions ranked as \emph{most important} by the victim LVLM, and (ii) regions ranked as \emph{least important} by an reference LVLM. 
As shown in Table~\ref{tab:adaptive_defense}, removing high-ranked regions can suppress the attack but severely degrades benign performance, since such regions typically encode task-critical information. 
Estimating importance using the victim model (LLaVA) is unreliable because the ranking has already been manipulated by the attack. 
Using an auxiliary LVLM (Qwen-VL) also fails due to substantial cross-model disagreement in least-informative regions, consistent with Section~\ref{sec:transfer_attack}. 
Overall, these results expose an inherent trade-off: removing high-ranked regions sacrifices utility, whereas removing low-ranked regions fails to intercept the attack.

\section{Conclusion}
\label{sec:conclusion}
In this work, we present the first systematic study of robustness risks introduced by visual token compression in LVLMs. We show that compression fundamentally reshapes the failure landscape by introducing stealthy vulnerabilities that manifest only under compressed inference and evade standard robustness evaluation. To characterize this risk, we propose Compression-Aware Attack (CAA), which exploits the fragility of token ranking to induce compression-specific failures while largely preserving uncompressed inference behavior. Our results demonstrate that these vulnerabilities persist in realistic black-box settings and that existing defenses provide limited protection, exposing an overlooked efficiency–security trade-off in efficient LVLMs.

Our findings underscore the need for security-aware compression mechanisms that balance efficiency and robustness. Moreover, the compression-related failure modes identified in this work may be further explored to construct stealthy backdoors, posing potential risks to the supply chain of efficient LVLM deployment.
\section*{Acknowledgments}
This paper was edited for grammar using ChatGPT.

\bibliographystyle{ACM-Reference-Format}
\bibliography{ref}

\appendix

\section{Ethical Considerations}
\label{sec:ethical}
This work examines the security and robustness implications of visual token compression in Large Vision-Language Models. Although we introduce adversarial techniques, our goal is strictly defensive: to identify previously overlooked risks and inform the design of more secure and reliable LVLM systems.
The vulnerabilities discussed in this paper stem from compression mechanisms that are already widely deployed in practice. By showing that visual token compression can introduce stealthy and hard-to-diagnose failure modes, we highlight compression as a security-sensitive component rather than a purely efficiency-oriented optimization. Understanding these risks is essential for deploying LVLMs in safety-critical and security-sensitive settings.
To mitigate potential misuse, we present our methods at an abstract level appropriate for academic analysis. Our evaluation further indicates that existing defenses are insufficient, underscoring the need for future research on robustness-aware compression mechanisms and principled mitigation strategies.
Overall, we aim to encourage responsible AI development by promoting the consideration of security and robustness alongside efficiency when designing and deploying compressed LVLM systems.

\section{Code Open Source}
\label{sec:code}
In the spirit of open science, we will release the code for this work to facilitate reproducibility and further research. The codebase includes implementations of Compression-Aware Attack (CAA) and its black-box extension T-CAA, evaluation scripts for multiple LVLMs and visual token compression methods, and tools for reproducing the main experimental results reported in this paper. We will also provide example configurations and documentation to support evaluation under differen settings. The code will be made publicly available upon publication at: \url{https://anonymous.4open.science/r/Compression_Aware_Attack-368D}


\section{Related Work}
\label{sec:related_work}
\subsection{Vision Token Compression}
\label{subsec:bg_token_comp}
Vision token compression has emerged as a practical inference-time optimization for large vision–language models (LVLMs), aiming to reduce computational cost by selectively retaining a subset of visual tokens while discarding redundant ones.

From the perspective of prompt dependency, existing approaches can be broadly categorized into text-agnostic and text-guided methods.
Text-agnostic approaches exploit spatial redundancy in visual information, with compression typically performed within the vision encoder. Representative methods such as VisionZip \cite{Yang2024VisionZipLI} and LLaVA-PruneMerge \cite{shang2025llava} select tokens based on attention scores between the \texttt{[CLS]} token and visual tokens; ZipVL \cite{He_2025_ICCV} further supports layer- or task-adaptive sparsity, while HoloV \cite{zoudon} adopts a holistic strategy to ensure global visual coverage. 
Text-guided methods leverage the semantic information in the language prompt to assess the task relevance of visual tokens, enabling more aggressive compression while preserving performance. Representative approaches including FastV \cite{Chen2024AnII}, PDrop \cite{Xing2024PyramidDropAY}, and SparseVLM \cite{Zhang2024SparseVLMVT} prune tokens based on attention between text and visual tokens, adopting early one-shot compression, progressive depth-wise pruning, and multi-text-reference guidance, respectively. Other methods such as $ST^3$ \cite{Zhuang2025ST3AM} perform cross-layer progressive pruning, while G-Search \cite{Zhao2024AcceleratingML} and FitPrune \cite{ye2025fit} determine layer-wise retention via search or optimization.
From the perspective of importance scoring, attention-based methods dominate current practice \cite{deepseek32, deepseek32f}. Attention scores naturally arise during inference, directly reflect each visual token’s contribution to the model’s response under a given query \cite{yuan2025native, tzachristas2025mathematical}, and all approaches discussed above fall into this category. Beyond this method, some approaches adopt embedding-similarity strategies, merging visually similar tokens based on feature distance or formulating token selection as a diversity-maximization problem \cite{alvar2025divprune}. However, these methods typically require explicit similarity computation, introducing additional overhead.
Accordingly, this work focuses on attention-based, text-guided token compression.


\subsection{Adversarial Attack}
\label{subsec:bg_adv_attack}
Adversarial examples are inputs perturbed with imperceptible noise to induce incorrect model predictions. Depending on the attacker’s knowledge and access to the target model, adversarial attacks are commonly categorized into white-box attacks, query-based black-box attacks, and transfer attacks.

White-box attacks assume full access to the model and exploit gradient information to optimize adversarial perturbations. Classic methods such as FGSM \cite{goodfellow2014explaining} and PGD \cite{madry2018towards} achieve high attack success under constrained distortion and are widely used as benchmarks for evaluating model robustness, representing an upper bound on adversarial vulnerability.
Query-based attacks craft adversarial examples solely through interaction with the model. Representative query-based methods include HSJA \cite{chen2020hopskipjumpattack}, which estimates gradient directions at the decision boundary by using binary feedback, and RayS \cite{chen2020rays}, which efficiently searches for the nearest boundary along discrete rays.  
While effective, such attacks typically require a large number of queries, often exceeding tens of thousands \cite{xu2025one}, making them impractical under realistic access restrictions and security auditing \cite{gemini3flash}. 
Transfer attacks assume no access to the target model but rely on a surrogate model with similar architecture or training distribution. 
Adversarial examples are generated on the surrogate and then transferred to the target model.
Typical strategies to improve transferability include applying input transformations to increase perturbation diversity and ensembling multiple surrogate models to reduce overfitting \cite{li2025frustratingly}. Such attacks are commonly used to evaluate robustness under the most restrictive attacker assumptions. 

However, existing adversarial attacks are tailored for standard, uncompressed models and remain agnostic to the compression mechanism. Moreover, the compression process often introduces non-differentiable decision boundaries, making traditional gradient-based attacks difficult to apply directly to the compressed state. Simultaneously, since real-world LVLM deployments typically impose strict query limits, query-intensive black-box attacks face significant practical challenges in this scenario. Consequently, evaluating the security vulnerabilities specifically introduced by compression requires a novel attack.

\section{Compression and Robustness}
\label{sec:comp_and_robust}

\subsection{The Impact of Token Ranking on Robustness}
\label{subsec:critical_ranking_role_supp}
\begin{figure}[t]
    \centering
    \begin{subfigure}{0.43\linewidth}
        \centering
        \includegraphics[width=\linewidth]{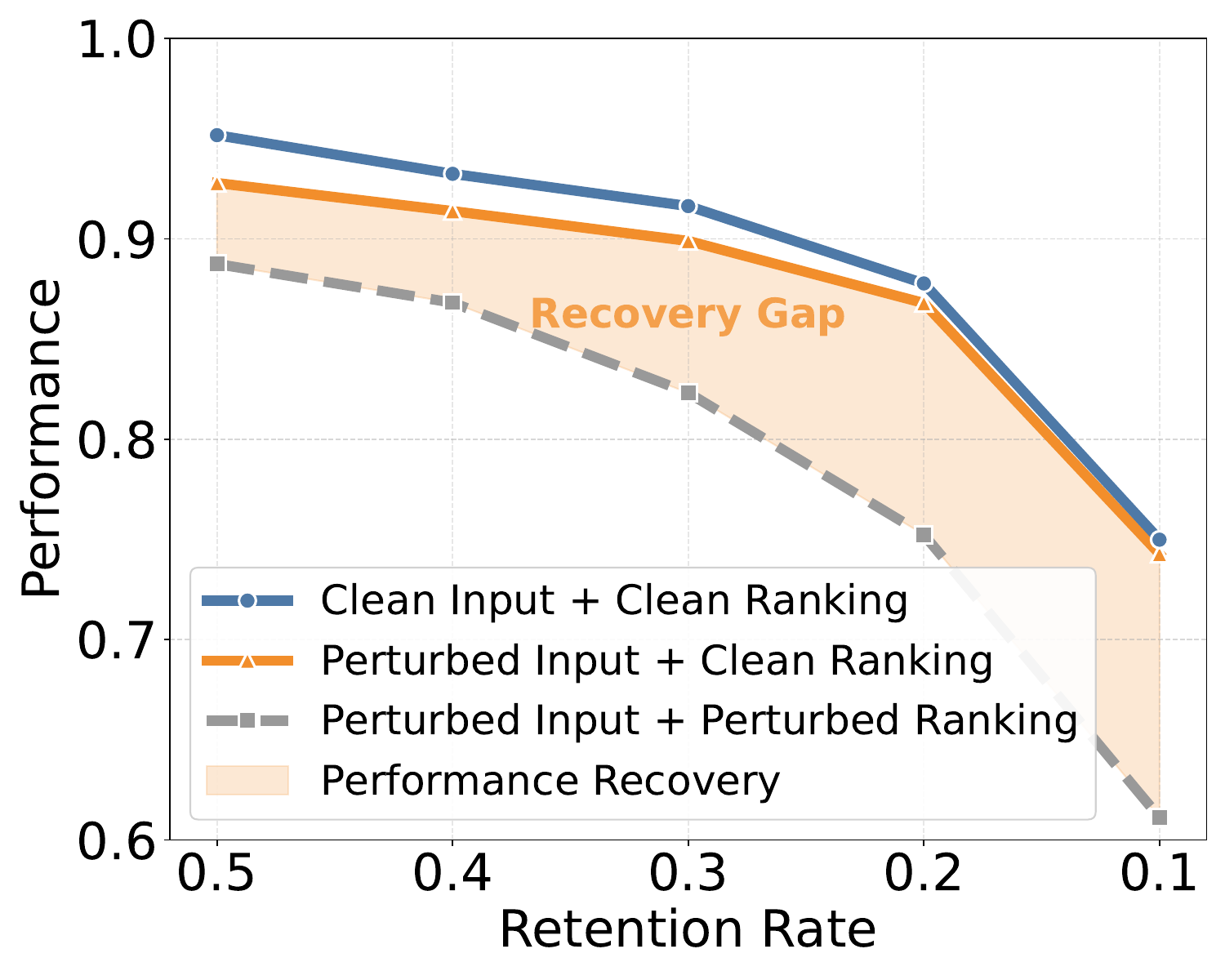}
        \caption{LLaVA $\epsilon = 64/255$}
    \end{subfigure}
        \begin{subfigure}{0.43\linewidth}
        \centering
        \includegraphics[width=\linewidth]{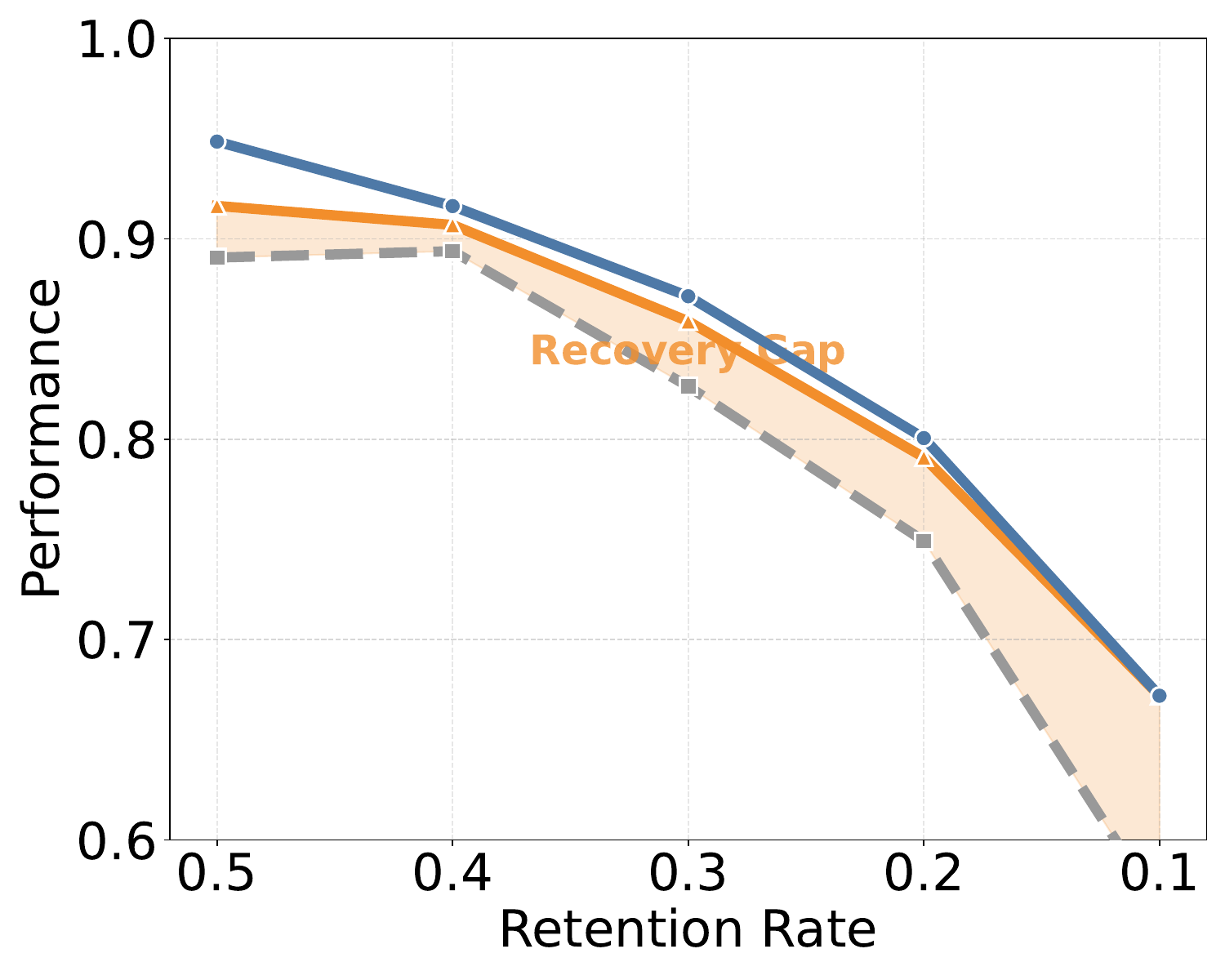}
        \caption{Qwen-VL $\epsilon = 64/255$}
    \end{subfigure}
    \caption{
    Effect of token importance ranking on robustness under compressed inference.
    The shaded region highlights the significant performance recovery achieved by restoring correct rankings, suggesting that ranking instability plays an important role in robustness degradation under compression.}
    \label{fig:semantic_rank_role_64}
\end{figure}

We provide additional results under a larger perturbation budget ($\epsilon=64/255$) to examine whether the observations in Section~\ref{sec:comp_and_roubst} remain consistent. Fig.~\ref{fig:semantic_rank_role_64} reports model performance under the same three inference configurations as in Fig.~\ref{fig:semantic_rank_role}.
Restoring the clean token-importance ranking (configuration (c)) leads to substantial performance recovery compared to using the perturbed ranking (configuration (b)), and the performance under configuration (c) remains close to the clean baseline (configuration (a)). This indicates that robustness degradation under compression is dominated by instability in token importance ranking.

\subsection{Ranking Instability}
\label{subsec:ranking_instab_supp}
\begin{figure}[t]
    \centering
    \begin{subfigure}{0.43\linewidth}
        \centering
        \includegraphics[width=\linewidth]{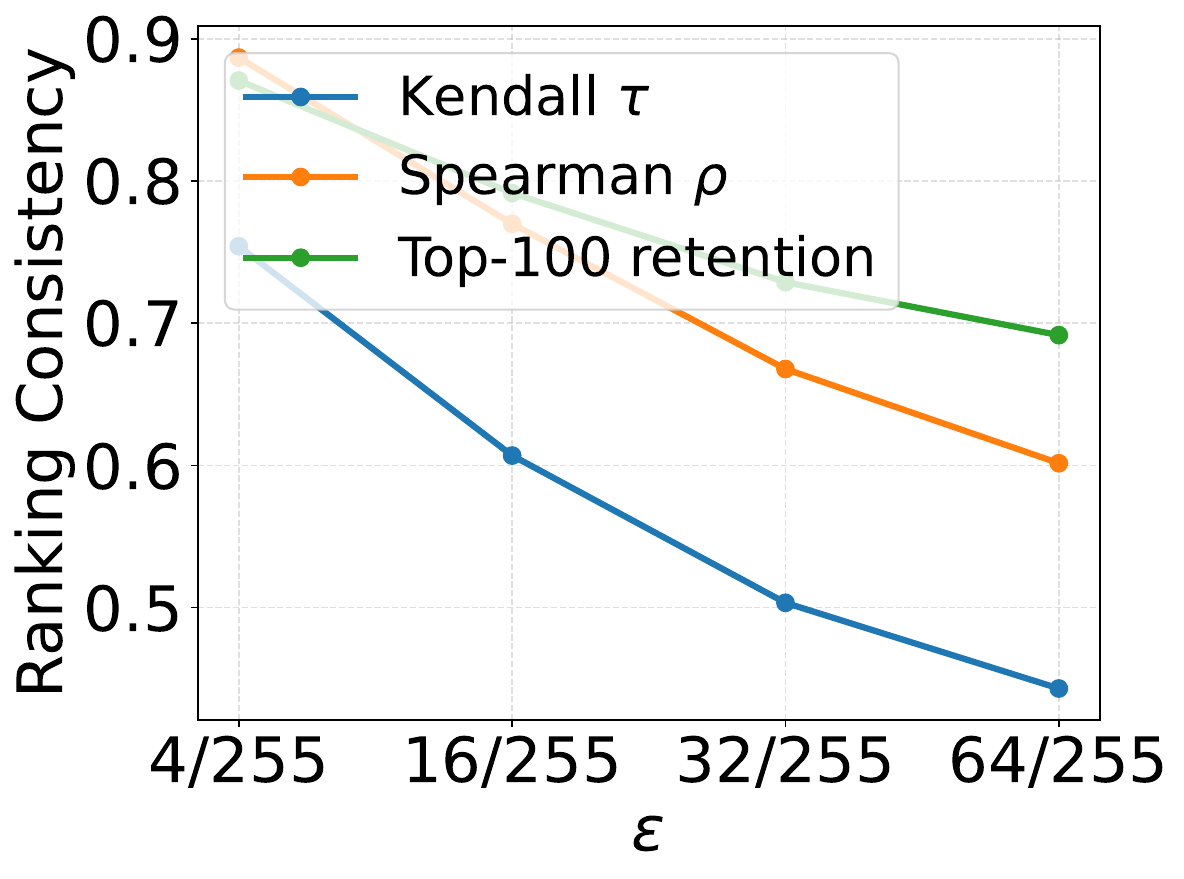}
        \caption{Qwen-VL}
        \label{fig:o2_1_ranking_qwen}
    \end{subfigure}
    \begin{subfigure}{0.43\linewidth}
        \centering
        \includegraphics[width=\linewidth]{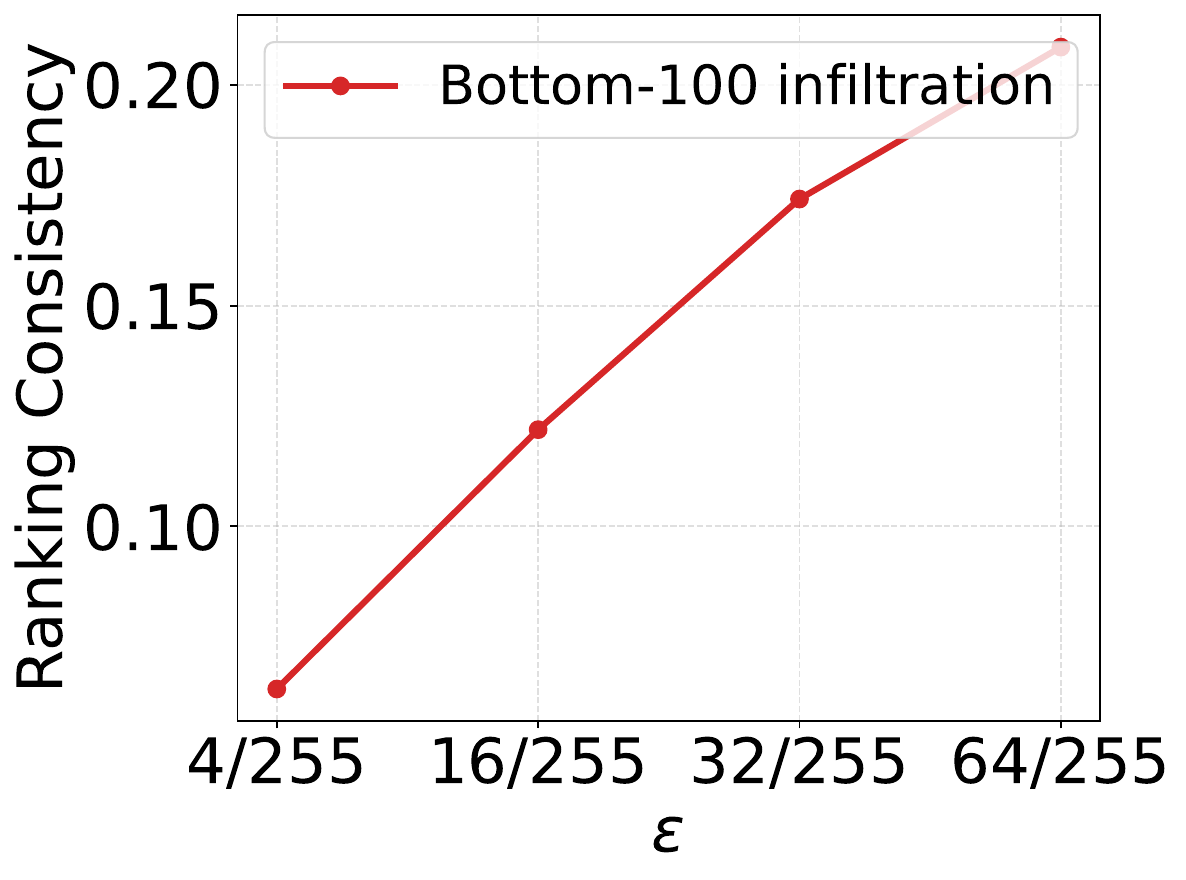}
        \caption{Qwen-VL}
        \label{fig:02_ranking_bot_qwen}
    \end{subfigure}

    \caption{
    Ranking stability under random $\ell_\infty$-bounded noise with different budgets.
    Global stability is measured using Kendall's~$\tau$ and Spearman's~$\rho$.
    Local stability is evaluated via Top-100 Preservation (how many top-important tokens remain stable) and Bottom-100 Infiltration (how much low-importance tokens enters the top ranks). Larger perturbations lead to progressively more severe ranking inconsistency.
    }
    \label{fig:o2_ranking_llava_qwen}
\end{figure}

Token importance ranking is highly sensitive to input perturbations. Fig.~\ref{fig:o2_ranking_llava_qwen} illustrates the ranking instability of Qwen-VL under random noise by comparing the token importance ordering induced by perturbed inputs with that of clean inputs. As the noise magnitude increases, the ranking becomes progressively more disordered: tokens that originally ranked among the bottom 100 increasingly enter the top-100 set. This misranking causes uninformative tokens to be retained while task-critical ones are discarded, ultimately degrading model performance.

\section{Victim Models and Datasets} 
\label{sec:victim_model_dataset}
Following prior work~\cite{zhao2025q, Zhang2024SparseVLMVT, Xing2024PyramidDropAY}, we evaluate our attacks on three state-of-the-art large vision–language models: 
\textbf{LLaVA}~\cite{liu2023visual}, a representative LVLM architecture; \textbf{LLaVA-Next}~\cite{liu2024llavanext}, which extends LLaVA with higher-resolution inputs and flexible aspect ratios; and \textbf{Qwen2.5-VL}~\cite{bai2025qwen2}, a highly competitive open-source LVLM.
We evaluate our method on three widely used benchmarks. 
\textbf{POPE}~\cite{li2023evaluating} evaluates object hallucination via binary visual question answering. 
\textbf{MME}~\cite{liu2024mmbench} assesses multimodal perception and reasoning with multi-domain multiple-choice questions. 
\textbf{TextVQA}~\cite{singh2019towards} measures a model’s ability to recognize and reason about embedded text in images. For TextVQA, we adopt the standard \textbf{VQA accuracy} metric~\cite{antol2015vqa}, which evaluates model predictions against human annotations. An answer receives full credit if it matches responses from at least three annotators, formally defined as
$\text{VQA\_ACC} =
\min\!\left(
\frac{\#~\text{matching human answers}}{3},\, 1
\right).$

Some benchmark samples can be correctly answered without visual input (e.g., in TextVQA). We therefore retain only samples for which the model’s prediction differs with and without the image, i.e., $f(I, T) \neq f(\emptyset, T)$, ensuring that the evaluation set requires genuine multimodal reasoning.

\section{White-box Attack}
\label{sec:white_box_attack_supp}
CAA injects adversarial perturbations into  unimportant regions of the image to manipulate the compression mechanism into retaining uninformative tokens, thereby causing the model to fail under compressed inference. The overall procedure is summarized in Algorithm~\ref{alg:caa}. 

\subsection{Vanilla Adversarial attacks}
\label{subsec:vanilla_attack_supp}
To account for lexical variability in LVLM outputs (e.g., “yes”, “Yeah”), we suppress a predefined set of correct answer variants $\mathcal{Y}{\mathrm{cor}}$. Since this set cannot cover all semantically equivalent responses, we additionally promote a set of incorrect anchor answers $\mathcal{Y}{\mathrm{wro}}$ to explicitly steer the output toward incorrect predictions. Let $p(I+\delta, T)$ denote the output distribution over the first generated token. The attack loss is defined over the full image as follows:
\begin{equation}
\begin{aligned}
\mathcal{L}_{\text{vanilla}}(\delta)
= & -\frac{1}{|\mathcal{Y}_{wro}|}\sum_{y_a \in \mathcal{Y}_{wro}} \log p_{y_w}(I+\delta,T) \\
& +\frac{1}{|\mathcal{Y}_{cor}|}\sum_{y_g \in \mathcal{Y}_{cor}} \log p_{y_c}(I+\delta,T),
\end{aligned}
\label{eq:vanilla_attack}
\end{equation}
 where $\|\delta\|_\infty \le \epsilon$, with $\epsilon$ set to $32/255$ by default.

\subsection{Main Results for CAA.}
\label{subsec:main_caa_supp}
Table~\ref{tab:caa_mian_reulst_supp} reports the performance of CAA under the PDrop compression method, while Table~\ref{tab:caa_other_rate_textvqa} further evaluates CAA across different retention rates on TextVQA. As shown, CAA consistently achieves high CSG values under all tested settings, indicating that it preserves performance in the non-compressed state while inducing substantial degradation under compression. These results confirm that CAA effectively exposes vulnerabilities introduced by the compression mechanism and remains effective across varying compression strengths.

\begin{algorithm}[t]
\SetAlgoLined
\SetKwInOut{Input}{Input}
\SetKwInOut{Output}{Output}

\caption{\textsc{Compression-Aware Attack (CAA)}}
\label{alg:caa}

\Input{Model $f$; image $I$; prompt $T$; target layer $\tilde{l}$; step size $\eta$; iterations $N$; $\ell_\infty$ constrain $\epsilon$; weights $\{\alpha, \beta, \lambda_e,\lambda_k\}$; hierarchy depth $n_g$}
\Output{Adversarial image $\hat{I}$}

\BlankLine
Forward $(I,T)$ to layer $\tilde{l}$ to get $\{v_j^{(\tilde{l})}\}_{j=1}^{n_V}$\;
Compute clean importance $\{s_j^{(\tilde{l})}\}$ (Eq.~\ref{eq:importance_score}) and split indices into $\Omega_{\mathrm{most}},\Omega_{\mathrm{least}}$ and partition $\Omega_{\mathrm{least}}$ into $n_g$ groups and form $\mathcal{P}_{\mathrm{LM}}^{(g)},\mathcal{P}_{\mathrm{LL}}^{(b,a)}$ (Eq.~\ref{eq:pair_sets})\;

\BlankLine
Initialize $\delta_j \sim Gaussian(0, 1)$ for $j \in \Omega_{\mathrm{least}}$ and $\delta_j = 0$ otherwise;

\BlankLine
\For{$n=1$ \KwTo $N$}{
    $\hat{I}\leftarrow I+\delta$\;

    Forward $(\hat{I},T)$ to layer $\tilde{l}$ and compute $\{\hat{s}_j^{(\tilde{l})}\}$ via Eq.~\ref{eq:importance_score}\;
    Compute $\{\hat{q}_i^{(\tilde{l})}\}_{i\in\mathcal{I}_{\mathrm{ref}}}$ and $\{\hat{k}_j^{(\tilde{l})}\}_{j\in\Omega_{\mathrm{least}}}$ via Eq.~\ref{eq:qk_vector}\;

    $\mathcal{L}_{\mathrm{total}}\leftarrow 
    \mathcal{L}_{\mathrm{rank}}
    +\lambda_e\mathcal{L}_{\mathrm{erase}}
    +\lambda_k\mathcal{L}_{\mathrm{key}}$ via Eqs.~\ref{eq:hierarchical_rank},\ref{eq:semantic_erasure},\ref{eq:key_align};

    $\delta \leftarrow \mathrm{clip}(\delta-\eta\cdot\mathrm{sign}(\nabla_{\delta}\mathcal{L}_{\mathrm{total}}),-\epsilon,\epsilon\big)$, with $\delta_j=0$ for $j\notin\Omega_{\mathrm{least}}$\;
}

\Return{$\hat{I}$};
\end{algorithm}

\begin{table*}[htbp]
  \centering
  \caption{Comparison of CAA with two baseline attacks under the PDrop compression method.}
  \resizebox{0.99\linewidth}{!}{
    \begin{tabular}{ll|cc|ccccc|ccccc|ccccc}
    \toprule
    \multicolumn{1}{c}{\multirow{2}[2]{*}{\textbf{Dataset}}} & \multicolumn{1}{c|}{\multirow{2}[2]{*}{\textbf{Model}}} & \multirow{2}[2]{*}{\textbf{M(cl,nc)}} & \multirow{2}[2]{*}{\textbf{M(cl,c)}} & \multicolumn{5}{c|}{\textbf{Vanilla Attack}} & \multicolumn{5}{c|}{\textbf{Random Attack}} & \multicolumn{5}{c}{\textbf{Compression-Aware Attack}} \\
          &       &       &       & \textbf{M(adv,nc)} & \textbf{M(adv,c)} & \textbf{UPR} & \textbf{CAE} & \textbf{CSG} & \textbf{M(adv,nc)} & \textbf{M(adv,c)} & \textbf{UPR} & \textbf{CAE} & \textbf{CSG} & \textbf{M(adv,nc)} & \textbf{M(adv,c)} & \textbf{UPR} & \textbf{CAE} & \textbf{CSG} \\
\cmidrule{1-7}\cmidrule{7-19}    \multirow{3}[2]{*}{POPE} & LLaVA & 0.9775 & 0.8392 & 0.1210 & 0.1529 & 0.1238 & 0.8178 & -0.0584 & 0.9550 & 0.7621 & 0.9770 & 0.0919 & 0.0689 & 0.9363 & 0.2038 & 0.9579 & 0.7571 & \textbf{0.7150} \\
          & LLaVA-NEXT & 0.9003 & 0.7878 & 0.2866 & 0.2229 & 0.3183 & 0.7171 & 0.0354 & 0.8650 & 0.7331 & 0.9608 & 0.0694 & 0.0302 & 0.9363 & 0.2038 & 1.0400 & 0.7413 & \textbf{0.7813} \\
          & Qwen-VL & 0.9550 & 0.7781 & 0.1465 & 0.3057 & 0.1534 & 0.6071 & -0.2395 & 0.8534 & 0.6116 & 0.8936 & 0.2140 & 0.1076 & 0.7618 & 0.3611 & 0.9579 & 0.5359 & \textbf{0.3336} \\
    \midrule
    \multirow{3}[1]{*}{TextVQA} & LLaVA & 0.8817 & 0.7222 & 0.1306 & 0.1656 & 0.1481 & 0.7707 & -0.0812 & 0.7473 & 0.6099 & 0.8476 & 0.1555 & 0.0031 & 0.7296 & 0.1637 & 0.8275 & 0.7733 & \textbf{0.6008} \\
          & LLaVA-NEXT & 0.6318 & 0.5563 & 0.1541 & 0.1503 & 0.2439 & 0.7298 & -0.0263 & 0.5695 & 0.5008 & 0.9014 & 0.0998 & 0.0012 & 0.5633 & 0.3270 & 0.8916 & 0.4122 & \textbf{0.3038} \\
          & Qwen-VL & 0.9135 & 0.6363 & 0.3299 & 0.3331 & 0.3611 & 0.4765 & -0.1624 & 0.8534 & 0.5939 & 0.9342 & 0.0666 & 0.0008 & 0.7618 & 0.3311 & 0.8339 & 0.4796 & \textbf{0.3136} \\
    \midrule
    \multirow{3}[1]{*}{MME} & LLaVA & 0.8714 & 0.7910 & 0.4713 & 0.4395 & 0.5409 & 0.4444 & -0.0148 & 0.8489 & 0.7690 & 0.9742 & 0.0278 & 0.0020 & 0.8328 & 0.3826 & 0.9579 & 0.5163 & \textbf{0.4720} \\
          & LLaVA-NEXT & 0.8617 & 0.8296 & 0.4459 & 0.4268 & 0.5175 & 0.4855 & 0.0030 & 0.8424 & 0.8099 & 0.9776 & 0.0237 & 0.0013 & 0.8360 & 0.5481 & 0.9702 & 0.3393 & \textbf{0.3095} \\
          & Qwen-VL & 0.9085 & 0.8170 & 0.2115 & 0.3718 & 0.2328 & 0.5449 & -0.2223 & 0.8954 & 0.8036 & 0.9856 & 0.0164 & 0.0020 & 0.7314 & 0.4042 & 0.8051 & 0.5053 & \textbf{0.3103} \\
    \bottomrule
    \end{tabular}%
    }
  \label{tab:caa_mian_reulst_supp}%
\end{table*}%

\begin{table}[htbp]
  \centering
  \caption{CAA performance under different token retention rates on TextVQA.
  }
\resizebox{0.95\linewidth}{!}{
    \begin{tabular}{lc|ccccccc}
    \toprule
    \multicolumn{1}{c}{\textbf{Model}} & \textbf{Ratio} & \textbf{M(cl, nc)} & \textbf{M(cl,c)} & \textbf{M(adv,nc)} & \textbf{M(adv,c)} & \textbf{UPR} & \textbf{CAE} & \textbf{CSG} \\
    \midrule
    \multirow{5}[2]{*}{LLaVA} & 0.5   & \multirow{5}[2]{*}{0.8817 } & 0.8444  & 0.6876  & 0.4308  & 0.7799  & 0.4898  & 0.2697  \\
          & 0.4   &       & 0.8197  & 0.7214  & 0.4323  & 0.8182  & 0.4726  & 0.2908  \\
          & 0.3   &       & 0.7997  & 0.7204  & 0.4035  & 0.8171  & 0.4954  & 0.3125  \\
          & 0.2   &       & 0.7428  & 0.7354  & 0.2158  & 0.8341  & 0.7095  & 0.5435  \\
          & 0.1   &       & 0.6402  & 0.7027  & 0.1027  & 0.7970  & 0.8396  & 0.6366  \\
    \midrule
    \multirow{5}[2]{*}{Qwen-VL} & 0.5   & \multirow{5}[2]{*}{0.9135 } & 0.8495  & 0.7058  & 0.3058  & 0.7726  & 0.6400  & 0.4127  \\
          & 0.4   &       & 0.8034  & 0.7389  & 0.2816  & 0.8089  & 0.6495  & 0.4584  \\
          & 0.3   &       & 0.7585  & 0.7592  & 0.2322  & 0.8311  & 0.6939  & 0.5250  \\
          & 0.2   &       & 0.6662  & 0.8585  & 0.1916  & 0.9398  & 0.7124  & 0.6522  \\
          & 0.1   &       & 0.6135  & 0.8134  & 0.1879  & 0.8904  & 0.6937  & 0.5841  \\
    \bottomrule
    \end{tabular}%
    }
  \label{tab:caa_other_rate_textvqa}%
\end{table}%

\subsection{Ablation Study of Key Components in CAA}
\label{subsec:ab_caa_key_qwen}
We conduct the same ablation studies on Qwen-VL to verify the generality of our findings. As shown in Table~\ref{tab:ab-key-components_qwen}, removing any key component of CAA consistently degrades attack effectiveness under compressed inference. In particular, eliminating the hierarchical ranking constraint or the query-guided optimization leads to substantial drops in CSG, while removing semantic erasure further weakens the attack by allowing retained tokens to preserve usable visual content. These results closely mirror those observed on LLaVA, confirming that the effectiveness of CAA relies on all three components and generalizes across different LVLM architectures.

\begin{table}[htbp]
  \centering
  \caption{Ablation of CAA Components on Qwen-VL. Three variants are formed by progressively removing hierarchical ranking, semantic erasure, and query-guided alignment.
}
   \resizebox{0.95\linewidth}{!}{
    \begin{tabular}{l|rrr|rrr|rrr}
    \toprule
    \multirow{2}[2]{*}{\textbf{Attack}} & \multicolumn{3}{c|}{\textbf{POPE}} & \multicolumn{3}{c|}{\textbf{TextVQA}} & \multicolumn{3}{c}{\textbf{MME}} \\
          & \multicolumn{1}{c}{\textbf{UPR}} & \multicolumn{1}{c}{\textbf{CAE}} & \multicolumn{1}{c|}{\textbf{CSG}} & \multicolumn{1}{c}{\textbf{UPR}} & \multicolumn{1}{c}{\textbf{CAE}} & \multicolumn{1}{c|}{\textbf{CSG}} & \multicolumn{1}{c}{\textbf{UPR}} & \multicolumn{1}{c}{\textbf{CAE}} & \multicolumn{1}{c}{\textbf{CSG}} \\
    \midrule
    CAA w/o Hierarchy & 0.9654  & 0.8078  & 0.7732  & 0.9356  & 0.6219  & 0.5575  & 0.9552  & 0.5688  & 0.5240  \\
    CAA w/o Erasure & 0.9638  & 0.7734  & 0.7372  & 0.9571  & 0.6187  & 0.5758  & 0.8746  & 0.5087  & 0.3833  \\
    CAA w/o Query & 0.9823  & 0.7086  & 0.6909  & 0.9707  & 0.5587  & 0.5294  & 0.8648  & 0.6405  & 0.5053  \\
    CAA   & 0.9158  & 0.9255  & \textbf{0.8413}  & 0.9398  & 0.7124  & \textbf{0.6522}  & 0.9344  & 0.6548  & \textbf{0.5892}  \\
    \bottomrule
    \end{tabular}%
    }
  \label{tab:ab-key-components_qwen}%
\end{table}%

\subsection{Effectiveness to Configuration Mismatch}
\label{subsec:rate_layer_mismatch}
\begin{figure}[t]
    \centering
    \begin{subfigure}{0.3\linewidth}
        \centering       \includegraphics[width=\linewidth]{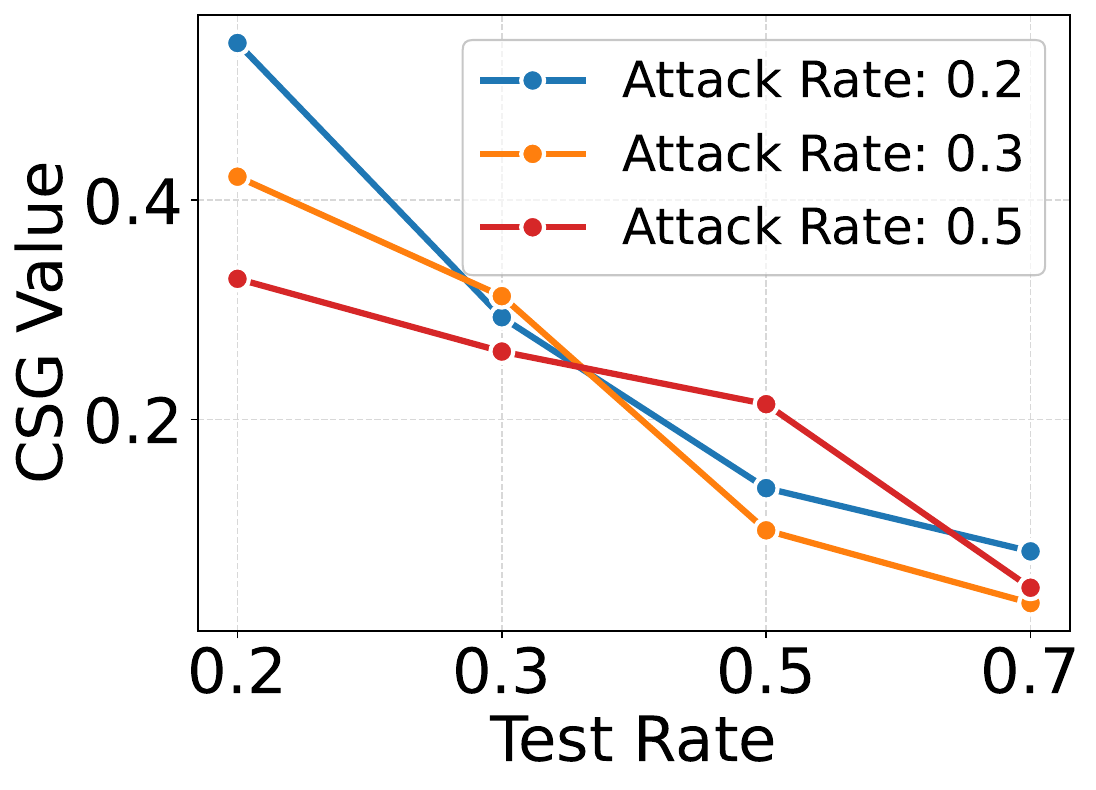}
        \caption{LLaVA-TextVQA}
    \end{subfigure}
        \begin{subfigure}{0.3\linewidth}
        \centering
   \includegraphics[width=\linewidth]{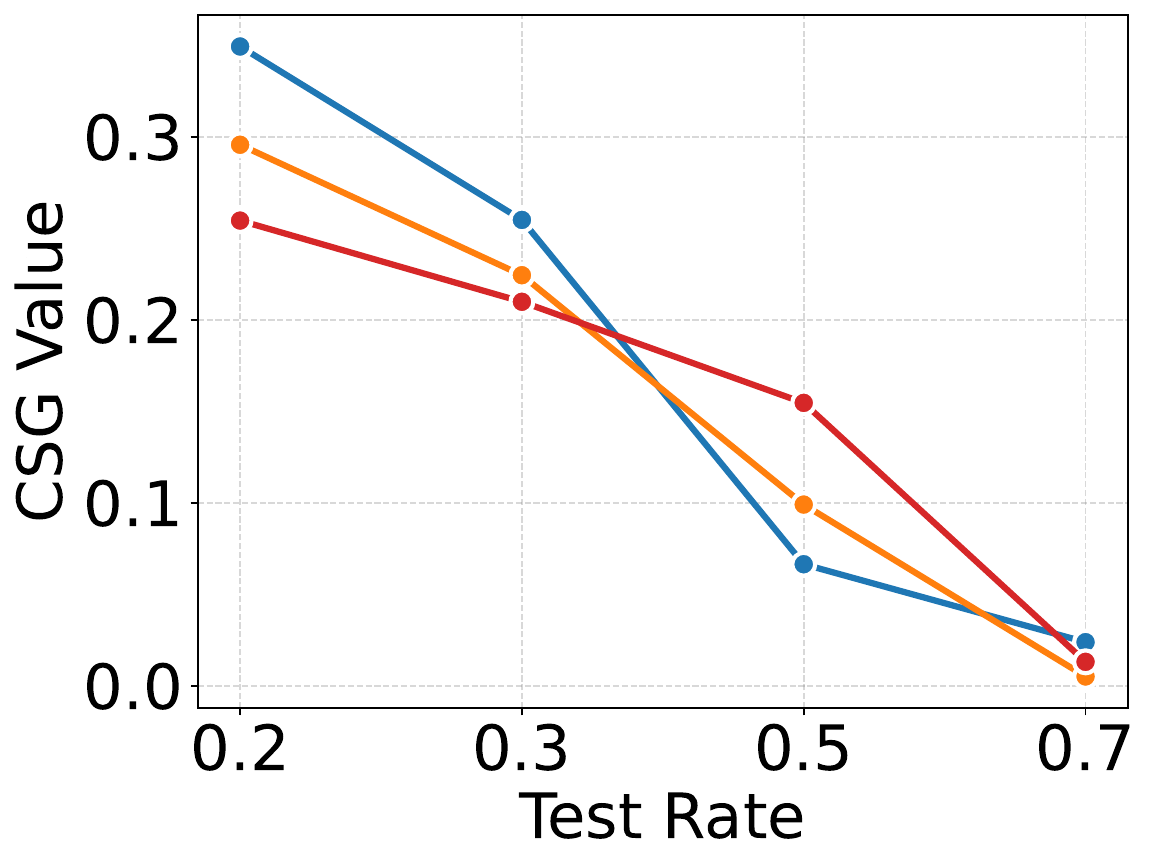} 
        \caption{LLaVA-MME}
    \end{subfigure}
    
    \begin{subfigure}{0.3\linewidth}
        \centering
   \includegraphics[width=\linewidth]{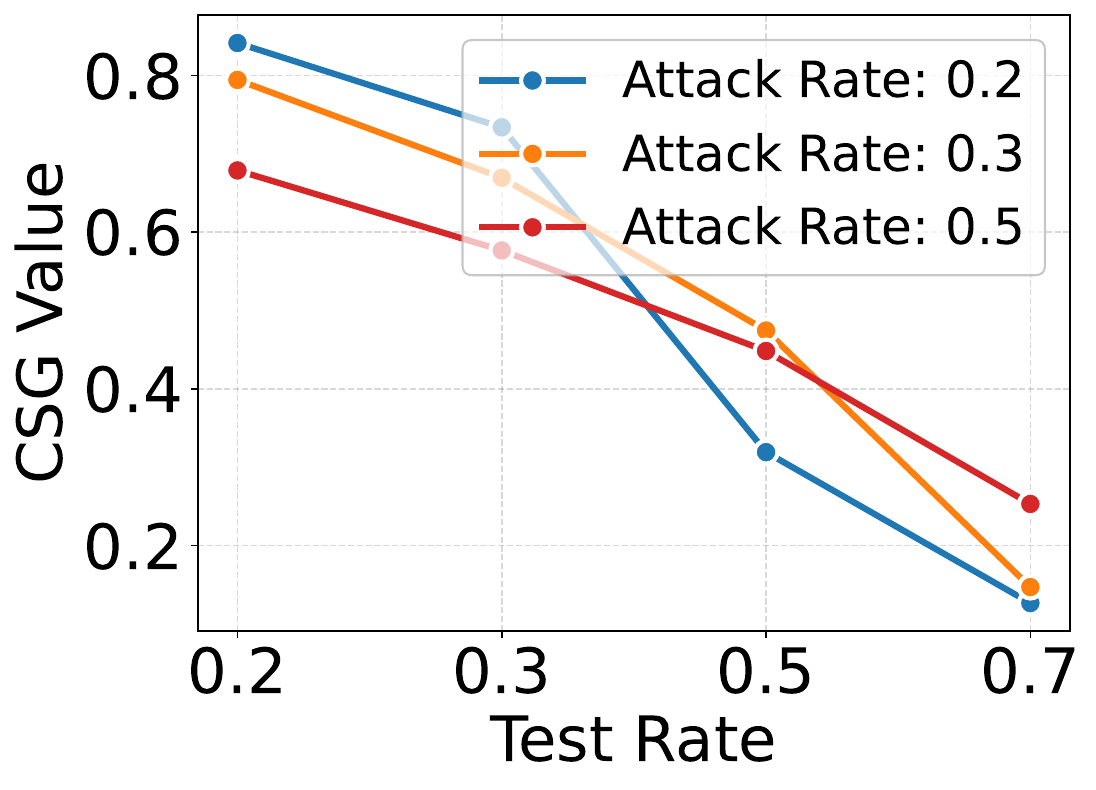} 
        \caption{Qwen-POPE}
    \end{subfigure}
    \begin{subfigure}{0.3\linewidth}
        \centering       \includegraphics[width=\linewidth]{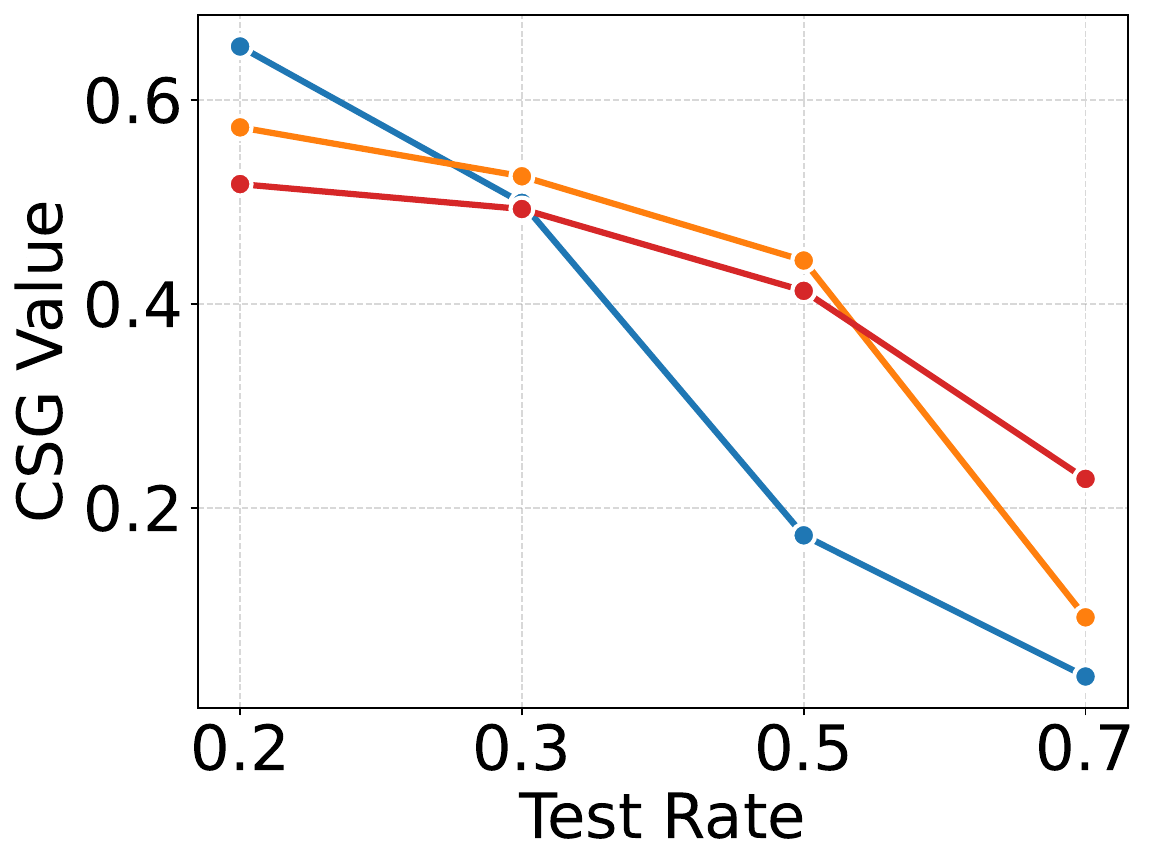}
        \caption{Qwen-TextVQA}
    \end{subfigure}
        \begin{subfigure}{0.3\linewidth}
        \centering
   \includegraphics[width=\linewidth]{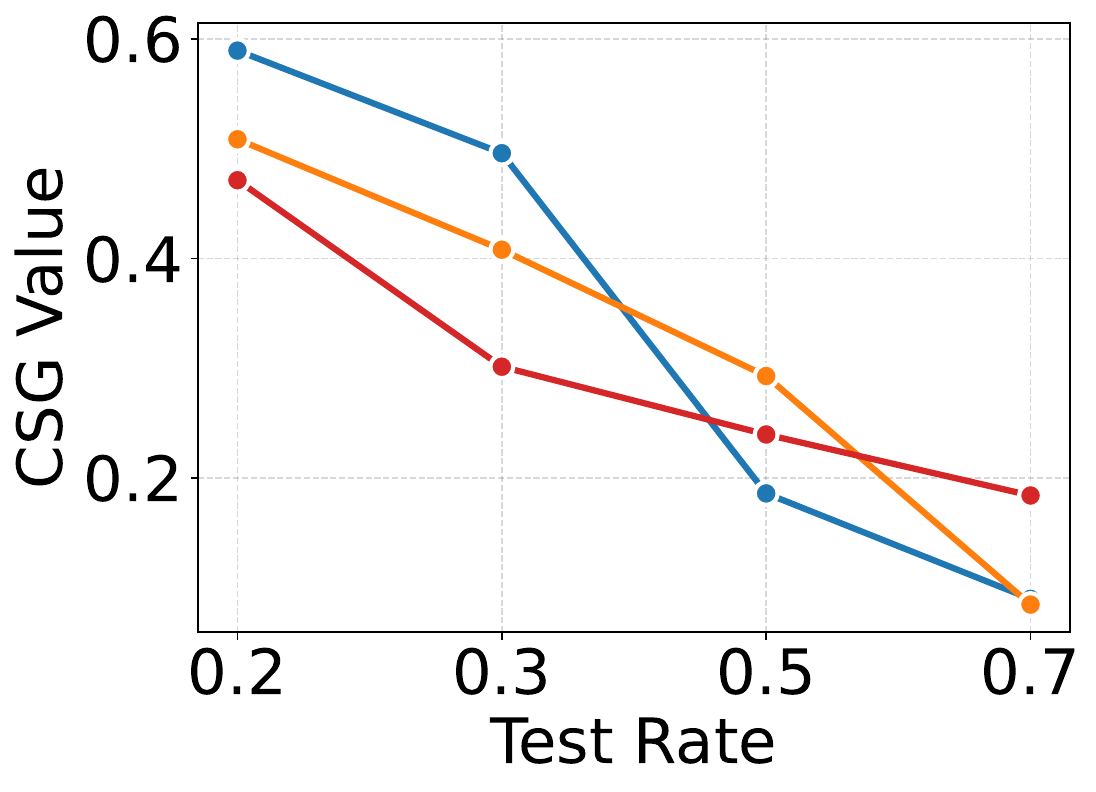} 
        \caption{Qwen-MME}
    \end{subfigure}

    \caption{Impact of mismatch between attack and test retention rates on attack effectiveness.}
    \label{fig:ab_attack_num_and_layer_Qwen}
\end{figure}

\noindent \textbf{Retention Rate Mismatch.} Fig.~\ref{fig:ab_attack_num_and_layer_Qwen} shows the attack performance when the number of least-important tokens targeted by the attack does not match the number of tokens retained by the compression mechanism. The results indicate that CAA remains highly effective when the mismatch between the two is not significant. 

\begin{figure}[t]
    \centering
    \begin{subfigure}{0.3\linewidth}
        \centering
        \includegraphics[width=\linewidth]{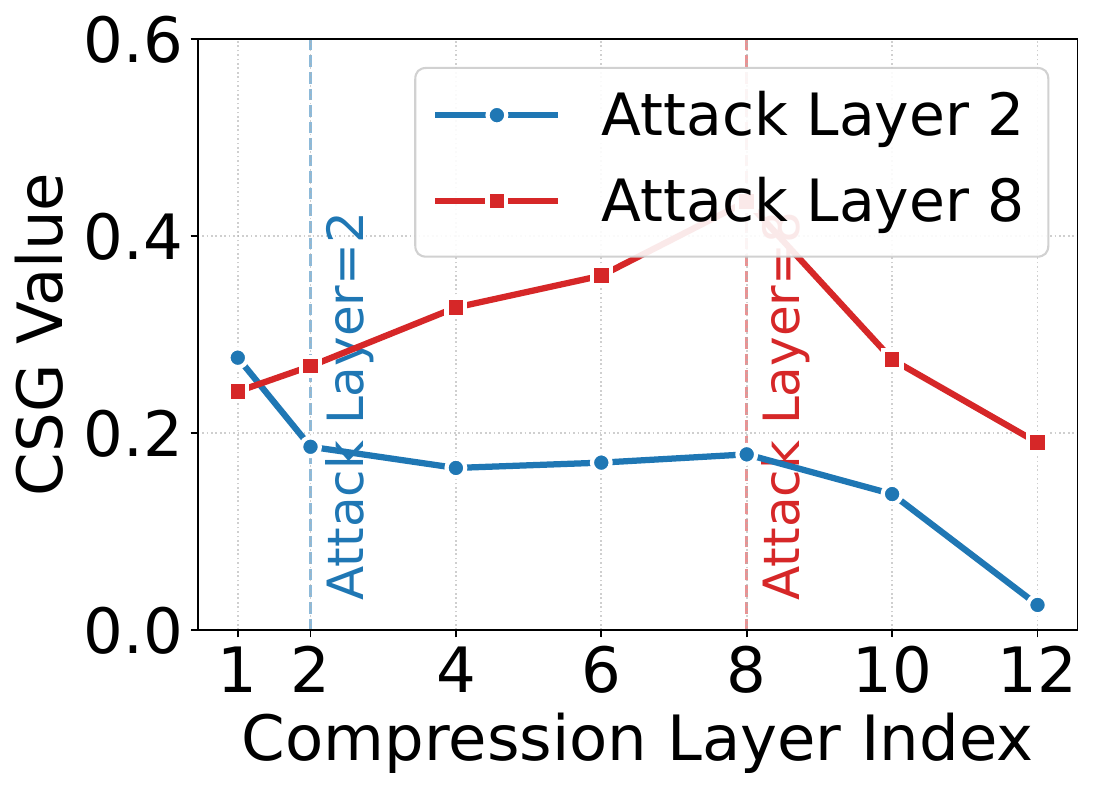}
        \caption{LLaVA-TextVQA}
    \end{subfigure}
    \begin{subfigure}{0.3\linewidth}
        \centering
        \includegraphics[width=\linewidth]{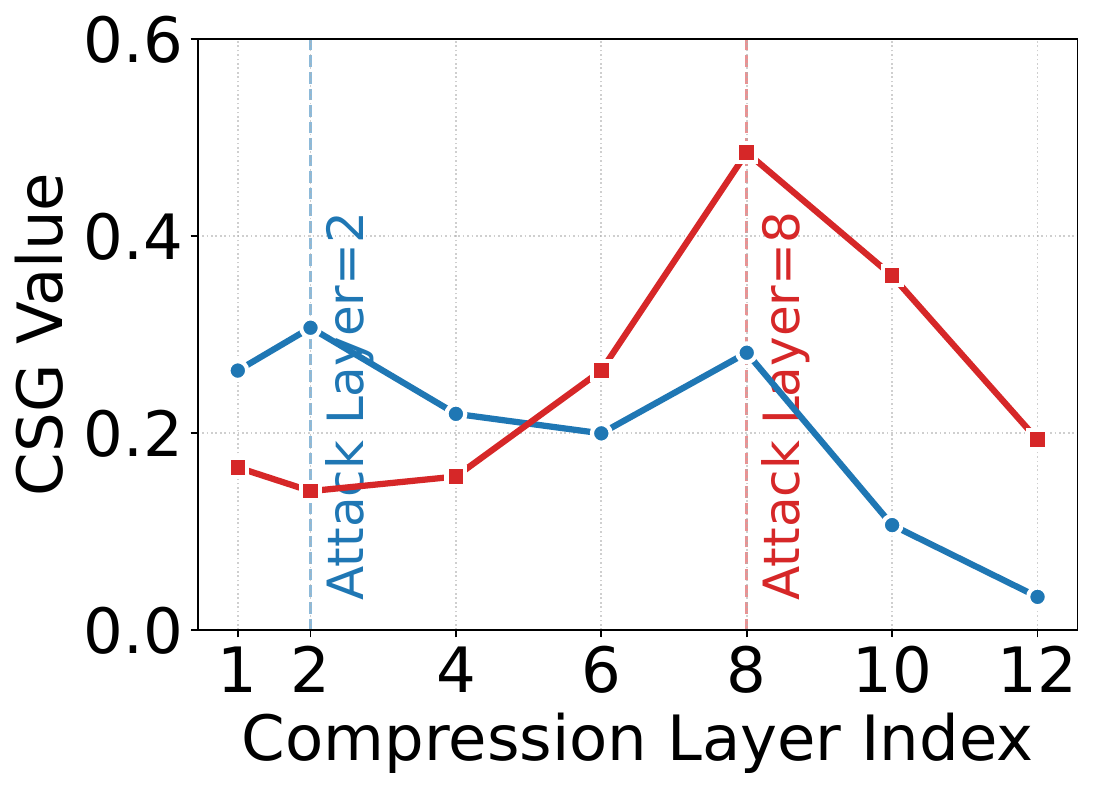}
        \caption{LLaVA-MME}
    \end{subfigure}
    
    \begin{subfigure}{0.3\linewidth}
        \centering
        \includegraphics[width=\linewidth]{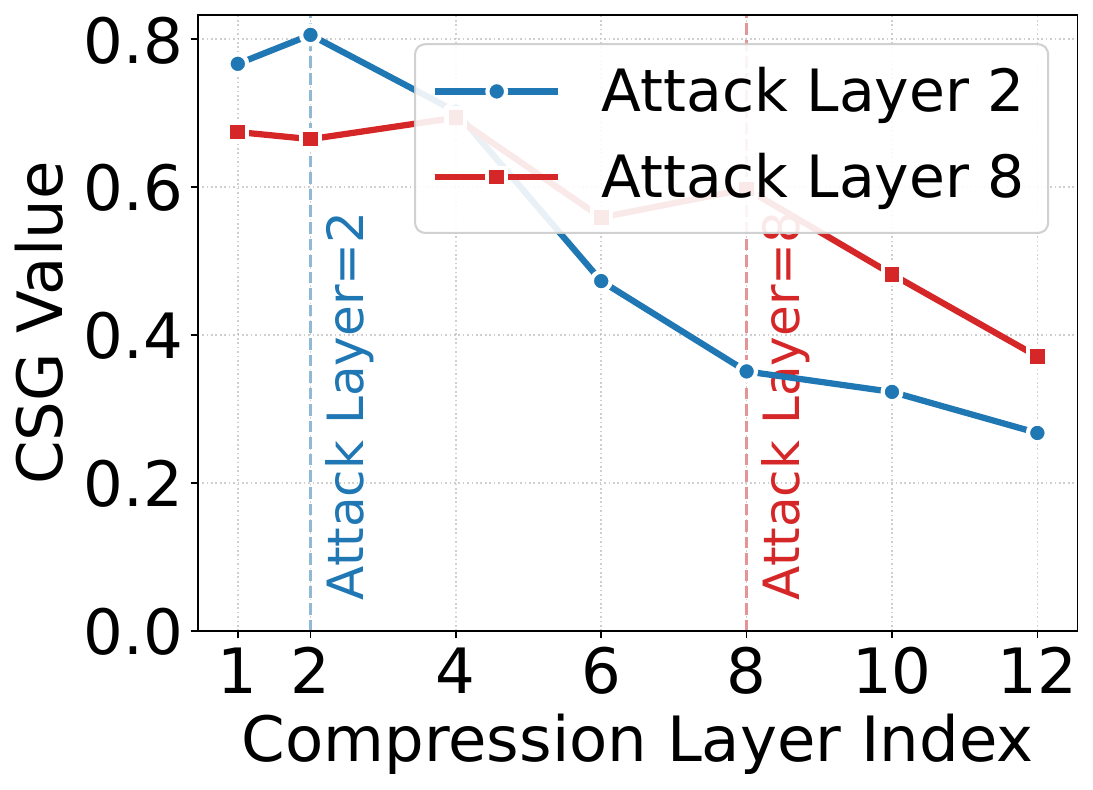} 
        \caption{Qwen-POPE}
    \end{subfigure}
    \begin{subfigure}{0.3\linewidth}
        \centering
        \includegraphics[width=\linewidth]{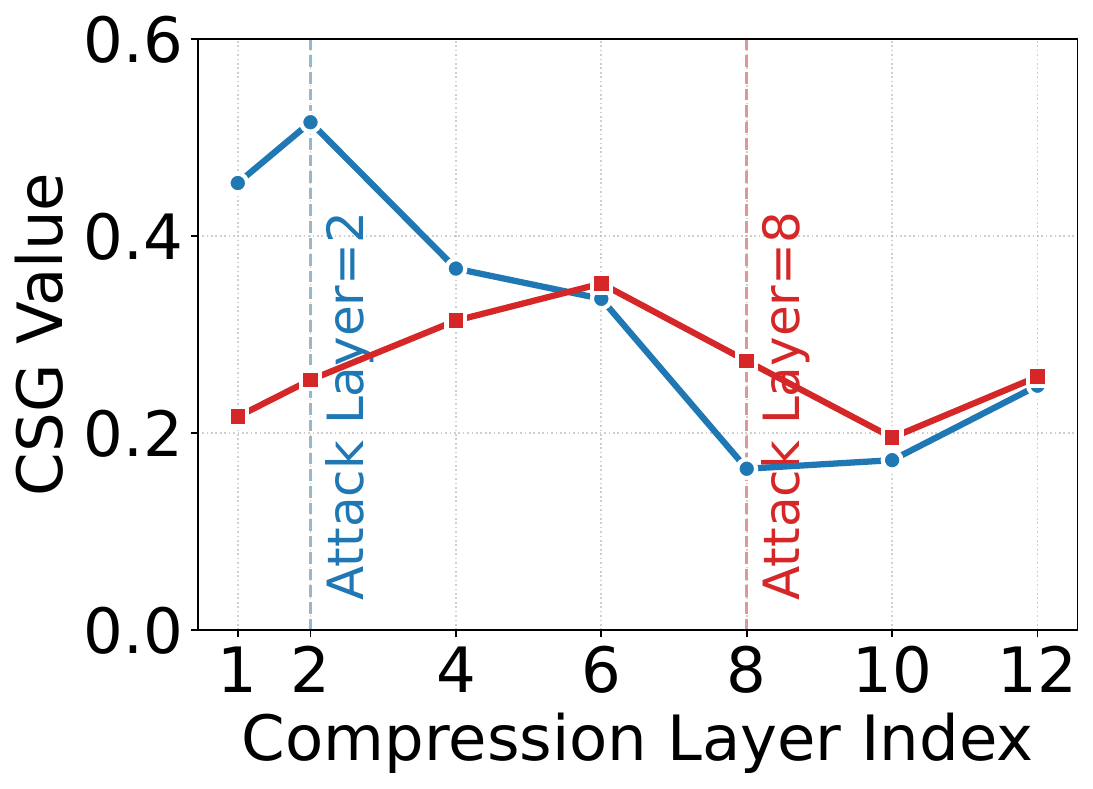}
        \caption{Qwen-TextVQA}
    \end{subfigure}
    \begin{subfigure}{0.3\linewidth}
        \centering
        \includegraphics[width=\linewidth]{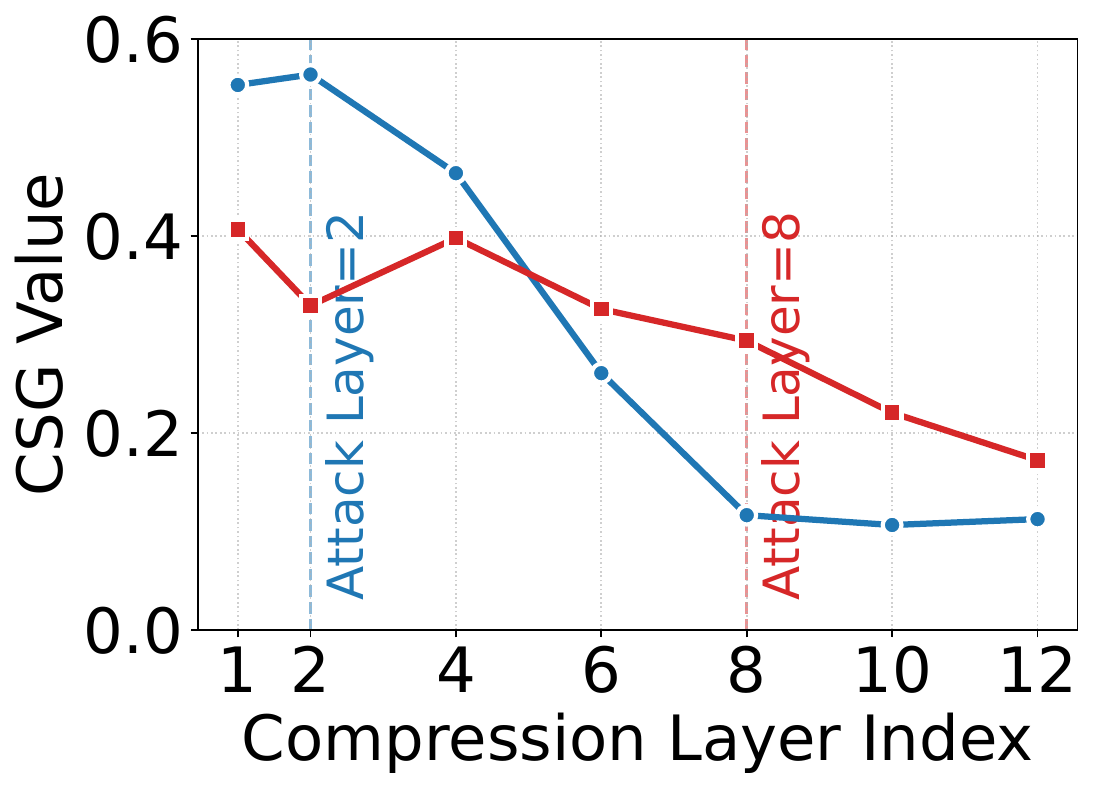}
        \caption{Qwen-MME}
    \end{subfigure}
    \caption{Attack effectiveness under mismatches between the attack layer and the compression layer.}
        \label{fig:ab_attack_num_and_layer_qwen}
\end{figure}

\noindent \textbf{Layer Mismatch.} 
Fig.~\ref{fig:ab_attack_num_and_layer_qwen} reports the attack performance when the attack layer does not align with the compression layer. The results show that CAA is robust to a certain degree of layer mismatch, and that attacks applied at later layers can still effectively influence compression applied at earlier layers.

\subsection{Impact of the Perturbation Budget}
\label{subsec:pert_budget}
\begin{figure}[t]
    \centering
    \begin{subfigure}{0.4\linewidth}
        \centering
        \includegraphics[width=\linewidth]{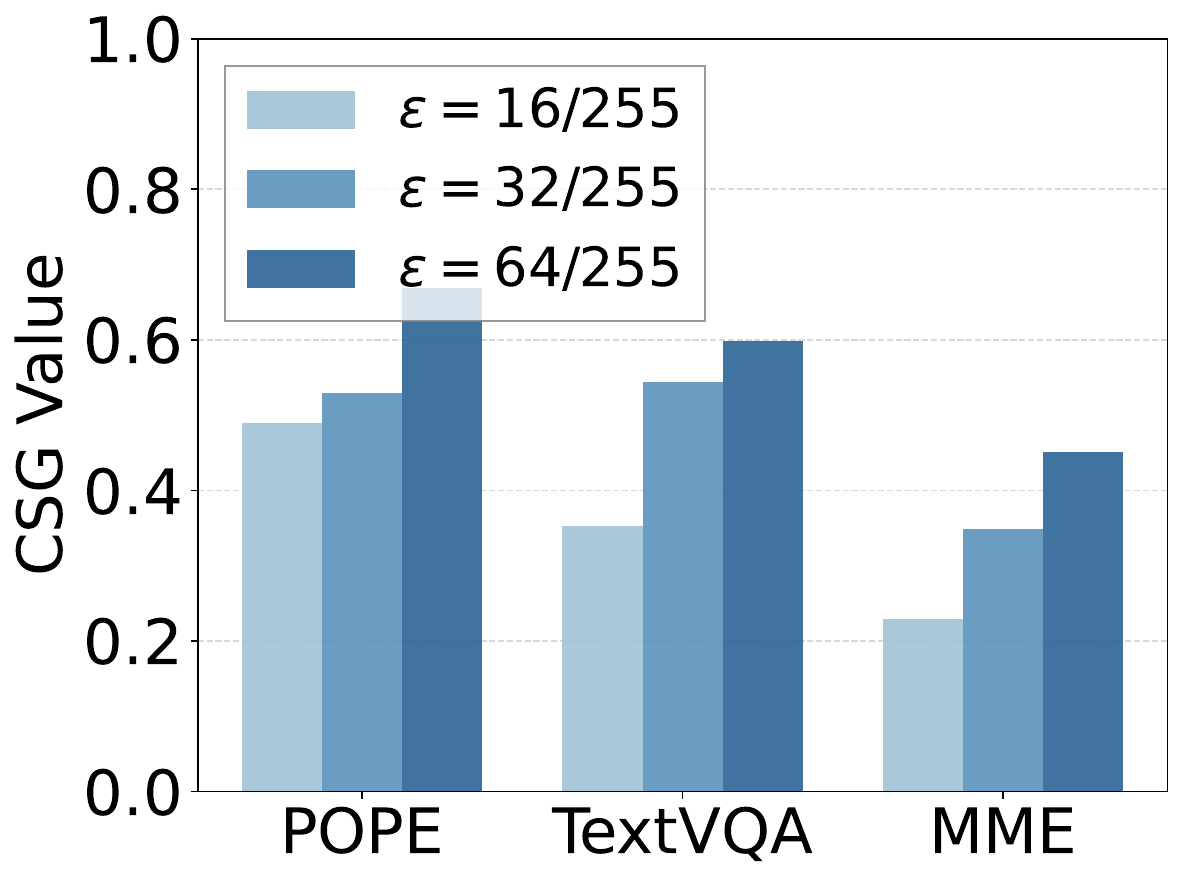} 
        \caption{LLaVA}
        \label{fig:pert_budget_llava}
    \end{subfigure}
    \begin{subfigure}{0.4\linewidth}
        \centering
        \includegraphics[width=\linewidth]{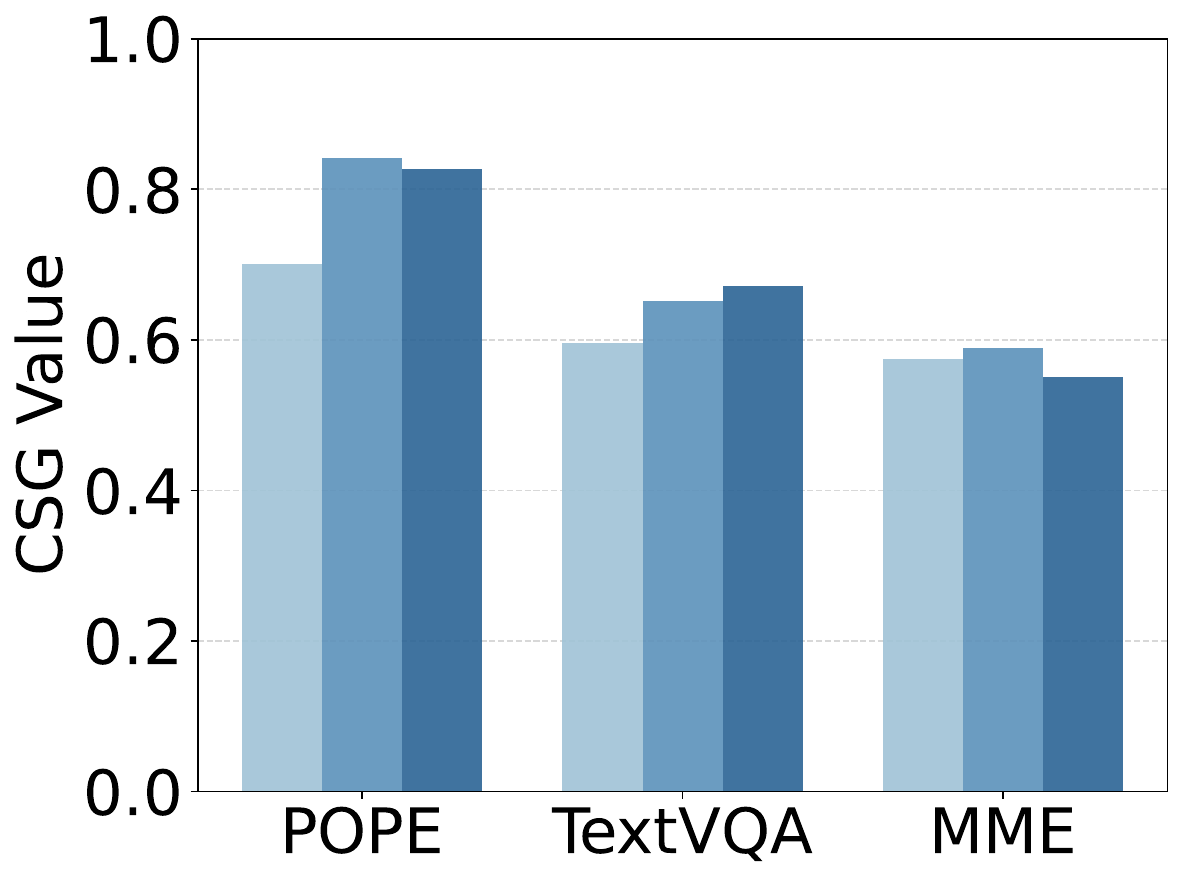}
        \caption{Qwen-VL}
        \label{fig:pert_budget_qwen}
    \end{subfigure}

    \caption{The impact of perturbation budgets on attack performance (CSG) across different datasets. 
    }
    \label{fig:pert_budget}
\end{figure}
We evaluate CAA under different perturbation budgets, with $\epsilon \in \{16/255, 32/255, 64/255\}$, and report CSG results in Fig.~\ref{fig:pert_budget}. since perturbations are restricted to a small portion of the image, the resulting visual distortion is largely imperceptible. 
As $\epsilon$ increases, attack effectiveness improves consistently across models and datasets. However, at larger budgets, perturbations begin to degrade uncompressed inference and introduce visible artifacts, leading to reduced overall CSG. This indicates a clear trade-off between attack strength and perceptual stealth, motivating the use of moderate perturbation budgets in practice.

\begin{figure}[t]
    \centering
    \begin{subfigure}{0.3\linewidth}
        \centering
        \includegraphics[width=\linewidth]{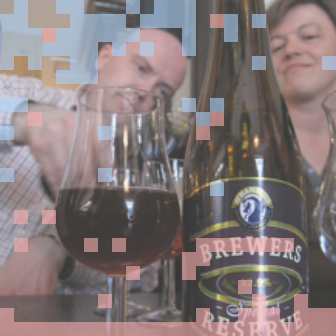} 
        \caption{LLaVA}
    \end{subfigure}
    \begin{subfigure}{0.3\linewidth}
        \centering
        \includegraphics[width=\linewidth]{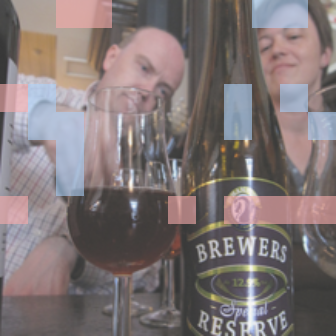}
        \caption{Qwen-VL}
    \end{subfigure}

    \caption{Visualization of the most important (red) and least important (blue) regions identified by LLaVA and Qwen-VL for the same text-visual prompt, illustrating model-dependent differences in region importance.}
    \label{fig:transer_least_example}
\end{figure}

\begin{table}[t]
  \centering
  \caption{Enumerated compression configurations ($\bar{r} \le 0.2$) with their average retention rates and corresponding (layer, rate) information.}
  \resizebox{0.95\linewidth}{!}{
    \begin{tabular}{cl|cl}
    \toprule
    Avg. Rate & Configuration (Layer, Rate) & Avg. Rate & \multicolumn{1}{l}{Configuration (Layer, Rate)} \\
    \midrule
    0.156 & [(2, 0.1)] & 0.188 & [(2, 0.2), (12, 0.1)] \\
    0.169 & [(2, 0.2), (6, 0.1)] & 0.191 & [(2, 0.2), (13, 0.1)] \\
    0.172 & [(2, 0.2), (7, 0.1)] & 0.194 & [(2, 0.2), (14, 0.1)] \\
    0.175 & [(2, 0.2), (8, 0.1)] & 0.197 & [(2, 0.2), (15, 0.1)] \\
    0.178 & [(2, 0.2), (9, 0.1)] & 0.197 & [(3, 0.2), (7, 0.1)] \\
    0.181 & [(2, 0.2), (10, 0.1)] & 0.200 & [(2, 0.2), (16, 0.1)] \\
    0.184 & [(3, 0.1)] & 0.200 & [(3, 0.2), (8, 0.1)] \\
    0.184 & [(2, 0.2), (11, 0.1)] &       &  \\
    \bottomrule
    \end{tabular}%
    }
  \label{tab:enum_example}%
\end{table}%

\begin{table}[htbp]
  \centering
  \caption{T-CAA performance across different models and compression configurations. Rate denotes the compression-layer retention rate, and Avg. Rate denotes the average per-layer visual-token retention rate.}
  \resizebox{0.99\linewidth}{!}{
    \begin{tabular}{ccc|cc|cc|cc}
    \toprule
    \multicolumn{2}{c}{\textbf{Configuration}} & \multirow{2}[2]{*}{\textbf{Avg. Rate}} & \multicolumn{2}{c|}{\textbf{LLaVA->LLaVA-NEXT}} & \multicolumn{2}{c|}{\textbf{LLaVA->Qwen-VL}} & \multicolumn{2}{c}{\textbf{Qwen-VL->LLaVA}} \\
    \textbf{Layer} & \textbf{Rate} &       & \textbf{CAE} & \textbf{CSG} & \textbf{CAE} & \textbf{CSG} & \textbf{CAE} & \textbf{CSG} \\
    \midrule
    \multirow{3}[2]{*}{4} & 0.3   & 0.388  & 0.3603  & 0.2502  & 0.4912  & 0.3516  & 0.3858  & 0.2005  \\
          & 0.2   & 0.300  & 0.4275  & 0.3174  & 0.6174  & 0.4778  & 0.4316  & 0.2463  \\
          & 0.1   & 0.213  & 0.4917  & 0.3816  & 0.7887  & 0.6491  & 0.7464  & 0.5611  \\
    \midrule
    \multirow{3}[2]{*}{6} & 0.3   & 0.300  & 0.3796  & 0.2695  & 0.3013  & 0.1617  & 0.3817  & 0.1964  \\
          & 0.2   & 0.200  & 0.4308  & 0.3207  & 0.4821  & 0.3426  & 0.4095  & 0.2242  \\
          & 0.1   & 0.100  & 0.5631  & 0.4530  & 0.5159  & 0.3763  & 0.6542  & 0.4688  \\
    \midrule
    \multirow{4}[2]{*}{2,6} & 0.5,0.25 & 0.328  & 0.2748  & 0.1647  & 0.4447  & 0.3051  & 0.3668  & 0.1814  \\
          & 0.4,0.2 & 0.275  & 0.3817  & 0.2716  & 0.4838  & 0.3442  & 0.3977  & 0.2123  \\
          & 0.3,0.15 & 0.222  & 0.4733  & 0.3632  & 0.6004  & 0.4609  & 0.4160  & 0.2306  \\
          & 0.2,0.1 & 0.169  & 0.5040  & 0.3939  & 0.6312  & 0.4916  & 0.5394  & 0.3540  \\
    \bottomrule
    \end{tabular}%
    }
  \label{tab:caa_transfer_other_conf}%
\end{table}%

\section{Transfer Attack}
\label{sec:transfer_attack_supp}

\subsection{Least Important Region Mismacth.}
\label{subsec:least_important_mismatch}
Fig.~\ref{fig:transer_least_example} shows that different models exhibit substantial disagreement in identifying the least-important regions for the same input. As a result, directly applying white-box attacks that perturb regions deemed least important by a surrogate model is suboptimal when transferred to a target model.

\subsection{Example Enumeration of Compression Configurations}
\label{subsec:enum_example}

We provide an illustrative example of compression configuration enumeration under the constraints described in Sec.~\ref{sec:exp_setup_tcaa}.
Given an estimated average retention rate $\bar{r}$, the attacker enumerates compression configurations consistent with common deployment practices used in prior work~\cite{Zhang2024SparseVLMVT, Xing2024PyramidDropAY, Chen2024AnII, shang2025llava, zhao2024stitch}. Specifically, we follow the same rate and interval constraints as summarize in Sec.~\ref{sec:exp_setup_tcaa} and enumerate all feasible configurations $(\tilde{L}, R)$ via depth-first search, where $\tilde{L}$ denotes the compression layers and $R$ denotes corresponding retention rates.
Under these constraints, average retention rates $\bar{r} \le 0.4$ yield several thousand feasible configurations. For clarity and space considerations, we report configurations with $\bar{r} \le 0.2$, shown in Table~\ref{tab:enum_example}. Our attack-effective regime is $ r \in [0.1, 0.4]$. Therefore, when $\bar{r} \le 0.2$, the selected attack layer range corresponds to layers 2–3, where the retention rate first falls within our attack-effective regime.
This example illustrates how candidate compression layer ranges can be systematically derived from an estimated $\bar{r}$.



\subsection{Main Results for Transfer Attack}
\label{subsec:transfer_attack_supp}
In the transfer setting, T-CAA jointly learns two such templates: one applied to an uninformative image border to amplify the importance of border tokens, and another applied to the original image content to mildly suppress its token importance. Together, these perturbations manipulate the relative ordering of vision-token importance, steering compression to retain uninformative tokens. 
Table~\ref{tab:caa_transfer_other_conf} reports the cross-model and cross-configuration performance of T-CAA on the POPE dataset under additional compression settings. The results show that T-CAA preserves model behavior under non-compressed inference, achieving an average Uncompressed Performance Retention (UPR) of 0.8550. At the same time, it induces severe performance degradation once compression is enabled, with an average Compressed Attack Effectiveness (CAE) of 0.4445, demonstrating the strong transferability and effectiveness of the proposed attack across models and compression configurations.

\begin{table}[t]
  \centering
  \caption{Ablation study of T-CAA components. Least-Pert perturbs least-important regions, Full-Opt optimizes the full border instead of universal templates, and T-CAA w/o Down removes the suppression template.}  \resizebox{0.95\linewidth}{!}{
    \begin{tabular}{ccccccc}
    \toprule
    \multicolumn{2}{c}{\textbf{Configuration}} & \multirow{2}[2]{*}{\textbf{Avg. Rate}} & \multirow{2}[2]{*}{\textbf{Attack }} & \multirow{2}[2]{*}{\textbf{UPR}} & \multirow{2}[2]{*}{\textbf{CAE}} & \multirow{2}[2]{*}{\textbf{CSG}} \\
    \textbf{Layer} & \textbf{Rate} &       &       &       &       &  \\
    \midrule
    \multirow{4}[2]{*}{2} & \multirow{4}[2]{*}{0.2} & \multirow{4}[2]{*}{0.250 } & Least-Pert & 0.9034  & 0.3392  & 0.2425  \\
          &       &       & Full-Opt & 0.6981  & 0.4476  & 0.1457  \\
          &       &       & T-CAA w/o Down & 0.9671  & 0.2542  & 0.2213  \\
          &       &       & T-CAA  & 0.8604  & 0.5338  & 0.3943  \\
    \midrule
    \multirow{4}[2]{*}{2,8} & \multirow{4}[2]{*}{0.3,0.15} & \multirow{4}[2]{*}{0.231 } & Least-Pert & 0.9034  & 0.3098  & 0.2131  \\
          &       &       & Full-Opt & 0.6981  & 0.5032  & 0.2013  \\
          &       &       & T-CAA w/o Down & 0.9671  & 0.1125  & 0.0796  \\
          &       &       & T-CAA  & 0.8604  & 0.4573  & 0.3177  \\
    \midrule
    \multirow{4}[2]{*}{2,8,16} & \multirow{4}[2]{*}{0.4,0.2,0.1} & \multirow{4}[2]{*}{0.238 } & Least-Pert & 0.9034  & 0.2069  & 0.1103  \\
          &       &       & Full-Opt & 0.6981  & 0.5092  & 0.2073  \\
          &       &       & T-CAA w/o Down & 0.9671  & 0.0341  & 0.0013  \\
          &       &       & T-CAA  & 0.8604  & 0.3551  & 0.2155  \\
    \bottomrule
    \end{tabular}%
    }
  \label{tab:transfer_key_component}%
\end{table}%

\subsection{Ablation Study of Key Components in T-CAA}
\label{subsec:t_ab_attack_component}

We conduct ablation studies to analyze the three key design components of T-CAA in the black-box setting. Specifically, we evaluate:  
(1) the choice of attack region (border-based vs. surrogate-selected least regions),  
(2) the use of universal perturbation templates versus direct full-border optimization, and  
(3) the necessity of the down template applied to image content.

\noindent \textbf{Impact of Attack Region Selection on
Transferability.}
To evaluate the role of the artificially introduced border, we compare T-CAA with a variant that perturbs regions identified as least important by the surrogate model (denoted as the Least-Pert strategy). As shown in Table~\ref{tab:transfer_key_component}, the least-region strategy performs substantially worse than border-based T-CAA, with CSG dropping from 0.3091 to 0.1886. 
This performance gap reveals a key limitation of surrogate-based region selection in black-box settings. 
Due to model-specific differences in importance estimation, regions deemed unimportant by the surrogate may still encode task-relevant information for the target model, which can inadvertently degrade non-compressed inference and weaken attack effectiveness under compression.

\noindent \textbf{Effect of Full-Border Optimization.}
Next, we evaluate the importance of universal templates using a variant that directly optimizes the entire border (T-CAA\textsubscript{Full-Opt}).
As shown in Table~\ref{tab:transfer_key_component} T-CAA\textsubscript{Full-Opt} achieves substantially lower average CSG than T-CAA, indicating much weaker transferability. Directly optimizing border pixels yields surrogate-specific perturbations that fail to induce consistent token-ranking shifts on target models. In contrast, universal templates capture more model-agnostic patterns, enabling more reliable transfer.

\noindent \textbf{Necessity of the Down Template.}
Finally, we assess the role of the down template by removing it from T-CAA (T-CAA\textsubscript{w/o Down}).
As shown in Table~\ref{tab:transfer_key_component}, removing the down template substantially reduces attack effectiveness on compressed models (average CAE = 0.1336), indicating that boosting border-token importance alone is insufficient for reliable transfer. Although border tokens may outrank image content on the surrogate model, cross-model variations in attention often cause target models to still favor original content. By mildly suppressing content tokens while amplifying border tokens, the down template increases the likelihood that border tokens dominate the ranking across models, leading to more robust transfer attacks.


\end{document}